# Talkin' 'Bout AI Generation:
## Copyright and the Generative-AI Supply Chain

Katherine Lee[*]
A. Feder Cooper[†]
James Grimmelmann[‡]

*"Does generative AI infringe copyright?" is an urgent question. It is also a difficult question, for two reasons. First, "generative AI" is not just one product from one company. It is a catch-all name for a massive ecosystem of loosely related technologies, including conversational text chatbots like ChatGPT, image generators like Midjourney and DALL·E, coding assistants like GitHub Copilot, and systems that compose music and create videos. Generative-AI models have different technical architectures and are trained on different kinds and sources of data using different algorithms. Some take months and cost millions of dollars to train; others can be spun up in a weekend. These models are made accessible to users in very different ways. Some are offered through paid online services; others are distributed on an open-source model that lets anyone download and modify them. These systems behave differently and raise different legal issues. We therefore need the right framework — to dig deeper than the term "generative AI" — in order to reason precisely and clearly about the different legal issues at play.*

*The second problem is that copyright law is notoriously complicated, and generative-AI systems manage to touch on a great many corners of it. They raise issues of authorship, similarity, direct and indirect liability, fair use, and licensing, among much else. These issues cannot be analyzed in isolation, because there are connections everywhere. Whether the output of a generative-AI system is fair use can depend on how its training datasets were assembled.*

---

[*]. Co-founder, The Center for Generative AI, Law, and Policy Research. All authors contributed equally to this work. We presented an earlier version of this work at the 2023 Privacy Law Scholars Conference, and discussed the issues extensively with other participants in the Generative AI + Law Workshop at the 2023 International Conference on Machine Learning. Our thanks to the organizers and participants, and to Jack M. Balkin, Aislinn Black, Miles Brundage, Christopher Callison-Burch, Nicholas Carlini, Madiha Zahrah Choksi, Christopher A. Choquette-Choo, Christopher De Sa, Fernando Delgado, Jonathan Frankle, Deep Ganguli, Daphne Ippolito, Matthew Jagielski, Gautam Kamath, Mark Lemley, David Mimno, Niloofar Mireshghallah, Milad Nasr, Pamela Samuelson, Ludwig Schubert, Andrew F. Sellars, Florian Tramèr, Kristen Vaccaro, and Luis Villa.

[†]. Co-founder, The Center for Generative AI, Law, and Policy Research.

[‡]. Tessler Family Professor of Digital and Information Law, Cornell Law School and Cornell Tech.

*Whether the creator of a generative-AI system is secondarily liable can depend on the prompts that its users supply.*

*In this Article, we aim to bring order to the chaos. To do so, we make two contributions. First, we introduce the **generative-AI supply chain**: an interconnected set of stages that transform training data (millions of pictures of cats) into generations (a new, potentially never-seen-before picture of a cat that has never existed). Breaking down generative AI into these constituent stages reveals all of the places at which companies and users make choices that have legal consequences — for copyright and beyond. Second, we specifically apply the supply-chain framing to U.S. copyright law: this framing enables us to trace the effects of upstream technical designs on downstream uses, and to assess who in these complicated sociotechnical systems bears responsibility for infringement when it happens. Because we engage so closely with the technology of generative AI, we are able to shed more light on the copyright questions. We do not give definitive answers as to who should and should not be held liable. Instead, we identify the key decisions that courts will need to make as they grapple with these issues, and point out the consequences that would likely flow from different liability regimes.*

## INTRODUCTION

Generative artificial-intelligence (i.e., "generative-AI") systems like ChatGPT, Claude, Gemini, DALL·E, and Ideogram are capable of turning a user-supplied prompt like `"give three arguments why marbury v. madison was wrongly decided"` into a persuasive essay, or `"a robot cowboy riding a rocket ship"` into a work of digital art. Their unpredictability and complexity means that they break out of existing legal categories. In particular, the fact that generative-AI systems involve training on millions of examples of human creativity means that they raise serious copyright issues. These copyright issues have not gone unnoticed. Numerous groups of plaintiffs have sued leading generative-AI companies for copyright infringement, with potential damages reaching into the billions of dollars.

This Article is an attempt to think carefully and systematically about how copyright applies to generative-AI systems. Our first contribution is to be precise about what "generative AI" is. It is not just one product from one company. Instead, "generative AI" is a catch-all name for a massive ecosystem of loosely related technologies, including conversational text chatbots like ChatGPT, image generators like Midjourney and DALL·E, coding assistants like GitHub Copilot, and systems that compose music, create videos, and suggest molecules for new medical drugs. Generative-AI models have different technical architectures and are trained on different kinds and sources of data using different algorithms. Some take months and cost millions of dollars to train; others can be spun up in a weekend. These models are also made accessible to users in very different ways. Some are offered through paid online services; others are distributed open-source,[1] such that anyone could download and modify them.

This Article takes the complexity and diversity of generative-AI systems seriously. To provide a clear framework for thinking about the different kinds of generative-AI systems and the different ways they are created and used, the Article introduces what we call the **generative-AI supply chain**: an interconnected set of stages that transform training data (millions of pictures of cats) into generations (a new and hopefully never-seen-before picture of a cat that may or may not ever have existed). Breaking down generative AI into these constituent stages reveals all of the places at which companies and users make choices that have legal consequences — for copyright and beyond.

---

[1]. The use of the term "open-source" in generative AI is quite complicated. Some models are truly open-source, in the sense that their parameters and information about the training data are publicly released. Others, which are often called "open-source" models, only release the parameters, and no information about the training data. Prior literature tends to refer to this second case as "semi-closed." Closed models are those for which neither the model parameters nor information about the training data are available. For simplicity, we will elide this nuance; it is an important detail for understanding the generative-AI supply chain, but not for our purposes here concerning copyright. *See infra* Part I.A (regarding training data, model parameters, and models). *See infra* Part I.C (regarding the generative-AI supply chain). Milad Nasr, Nicholas Carlini & Jonathan Hayase et al., Scalable Extraction of Training Data from (Production) Language Models (2023) (unpublished manuscript) (for distinguishing closed, semi-closed, and open models). Stella Biderman, Hailey Schoelkopf & Quentin Gregory Anthony et al., *Pythia: A Suite for Analyzing Large Language Models Across Training and Scaling*, *in* 2023 Proc. 40th Int'l Conf. on Mach. Learning 2397—2430 (2023); Dirk Groeneveld, Iz Beltagy & Pete Walsh et al., OLMo: Accelerating the Science of Language Models (2024) (unpublished manuscript) (for examples of open models). Hugo Touvron, Thibaut Lavril & Gautier Izacard et al., LLaMA: Open and Efficient Foundation Language Models (2023) (unpublished manuscript), https://arxiv.org/pdf/2302.13971.pdf; Hugo Touvron, Louis Martin & Kevin Stone et al., Llama 2: Open Foundation and Fine-Tuned Chat Models (2023) (unpublished manuscript), https://arxiv.org/pdf/2307.09288.pdf (for examples of semi-closed models). OpenAI, *ChatGPT: Optimizing Language Models for Dialogue*, OpenAI (Nov. 30, 2022), https://web.archive.org/web/20221130180912/https://openai.com/blog/chatgpt/ (for an example of a system that embeds a closed model).

1. The supply chain starts with **creative works**: all of the books, artwork, software, and other products of human creativity that generative AI seeks to learn from and emulate.

2. Next, works and other information must be converted into **data**: digitally encoded files in standard formats.

3. Individual items of data are useless for AI training by themselves. Instead they must be compiled into **training datasets**: vast and carefully structured collections of related data. The process requires both extensive automation and thoughtful, human-curated decision-making.

4. To create a generative-AI **model**, its creator picks a technical architecture, assembles training datasets, and then runs a training algorithm to encode features of the training data in the model. Model training is both a science and an art, and it involves massive investments of time, money, computing resources, and (often) human monitoring and intervention.

5. The model that results from this initial training process is called a "base" or "pre-trained model," because it is often just a starting point. A model can also be **fine-tuned** to improve its performance or adapt it to a specific problem domain. This process, too, involves extensive choices — and it need not be carried out by the same entity that did the initial training.

6. A model by itself is an inert artifact. It can be used only by technical experts with substantial computing resources. To make a model usable by a wider userbase, it must be **deployed** as a service: embedded in some larger software system that provides a convenient interface. ChatGPT has a conversational text-box interface that allows users to interact with a GPT model hosted on OpenAI's servers. Midjourney is deployed as a Discord bot; users request images by sending messages to it. Other services are provided as downloadable apps, or released publicly for other developers to modify and deploy themselves.

7. A deployed service can be used to **generate** outputs: new creative works that are based on statistical patterns in the training dataset but combine them in new ways. An output — or "generation" — is based on a prompt supplied by the user: an input that describes the particular features they want the output to have. Generation is the only part of the supply chain that most users see directly.

8. The supply chain does not end with generation. Both before and after they are deployed, the developers of a generative-AI system can perform **alignment** by rating prompts and generations: further adjusting the model and the system it is embedded in to better achieve users' (and their own) needs. Those needs can include safety, helpfulness, and legal compliance. In this way — as in many others — the supply chain feeds back into itself. It is not a simple cascade from data to generations. Instead, each stage is regularly adjusted to better meet the needs of the others.

Breaking down generative AI into these constituent stages reveals all of the places at which companies and users make choices that have legal consequences. In our analysis, we specifically explore the copyright consequences.

Next, the Article works systematically through the copyright analysis of these different stages. Copyright law is notoriously complicated, and generative-AI systems manage to touch on a great many corners of it. They raise issues of authorship, similarity, direct and indirect liability, fair use, and licensing, among much else. These issues cannot be analyzed in isolation, because there are connections everywhere. Whether the output of a generative-AI system is fair use can depend on how its training datasets were assembled. Whether the creator of a generative-AI system is secondarily liable can depend on the prompts that its users supply. The Article traces the effects of upstream technical designs on downstream uses, and assesses who in these complicated sociotechnical systems bears responsibility for infringement when it happens. Because we engage so closely with the technology of generative AI, we are able to shed more light on the copyright questions. We do not give definitive answers as to who should and should not be held liable. Instead, we identify the key decisions that courts will need to make as they grapple with these issues, and point out the consequences that would likely flow from different liability regimes.

The Article proceeds in three Parts. It begins (Part I) by describing the generative-AI supply chain in detail. It leads with the necessary technical background on the broader field of **machine learning** (Part I.A), and then explains how generative AI both relates to and is distinct from more traditional machine learning (Part I.B). The heart of this section (Part I.C) is a detailed, step-by-step walkthrough of the supply chain, describing what happens at each stage, the diversity of variations on the basic theme, and the design choices that the various actors must make to create and use a generative-AI system.

Part II then provides the copyright analysis. This time, we proceed in order through the doctrinal stages of a typical copyright lawsuit: starting with authorship (Part II.A), and then covering infringement (Parts II.B through II.E), secondary liability (Part II.F), defenses (Parts II.G through II.J), and remedies (Part II.K). We ask *what* might possibly be an infringing technical artifact, *who* might be an infringing actor, and *when* infringement may occur. We also include discussion of three legal regimes that are not quite copyright, but are close enough that they raise similar issues: removal of copyright management information (Part II.L), the right of publicity (Part II.M), and hot-news misappropriation (Part II.N). This is where — we hope — our choice to detail the generative-AI supply chain proves its worth. Instead of asking discrete and insular questions like "are AI models fair use?" we can consider how the fair use analysis changes as one moves up and down the supply chain. We describe how the choices made by actors at one point in the supply chain affect the copyright risks faced by others; we show how copyright compliance depends on coordinated action by parties upstream and downstream from each other.

Part III, pulls back to provide broader lessons. We first (Part III.A) describe the options courts have — from no copyright liability at all to shutting down generative AI completely. We explain why courts may be drawn to various regimes, and what the risks and instabilities of those regimes are. Then (Part III.B) we offer some thoughts for how courts should conceptualize copyright and generative-AI. We argue that copyright pervades the generative-AI supply chain, that fair use is not a silver bullet, that the ordinary business of copyright litigation will continue even in a generative-AI age, and that courts should beware of metaphors that provide too-easy answers to the genuinely hard problems before them.

## I. Machine Learning and the Generative-AI Supply Chain

There are two kinds of AI-generated content that we consider in detail in this Article: text and images.[2] The terminology associated with the technology and processes for producing these types of content is numerous, overloaded, and sometimes perplexing. So, as a first step, we provide some background on data and machine learning,[3] and we rely on these details to be precise

---

**2.** These are two of the most common types of output content (also called *modalities*). There are, nevertheless, many other modalities — audio, video, code, etc. — and there are increasingly more applications that enable users to generate them. *See infra* Part I.B.2.

**3.** *See infra* Part I.A.

about what is new (and not-so-new) in generative AI.[4] We do not aspire for completeness. Instead, we highlight important concepts and observations that enable us to pinpoint the use of specific technologies at different stages of the generative-AI supply chain.[5] Readers familiar with the technical background on generative AI should feel free to skim the first two sections in this Part. Nevertheless, we will refer back to terms that we define here throughout the remainder of the Article. Our contributions in the third section, regarding the generative-AI supply chain, are essential for legal analysis that contends with generative AI. We exercise the supply-chain framing in our later treatment of copyright implications in Part II.

### A. Background on Machine Learning

To begin, we discuss data,[6] which are the fundamental (and hotly contested) inputs to *all* machine-learning algorithms. We then provide a brief primer on the aims of machine learning, with special attention paid to how techniques used for generation differ from methods used for more familiar tasks like prediction and classification.[7]

### 1. What is data?

In the context of AI and machine learning, **data** refers to quantified entities that have been compiled, produced, or derived from information about individuals, entities, events, materials, and physical phenomena that exist in the world. For example, US Census data reflect information about individual people and households in the US at a given period of time, where the information is composed of particular chosen **features** to collect, such as age, zip code, and income. Each person represented in a US Census has their own record of features. In general, such individual records are typically called data **examples**, and the collection of all examples comprises a **dataset**.

Such quantified data exist in many formats, including raw numbers, text, audio, images, and video. All of these formats must first be converted to numerical representations so that they can be stored, processed, and interpreted by a computer and, subsequently, by machine-learning models.[8] For

---

**4.** *See infra* Part I.B.

**5.** *See infra* Part I.C.

**6.** *See infra* Part II.A.1.

**7.** *See infra* Part II.A.2.

**8.** For simple examples of different types of data formats used in machine learning, see YASER S. ABU-MOSTAFA, MALIK MAGDON-ISMAIL, AND HSUAN-TIEN LIN, LEARNING FROM DATA: A SHORT COURSE, AMLBOOK 1–3 (2012); TREVOR HASTIE, ROBERT TIBSHIRANI, AND JEROME FRIEDMAN, THE ELEMENTS OF STATISTICAL LEARNING: DATA

example, text data is often represented as **word embeddings**, which, typically, are ordered lists of numbers (i.e., **vectors**) that reflect underlying information about the words they encode.[9]  Common embedding strategies capture semantic similarity, where vectors with similar numerical representations (as measured by a chosen distance metric) reflect words with similar meanings.[10]

Needless to say, such quantified data are not identical to the entities that they reflect.  *However*, they can capture certain useful information about said entities *and* even be used interchangeably with them, as might be the case

---

MINING, INFERENCE AND PREDICTION, SPRINGER 1–6 (2009); KEVIN P. MURPHY, PROBABILISTIC MACHINE LEARNING: AN INTRODUCTION, THE MIT PRESS 2–4 (2022).

**9.** *See* MURPHY, *supra* note 8, at 26 (providing a short definition of word embeddings); *id.* at 703–10 (providing a summary of different types of popular word embeddings); Vicki Boykisi, What are embeddings? (June 2023) (unpublished manuscript), https://github.com/veekaybee/what_are_embeddings (for an accessible treatment of the history of embeddings and discussion in relation to modern-day generative-AI models); Tomas Mikolov, Kai Chen, Greg Corrado & Jeffrey Dean, *Efficient Estimation of Word Representations in Vector Space, in* 2013 INT'L CONF. ON LEARNING REPRESENTATIONS (2013) (discussing word2vec, a common neural-network-based approach for producing embeddings); Tomas Mikolov, Ilya Sutskever & Kai Chen et al., *Distributed Representations of Words and Phrases and their Compositionality, in* 26 ADVANCES NEURAL INFO. PROCESSING SYS. (2013) (for influential follow-on work to word2vec).

**10.** A neat intuition for word embeddings (that should not be taken generally, as it does not always extend to other examples) is that you can take the word embedding for `"king"` (a list of numbers) subtract the word embedding representing `"man"`, add the word embedding representing `"woman"`, and get the word embedding for `"queen"`. *See* Ekaterina Vylomova, Laura Rimell, Trevor Cohn & Timothy Baldwin, *Take and Took, Gaggle and Goose, Book and Read: Evaluating the Utility of Vector Differences for Lexical Relation Learning* 1671, *in* 1 PROC. 54TH ANN. MEETING ASS'N FOR COMPUT. LINGUISTICS 1671 (2016).  There are many ways to compute word embeddings.  A common embedding strategy that quantifies word importance involves computing word frequency (term frequency, TF) for a particular document in corpus, and scaling it by word rarity (inverse document frequency, or IDF) across documents in the corpus.  For more on TD-IDF, see generally Karen Sparck Jones, *A Statistical Interpretation of Term Specificity and Its Application in Retrieval,* 1988 DOCUMENT RETRIEVAL SYS. 132; Gerard Salton & Christopher Buckley, *Term-weighting approaches in automatic text retrieval,* 24 INFO. PROCESSING & MGMT. 513, 516 (1988). By relying strictly on frequencies, this type of embedding does not capture any semantic information in the encoded words.  More sophisticated techniques involve learning word embeddings from data.  For example, the BERT language model uses deep learning and a transformer architecture to encode word embeddings.  *See generally* Jacob Devlin, Ming-Wei Chang, Kenton Lee & Kristina Toutanova, *BERT: Pre-training of Deep Bidirectional Transformers for Language Understanding, in* 1 PROC. 2019 CONF. N. AMERICAN CHAPTER ASS'N FOR COMPUT. LINGUISTICS: HUM. LANGUAGE TECHS. 4171 (2019). *See infra* Part I.B.3a (regarding the transformer architecture).

with digital formats of film recordings.[11]  For our purposes, an item like a painting or book is not itself data; rather, it can be processed computationally to be converted into data to be used in machine-learning applications.

## 2. What is machine learning?

**Algorithms** are computational procedures, typically implemented in software.  **Machine learning** is a subfield of computing that develops and applies algorithms to learn from data.[12]  These algorithms employ mathematical tools from probability and statistics to model (hopefully useful and interesting) patterns in the data.  Maching-learning scientists and practitioners may use these algorithms for different aims.

Two types of tasks that machine learning is commonly used for are **discriminative**[13] and **generative**[14] modeling.  Discriminative modeling includes classification (is this image of a cat or a dog?) and regression (how many ice cream cones can I expect to sell if the weather is $80°F$ today?),[15] whereas generative modeling can produce content, such as images or text.[16]  We discuss this split in the next two subsections, as it is useful for understanding the machine-learning methods used in generative AI, which we will address specifically in the following section.[17]

---

**11.** For a detailed treatment of how data serves as a proxy for entities in the world, see Dylan Mulvin, Proxies: The Cultural Work of Standing In 1–33 (2021).

**12.** *See supra* Part I.A.1.

**13.** *See infra* Part I.A.2a.

**14.** *See infra* Part I.A.2b.

**15.** While the examples we provide in the Article concern classification of inputs into discrete output categories, regression tasks that involve real numbers, such as predicting housing price given a set of features, are also discriminative. The distinction ultimately hinges on the modeling choice regarding underlying probabilities. *See generally* Dan Y. Rubinstein & Trevor Hastie, *Discriminative Versus Informative Learning, in* 1997 Proc. Third Int'l Conf. on Knowledge Discovery & Data Mining (1997) (using the term "informative" instead of "generative").

**16.** This is a simplification that is sufficient for our purposes. Generative modeling does not necessarily produce new content; it estimates probability distributions from which such content can be (but does not have to be) sampled. These probabilities can be useful for applications other than content generation. For example, the BERT language model employs generative techniques and can be used to produce word embeddings, but not content intended to be consumed or enjoyed directly by a human user. *See generally* Devlin, Chang, Lee & Toutanova, *supra* note 10.

**17.** *See infra* Part I.B.

*a. Discriminative modeling*

A common analogy for machine learning in legal literature is to think of a machine-learning **model** as a mathematical function that maps inputs to outputs.[18] We will discuss later in this section how this analogy does not hold for generative modeling. Nevertheless, revisiting this analogy is instructive for highlighting how generative modeling differs from discriminative modeling, which has been historically been more prevalent in legal discourse on machine learning.[19]

Consider a machine-learning model that classifies images as either cats or dogs. This model will serve as our example for the function analogy: it takes a computer-readable version of an image as input,[20] and returns a class **label** of either `cat` or `dog` as its output. In math, there is a function $f$ that maps images $\mathcal{X}$ onto a set of possible labels $\mathcal{Y}$, and, for any particular input image $x$, the function $f$ will *always* return the same label $y$ (Figure 1).[21]

To produce such a model, one chooses a **training algorithm** that takes data and a **model architecture** as input.[22] Extending our above example of an image classifier, the data could consist of images with corresponding labels of `cat` or `dog`, and a **neural network** could be used as our model architecture in $f$ for classifying images according to those labels.[23] Similar to data, the model architecture is also composed of vectors of numbers, which are typically called **parameters** or **weights**.[24]

---

18. For example, the function $f(x) = x + 1 = y$ simply adds 1 to the input $x$ and sets that equal to the output $y$.

19. *See generally* A. Feder Cooper, Jonathan Frankle & Christopher De Sa, *Non-Determinism and the Lawlessness of Machine Learning Code*, *in* 2022 Proc. 2022 Symposium on Comput. Sci. & L. 1 (2022) (discussing the prevalence of this view).

20. e.g., two-dimensional images can be saved as a set of numbers. Typically they are formatted as a **matrix** representing pixels, where each pixel is a vector of numbers in the range 0-255 that represents combinations of *red*, *blue*, and *green* (RGB) hues.

21. $f : \mathcal{X} \mapsto \mathcal{Y}$, where $f$ is the function, $\mathcal{X}$ is the set of possible inputs, $\mathcal{Y}$ is the set of possible class labels, in this example, $\{\texttt{cat}, \texttt{dog}\}$. It is an underlying assumption of this analogy is that the function $f$ is *deterministic*, meaning that $f(x) = y$, where the same $y$ is always returned for the same $x$. *See generally* Cooper, Frankle & De Sa, *supra* note 19 (discussing this assumption in the legal literature on machine learning).

22. For a useful glossary of terms in machine learning and generative AI, see Appendix A, A. Feder Cooper, Katherine Lee, James Grimmelmann & Daphne Ippolito et al., Report of the 1st Workshop on Generative AI and Law (2023) (unpublished manuscript), https://arxiv.org/abs/2311.06477.

23. Of course, the input image could be of anything. Performing classification involves manipulating numbers under the hood — typically, linear algebra operations on vectors and matrices that contain the model parameters and the new data example. So, one could provide, for example, an image of an airplane as input, and the model would still output a classification of either `cat` or `dog`.

Figure 1: Depicting the analogy of a machine-learned model as a function, where a classifier $f$ takes an image $x$ as input and returns the class label $y =$ dog. (Image: "Arabela, The Venus of Evanston." Source: Fernando Delgado, reprinted with permission.)

Different model architectures vary widely in size and complexity, and in turn have different capabilities for encoding relationships in the data. Simpler, more traditional statistical models like linear regression have relatively few parameters, while modern-day deep neural networks can have *billions* of parameters (with *trillions* of connections between them).[25] During the execution of the training algorithm, the model architecture is trained on a subset of

---

**24**. Model architectures and training algorithms also include **hyperparameters**. Hyperparameters are parameters that traditionally are not learned; they are often set by a human. For the model, they can dictate the number of parameters, connections, and layers. For the training algorithm, they dictate properties of how training is run. For example, a hyperparameter called the "learning rate" determines how fast or slow model training should proceed. *See* A. Feder Cooper, Yucheng Lu, Jessica Zosa Forde & Christopher De Sa, *Hyperparameter Optimization Is Deceiving Us, and How to Stop It*, *in* 34 Advances Neural Info. Processing Sys. (2021). (regarding the effects of hyperparameter choices on resulting learned models, and citations therein)

**25**. Consider three current examples. First, PaLM, a language model built by Google, has 540 billion parameters. Aakanksha Chowdhery, Sharan Narang & Jacob Devlin et al., *PaLM: Scaling Language Modeling with Pathways*, 24 J. Mach. Learning Rsch. 1–113 (2023). Second, the largest Llama 2 model, a semi-closed model released by Meta, has 70 billion parameters. (Llama 2 is a **family** of models that come in different sizes. It is common today for open- and semi-closed models to come in a variety of differently sized architectures, with larger models tending to produce higher quality generations for a larger cost in compute resources. Llama 2 is just one example of such a model family.) *See supra* note 1 and accompanying text (regarding Llama 2 and the distinction between open- and semi-closed models). *See infra* Part I.B.4. Third, GLM-130B, a bilingual Chinese and English model, has 130 billion parameters. Aohan Zeng, Xiao the available data, called the **training dataset**. This **model training** typically involves running an optimization-based routine, which iteratively processes the input data to update (i.e., **train**) the model parameters.[26] After training is complete, we can evaluate the resulting model by running it on new data examples and seeing how well it classifies them as either `cat` or `dog`.[27]

The above describes a sketch of machine learning that is familiar in legal scholarship. This work has scrutinized the implications of machine-learning-based decision-making in a variety of areas, such as whether or not to interview or hire a job candidate, grant an applicant a loan,[28] or, as in the case of the infamous Northpointe COMPAS system, to predict prison recidivism.[29] These types of yes/no decision-making tasks generally fall under the heading of **discriminative** machine learning: a type of machine learning that attempts to draw boundaries in available data, and that is often used for making predictions. As we stated at the beginning of this section, discriminative machine-learning tasks typically involve classification or regression.

---

Liu & Zhengxiao Du et al., GLM-130B: An Open Bilingual Pre-trained Model (2022) (unpublished manuscript), https://arxiv.org/abs/2210.02414.

**26**. There are many different optimization methods used in deep learning. *See generally* Robin M. Schmidt, Frank Schneider & Philipp Hennig, *Descending through a Crowded Valley - Benchmarking Deep Learning Optimizers*, *in* 139 Proc. 38th Int'l Conf. on Mach. Learning 9367—9376 (2021). The most common is an optimization method called Adam (and variants thereof). *See generally* Diederik P. Kingma & Jimmy Lei Ba, *Adam: A Method for Stochastic Optimization*, *in* 2015 Int'l Conf. on Learning Representations (2015). However, optimization algorithms for machine learning, and for training generative-AI models, remains an active area of research. *See, e.g.*, Pierre Foret, Ariel Kleiner, Hossein Mobahi & Behnam Neyshabur, *Sharpness-aware Minimization for Efficiently Improving Generalization*, *in* 2021 Int'l Conf. on Learning Representations (2021); Dara Bahri & Hossein Mobahi author, *Sharpness-Aware Minimization Improves Language Model Generalization*, *in* 2022 Proc. 60th Ann. Meeting Ass'n for Comput. Linguistics (Volume 1: Long Papers) 7360—7371 (2022).

**27**. To evaluate models reliably, it is important to execute them on a **test dataset**. Test datasets are made up of reserved data examples that are not a part of training. They are ostensibly from the same *distribution* as the training data, but the model has not seen them before being evaluated. *See* Abu-Mostafa, *supra* note 8, at 39–69.

**28**. Danielle Keats Citron & Frank A. Pasquale, *The Scored Society: Due Process for Automated Predictions*, 89 Wash. L. Rev. 655 (2014).

**29**. *See generally* Jeff Larson, Surya Mattu, Lauren Kirchner & Julia Angwin, *How We Analyzed the COMPAS Recidivism Algorithm*, ProPublica (May 16, 2016), https://www.propublica.org/article/how-we-analyzed-the-compas-recidivism-algorithm/ (for the original study indicating algorithmic bias in this system).

*b. Generative modeling*

Discriminative tasks are only one type of machine-learning modeling. Another paradigm is called **generative** machine learning.[30] Whereas discriminative machine-learning problems return a *single*[31] output $y$ from a set of possible outputs $\mathcal{Y}$,[32] generative machine learning has *multiple possible reasonable outputs* for a given input to a particular generative model. For example, there are many reasonable images that match the caption: `"cat in a red and white striped hat"` (Figure 2). Similarly, a generative model for text could have many reasonable completions to the following sentence: `"In the summer, I like to go to the [blank]"`, such as: `"beach"`, `"park"`, `"pool"`, or `"mountains"`.

From this example, we start to see how the analogy of machine learning as a function, which provides a useful intuition for discriminative modeling, does not extend to generative modeling. Instead of a single output $y$ for a given input $x$, for generative modeling there are many reasonable outputs for a given input. Choosing among these possible outputs involves some *randomness*, which means different outputs could be generated when a model is run on the same input.

In more detail, generative models learn from the training data which outputs are more likely. As a result, for the sentence `"In the summer, I like to go to the [blank]"`, the word `"beach"` is a more likely com-

---

**30**. Deep generative models, such as OpenAI's CLIP (which can be used to generate text descriptions of images) , Midjourney, or Stability AI's Stable Diffusion, are not the only form of generative machine learning. Generative machine learning is often subdivided into probabilistic graphical models and deep generative models. *See generally* OpenAI, *CLIP: Connecting text and images*, OpenAI (Jan. 5, 2021), https://openai.com/research/clip (regarding OpenAI's CLIP model). *See generally Midjourney* (2023), https://midjourney.com/ (regarding Midjourney). *See generally Stable Diffusion XL*, Stability AI (2023), https://stability.ai/stablediffusion; Robin Rombach, Andreas Blattmann & Dominik Lorenz et al., *High-Resolution Image Synthesis with Latent Diffusion Models*, *in* 2022 2022 IEEE Conf. on Comput. Vision & Pattern Recognition (2022) (regarding Stable Diffusion). *See generally* Daphne Koller & Nir Friedman, Probabilistic Graphical Models: Principles and Techniques (2009) (for a canonical textbook treatment on probabilistic graphical models). *See generally* Jakub M. Tomczak, Deep Generative Modeling (2022) (for details on different techniques for generative modeling in machine learning).

**31**. These single outputs can nevertheless have differing degrees of uncertainty associated with them. *See generally* A. Feder Cooper, Katherine Lee & Madiha Choksi et al., *Arbitrariness and Prediction: The Confounding Role of Variance in Fair Classification*, *in* 2024 Proc. 38th Ann. AAAI Conf. on A.I. (2024).

**32**. In the running classification example above, every input image must be labeled either as a $y = $ `cat` or $y = $ `dog`.

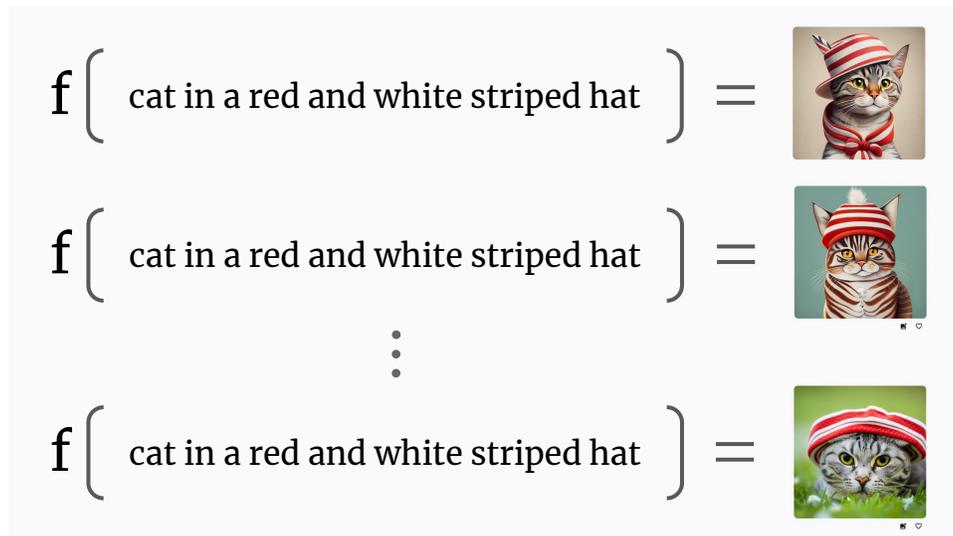

Figure 2: Images of `"cat in a red and white striped hat"` generated with Ideogram (Ideogram.AI 2023 `https://ideogram.ai/`). Running the model ($f$) multiple times on the same input can generate different outputs.

pletion than `"slopes"`. While the words `"summer"` and `"beach"` are often associated together in writing (and thus the training data), this is not the case for `"summer"` and `"slopes"`.[33] But, `"beach"` and `"pool"` might be just as likely as the other. So, the model's choice between `"beach"` and `"pool"` is made with some degree of randomness.[34]

---

**33.** The model captures the *conditional probability* of the next word $x$ *given* having already seen a prior sequence of words $a$. In the above example, we could consider the probability of the next word being $x$ = `"beach"` given that $a$ = `"In the summer, I like to go to the [blank]"`.

**34.** Discriminative and generative modeling can be related to each other mathematically. Under the hood, both approaches model *conditional probabilities*, but this observation gets abstracted away in typical discussions that analogizes discriminative models to functions. *See generally* Rubinstein & Hastie, *supra* note 15. *See also* Andrew Y. Ng & Michael I. Jordan, *On Discriminative vs. Generative Classifiers: A comparison of logistic regression and naive Bayes* p. 2, *in* 14 Advances Neural Info. Processing Sys. (2001). (describing how the two approaches can be related to each other using Bayes' rule).

### B. What Is "Generative AI"?

In the previous section, we introduced concepts and terminology concerning data[35] and machine learning[36] because they are the building blocks for technology that we refer to today as "generative AI." Given this background, we can now be more precise about what constitutes "generative AI." Generative AI makes use of technical elements that overlap with traditional machine learning, but also involves technological innovations, which we introduce in this section, to power familiar generative-AI applications like OpenAI's ChatGPT[37] and Stability AI's DreamStudio.[38]

Generative-AI models can take in a variety of inputs, typically expressive content like text or an image, and can produce expressive content as their outputs. The inputs are often (though do not have to be) user-generated;[39] this is why a user of an application like ChatGPT or DreamStudio is said to provide a **prompt**, for which the application produces an output content **generation** in response.

With the exception of a few new terms, our description of generative AI sounds a lot like our discussion of generative modeling in more traditional machine learning.[40] Indeed, contemporary generative AI does involve generative modeling, including some traditional generative modeling techniques, but it also involves a lot more.

In the remainder of this section, we unpack four ways that generative AI is different and new. First, contemporary generative AI often involves multiple models, which rely on a mixture of training algorithms and modeling

---

**35.** *See supra* Part I.A.1.

**36.** *See supra* Part I.A.2.

**37.** OpenAI, *supra* note 1.

**38.** *DreamStudio* (2023), https://dreamstudio.ai/.

**39.** Synthetic data, rather than user-generated data, can also be supplied as inputs, both for prompting and as training data. Producing and leveraging synthetic data (that has been produced by generative-AI) for prompting and training is an active area of machine-learning research. *See, e.g.,* Aaron Gokaslan, A. Feder Cooper & Jasmine Collins et al., CommonCanvas: An Open Diffusion Model Trained with Creative-Commons Images (2023) (unpublished manuscript), https://arxiv.org/abs/2310.16825 (using generative-AI produced text captions to train a text-to-image diffusion model); Liang Wang, Nan Yang & Xiaolong Huang et al., Improving Text Embeddings with Large Language Models (2023) (unpublished manuscript) (producing text embeddings using only synthetic data); Avi Singh, John D. Co-Reyes & Rishabh Agarwal et al., Beyond Human Data: Scaling Self-Training for Problem-Solving with Language Models (2023) (unpublished manuscript) (generating synthetic fine-tuning data). *See infra* Part I.C.8.

**40.** One example of a generative model in that section, illustrated in Figure 2, takes the text input `"cat in a red and white striped hat"` and produces several reasonable images as output. *See supra* Part I.A.2b.

approaches.[41] These models are embedded within larger *systems*. For this reason, we discuss how it is often more appropriate to think of generative AI with respect to an overarching system, rather than in terms of a specific model.[42] Second, we focus our attention on systems that involve text and image data, in order to explain how the generative models in these systems are trained on web-scraped datasets of previously unprecedented scale.[43] Third, we describe recent technological developments — namely, the transformer architecture and diffusion-based modeling — that have contributed to the improved quality of generative models.[44] Last, we emphasize that each of these three observations have an overlapping theme: scale. Generative AI involves large-scale systems and the training of massive models on similarly massive datasets. Scale stands on its own as another reason why generative AI is different from more traditional generative modeling.[45]

The sections that follow rely heavily on the material we present here. Later, we will discuss how the process of development, evaluation, deployment, and evolution of generative-AI systems is best conceived of as a complex supply chain, composed of different stages and involving various people and organizations.[46] The supply-chain lens, in turn, is indispensable for analyzing potential legal issues. We will show this by turning our attention to copyright.[47]

## 1. Generative-AI Systems

Most users of generative AI do not interact with a model directly. Instead, they use an interface to a *system*, in which the model is just one of several embedded, inter-operating components.[48] For example, OpenAI hosts various

---

41. Historically, practitioners typically would have chosen to solve a particular problem with a particular modeling technique. For example, they would take either a discriminative or generative modeling approach, or use another modeling paradigm called **reinforcement learning**. Abu-Mostafa, *supra* note 8, at 11–14; Murphy, *supra* note 8, at 1–19 (for an intuition behind reinforcement learning). We introduce this concept in more detail when discuss **model alignment** in the generative AI supply chain. *See infra* Part I.C.7. Generative AI can involve all of these approaches.

42. *See infra* Part I.B.1.
43. *See infra* Part I.B.2.
44. *See infra* Part I.B.3.
45. *See infra* Part I.B.4.
46. *See infra* Part I.C.
47. *See infra* Part II.
48. *See generally* A. Feder Cooper & Karen Levy, *Fast or Accurate? Governing Conflicting Goals in Highly Autonomous Vehicles*, 20 Colo. Tech. L.J. 249 (2022). *See* A. Feder Cooper, Karen Levy & Christopher De Sa, *Accuracy-Efficiency Trade-Offs and Accountability in Distributed ML Systems* pp. 1–2, *in* 2021 Equity & Access Algorithms ways to access its latest GPT models. ChatGPT is a user interface, where the priced version is currently built on top of the GPT-4 model architecture.[49] OpenAI also has a developer API, which serves as an interface for programmers to access different models using code. There are additional components behind each of these interfaces, including possibly (according to rumor) as many as sixteen GPT-4 models, to which different prompts are routed.[50] As another example, consider Stable Diffusion, an open-source model for producing image generations.[51] Most users do not interact directly with the Stable Diffusion model;[52] rather, they typically access a version that is embedded in a larger system operated by Stability AI,[53] which has multiple components, including a web-based application called DreamStudio.[54]

In this Article, we focus on generative-AI systems, rather than just generative-AI models, to highlight how models are only one (however, important) component of an entire system. This focus is particularly important when we introduce our framing of the generative-AI supply chain.[55]

## 2. Generation Modalities

The input and output content types for generative-AI models are often referred to as **modalities**. For example, a chatbot that produces *text* generations when given a user-provided *text* prompt would use an underlying **text-to-text** model; this model operates in the text modality. Such a chatbot uses the same modality for the input and output, but this is not a requirement for generative AI more broadly. Many image-generation models (used in

---

Mechanisms & Optimization 1 (2021) (discussing the importance of such a systems framing in contemporary computing applications). OpenAI also emphasizes this point in their policy research work. For example, OpenAI has produced a GPT-4 *system* card (emphasis added), and this point was made at the *GenLaw 2023* workshop by Miles Brundage in his talk "Where and when does the law fit into AI development and deployment?." *See generally* OpenAI, GPT-4 System Card (Mar. 23, 2023) (unpublished manuscript), https://cdn.openai.com/papers/gpt-4-system-card.pdf (emphasizing systems, whcih contain models and other components).

**49.** OpenAI, *supra* note 1.

**50.** This rumor originated in a Twitter post. Maximilian Schreiner, *GPT-4 architecture, datasets, costs and more leaked*, THE DECODER (July 11, 2023), https://the-decoder.com/gpt-4-architecture-datasets-costs-and-more-leaked/.

**51.** Rombach, Blattmann & Lorenz et al., *supra* note 30.

**52.** At a minimum, using the model directly would involve downloading the model parameters, writing code to run the model, and executing that code.

**53.** *Stable Diffusion XL, supra* note 30.

**54.** *DreamStudio, supra* note 38.

**55.** *See infra* Part I.C.

systems like Stable Diffusion[56] DALL·E-2,[57] and ChatGPT Plus and Enterprise[58] etc.) take a text description as input and produce an image generation as output. These models are **multimodal**, **text-to-image** models.

Generative AI models are trained on data in both their input and output modalities. Throughout this Article, we focus on text and image modalities: systems that predominantly take text as input, and either produce text or image generations. In this section, we detail some popular systems that involve text[59] and image[60] training data and generations, and briefly describe other modalities[61] for which generative-AI technology is being put to use.

### a. Text data and generations

At time of writing, ChatGPT is a system that takes in text inputs and (usually)[62] produces text outputs. ChatGPT is built on top of multiple models, including several different text-to-text model architectures trained on massive amounts of text data, e.g., GPT-3.5 and GPT-4.[63] During training, each of these text-to-text models is shown text sequences and, for every sequence, it is trained to predict the next word given all of the previous words. For example, if the sentence `"In the summer, I like to go to the beach"` were in the training data, then the model would first be shown `"In"` and trained to predict `"the"`, then given `"In the"` and trained to predict `"summer"`, and so on.

Text data is in many ways easier to collect than other modalities[64] because it is readily available on the Internet. Common data sources include data scraped from the web, books (both copyrighted and in the public domain),

---

**56.** *See* Rombach, Blattmann & Lorenz et al., *supra* note 30 (describing the model); *Stable Diffusion XL*, *supra* note 30 (describing the product).

**57.** *See* Aditya Ramesh, Prafulla Dhariwal & Alex Nichol et al., Hierarchical Text-Conditional Image Generation with CLIP Latents (2022) (unpublished manuscript), https://arxiv.org/abs/2204.06125 (describing the model); *DALL·E 2*, OᴘᴇɴAI (2022), https://openai.com/dall-e-2 (describing the product).

**58.** *See DALL·E 3 is now available in ChatGPT Plus and Enterprise*, OᴘᴇɴAI (Oct. 19, 2023), https://openai.com/blog/dall-e-3-is-now-available-in-chatgpt-plus-and-enterprise (announcing the integration of DALL·E-3 text-to-image functionality into the paid versions of the ChatGPT chatbot system).

**59.** *See infra* Part I.B.2a.

**60.** *See infra* Part I.B.2b.

**61.** *See infra* Part I.B.2c.

**62.** *See DALL·E 3 is now available in ChatGPT Plus and Enterprise*, *supra* note 58. *See infra* Part I.B.2b.

**63.** OpenAI, *supra* note 1.

**64.** E.g., music. *See infra* Part I.B.2c.

and news articles,[65] as well as data produced through user interactions with a product.[66] Web data may include structured text like product reviews, and free-form social-media posts and blogs.[67]

It is important to note that generative text models are used extensively beyond chatbot systems like ChatGPT.[68] For example, generative text models also play an important role in translation systems[69] and in scientific appli-

---

**65**. *See* Katherine Lee, Daphne Ippolito & A. Feder Cooper, The Devil is in the Training Data (2023) (unpublished manuscript), *in* Katherine Lee, A. Feder Cooper, James Grimmelmann & Daphne Ippolito, AI and Law: The Next Generation 5 (2023) (unpublished manuscript), https://www.researchgate.net/profile/A-Cooper-2/publication/372251056_AI_and_Law_The_Next_Generation_An_explainer_series/links/64ad12b7b9ed6874a51152ec/AI-and-Law-The-Next-Generation-An-explainer-series.pdf (discussing training data sources). *See generally* Tom B. Brown, Benjamin Mann & Nick Ryder et al., Language Models are Few-Shot Learners (2020) (unpublished manuscript), https://arxiv.org/abs/2005.14165; Leo Gao, Stella Biderman & Sid Black et al., The Pile: An 800GB Dataset of Diverse Text for Language Modeling (2021) (unpublished manuscript), https://arxiv.org/abs/2101.00027; Colin Raffel, Noam Shazeer, Adam Roberts & Katherine Lee et al., *Exploring the Limits of Transfer Learning with a Unified Text-to-Text Transformer*, 21 J. Mach. Learning Rsch. 1 (2020).

**66**. For example, it is widely believed (though unconfirmed) that user data ingested by the ChatGPT interface is used to train the underlying model(s). *See New Ways to Manage Your Data in ChatGPT*, OpenAI (2023), https://openai.com/blog/new-ways-to-manage-your-data-in-chatgpt (describing only the cases in which user data is *not* used to train the ChatGPT system).

**67**. *See* Kevin Schaul, Szu Yu Chen & Nitasha Tiku, *Inside the secret list of websites that make AI like ChatGPT sound smart*, Washington Post (Apr. 19, 2023), https://www.washingtonpost.com/technology/interactive/2023/ai-chatbot-learning/. *See generally* Jesse Dodge, Maarten Sap & Ana Marasović et al., *Documenting Large Webtext Corpora: A Case Study on the Colossal Clean Crawled Corpus*, *in* 2021 Proc. 2021 Conf. on Empirical Methods Nat. Language Processing 1286 (2021) (a paper from the same researchers).

**68**. *See* Alec Radford, Karthik Narasimhan, Tim Salimans & Ilya Sutskever, Improving Language Understanding by Generative Pre-training (2018) (unpublished manuscript), https://cdn.openai.com/research-covers/language-unsupervised/language_understanding_paper.pdf; Alec Radford, Jeffrey Wu & Rewon Child et al., Language Models are Unsupervised Multitask Learners (2019) (unpublished manuscript), https://d4mucfpksywv.cloudfront.net/better-language-models/language_models_are_unsupervised_multitask_learners.pdf; Raffel, Shazeer, Roberts & Lee et al., *supra* note 65; Devlin, Chang, Lee & Toutanova, *supra* note 10 (which all use generative text models to perform a variety of text tasks including translation, question answering, summarization, and text classification).

**69**. e.g., Google Translate uses generative AI to produce translated text given an input in another language. *See* Isaac Caswell, *Bowen Liang, Recent Advances in Google Translate*, Google Rsch. (June 8, 2020), https://blog.research.google/2020/06/recent-advances-in-google-translate.html. (describing the Google Translate system in 2020, which uses cations.[70]  The training data for these different types of applications tend to differ according to use case, e.g., translation-model training datasets include information from multiple languages, and chat-model training datasets include dialog.[71]

### b. *Image data and generations*

We also consider examples of image data and generations in the context of multimodal text-to-image[72] systems like DALL·E,[73] DALL·E-2,[74] DALL·E-3 (which is embedded within ChatGPT[75]),[76] Ideogram[77] Midjourney,[78] and Stability AI's DreamStudio[79] (built on top of Stable Diffusion[80]).  The generative-AI models in these systems are trained on huge amounts of image-text pairs, where the text is a caption that describes the image.  Similar to the collection of text data, described above, these datasets are also often scraped from the Internet, and can include both copyrighted and public-domain im-

---

a transformer model in conjunction with another type of model called a Recurrent Neural Network).

**70**. *See infra* Part I.B.2c.

**71**. *See generally* Romal Thoppilan, Daniel De Freitas & Jamie Hall et al., LaMDA: Language Models for Dialog Applications (2022) (unpublished manuscript), https://arxiv.org/pdf/2201.08239.pdf (discussing the inclusion of dialogue in the training of a chat model).

**72**. There are also unimodal image-to-image models and systems, like the one owned and operated by Runway.  *See* Runway, *Image to Image* (2023), https://runwayml.com/ai-magic-tools/image-to-image/.

**73**. *See generally* Aditya Ramesh, Mikhail Pavlov & Gabriel Goh et al., *Zero-Shot Text-to-Image Generation*, *in* 2021 Proc. 38th Int'l Conf. on Mach. Learning 8821 (2021) (the original DALL·E model paper); Alec Radford, Jong Wook Kim & Chris Hallacy et al., *Learning Transferable Visual Models From Natural Language Supervision*, *in* 2021 Proc. 38th Int'l Conf. on Mach. Learning 8748 (2021) (the critic model used to rank DALL·E generation outputs for a given prompt).  Both components are part of the OpenAI DALL·E system.  *See generally* OpenAI, *DALL·E: Creating images from text*, OpenAI (Jan. 5, 2021), https://openai.com/research/dall-e.

**74**. *See generally* Ramesh, Dhariwal & Nichol et al., *supra* note 57 (the original DALL·E-2 research paper); *DALL·E 2*, *supra* note 57 (the DALL·E-2 OpenAI system).

**75**. *DALL·E 3 is now available in ChatGPT Plus and Enterprise*, *supra* note 58.

**76**. *See generally* James Betker, Gabriel Goh & Li Jing et al., Improving Image Generation with Better Captions (2023) (unpublished manuscript), https://cdn.openai.com/papers/dall-e-3.pdf (the original DALL·E-3 research paper); OpenAI, *DALL·E 3* (2023), https://openai.com/dall-e-3 (describing the functionality of DALL·E-3 in OpenAI products).

**77**. *See generally Ideogram.AI*, Ideogram.AI (2023), https://ideogram.ai/.

**78**. *Midjourney*, *supra* note 30.

**79**. *DreamStudio*, *supra* note 38.

**80**. Rombach, Blattmann & Lorenz et al., *supra* note 30.

ages and captions.[81]  In some cases, only the images are scraped from the Internet, and the corresponding captions are produced using machine learning[82] — for example, using an image-to-text generative-AI model or system to produce synthetic captions.[83]

The text-to-image models trained on these datasets can use different underlying architectures and training processes, which we discuss below.[84]  Nevertheless, regardless of the specific implementation, model training serves to find relationships between the text and images in the training data.  Trained models leverage these learned relationships at generation time:  when supplied with a text prompt as input, they generate image outputs to match the prompt.[85]  Today's text-to-image models can produce generations that span a variety of artistic styles — from cartoons to photorealistic images — and

---

**81.** It is possible for one item in the pair to be copyrighted and the other to be in the public domain, such as a copyrighted image with a public-domain caption. *See infra* Part II.B.

**82.** We again refer to Katherine Lee, Daphne Ippolito & A. Feder Cooper, The Devil is in the Training Data (2023) (unpublished manuscript), *in* Lee, Cooper, Grimmelmann & Ippolito, *supra* note 65, at 5. (discussing training data sources). One common source is LAION-5B, a dataset constructed from images and alt-text from the Common Crawl corpus. *See generally* Romain Beaumont, *LAION-5B: A New Era of Large-Scale Multi-Modal Datasets*, LAION (Mar. 31, 2022), https://laion.ai/blog/laion-5b/ (describing the LAION-5B dataset). *See generally* Christoph Schuhmann, Romain Beaumont & Richard Vencu et al., *LAION-5B: An open large-scale dataset for training next generation image-text models*, *in* 2022 Thirty-sixth Conf. on Neural Info. Processing Sys. Datasets & Benchmarks Track (2022) (for the *NeurIPS* conference datasets track paper on LAION-5B). *See generally* Dodge, Sap & Marasović et al., *supra* note 67 (regarding the Common Crawl corpus; also see citations therein). LAION-5B was recently removed from HuggingFace and other public hosting services due to identification of CSAM in images at its linked URLs. *See generally* Abeba Birhane, Vinay Uday Prabhu & Emmanuel Kahembwe, Multimodal datasets: misogyny, pornography, and malignant stereotypes (2021) (unpublished manuscript) (for one of the first studies documenting pornography in LAION-linked images). *See generally* Emilia David, *AI image training dataset found to include child sexual abuse imagery*, The Verge, Dec. 20, 2023, https://www.theverge.com/2023/12/20/24009418/generative-ai-image-laion-csam-google-stability-stanford (regarding the Stanford study that prompted LAION's removal).

**83.** *See* Junnan Li, Dongxu Li, Silvio Savarese & Steven Hoi, BLIP-2: Bootstrapping Language-Image Pre-training with Frozen Image Encoders and Large Language Models (2023) (unpublished manuscript), https://arxiv.org/abs/2301.12597 (for the BLIP-2 model, which can be used for synthetic captioning of images); Gokaslan, Cooper & Collins et al., *supra* note 39 (for an example application of caption generation using BLIP-2).

**84.** Some models use diffusion, like Stable Diffusion. Other models mix transformer-based architectures and diffusion techniques for different parts of training, like DALL·E-2. *See infra* Part I.B.3b.

**85.** Of course, many such generations can match the prompt; there are multiple reasonable outputs for the same input. *See supra* Part I.A.2b. Some generative-AI systems include can incorporate different abstract concepts and concrete elements. For an example of such a generation, see Figure 2.

### c. Other modalities

While we focus on generative-AI systems that involve text and image inputs and outputs, there are many other modalities which generative AI can be applied to, such as computer code, audio (music), video, and molecular structures. Text-to-code models, which are designed specifically take in natural language as input and generate code snippets as output,[86] include OpenAI Codex[87] and Code Llama[88] from Meta.[89] Notably, Codex is the generative-AI model embedded in the GitHub Copilot system,[90] which is named in active lawsuits regarding copyright infringement.[91] Google Deep-

---

models that rank match quality. *See generally* Radford, Kim & Hallacy et al., *supra* note 73 (discussing the CLIP-model-based ranking methodology used in DALL·E).

**86.** ChatGPT can also produce code snippets, but is a chatbot system with other functionality. *See* OpenAI, *supra* note 1.

**87.** Wojciech Zaremba, *Greg Brockman, and OpenAI, OpenAI Codex*, Open AI (Aug. 10, 2021), https://openai.com/blog/openai-codex. (describing the Codex model). OpenAI, *Powering next generation applications with OpenAI Codex*, Open AI (May 24, 2022), https://openai.com/blog/codex-apps. (discussing applications using Codex). Mark Chen, Jerry Tworek & Heewoo Jun et al., Evaluating Large Language Models Trained on Code (2021) (unpublished manuscript), https://arxiv.org/abs/2107.03374. (for the technical report detailing the original Codex model).

**88.** Meta, *Introducing Code Llama, an AI Tool for Coding*, Meta News (Aug. 24, 2023), https://about.fb.com/news/2023/08/code-llama-ai-for-coding/. (announcing Code Llama). Meta, *Introducing Code Llama, a state-of-the-art large language model for coding*, Meta Rsch. Blog (Aug. 24, 2023), https://ai.meta.com/blog/code-llama-large-language-model-coding/. (describing Code Llama in a technical blog post). Baptiste Rozière, Jonas Gehring & Fabian Gloeckle et al., Code Llama: Open Foundation Models for Code (2023) (unpublished manuscript), https://arxiv.org/abs/2308.12950. (for the technical report detailing the Code Llama model).

**89.** Both of these models use transformer-based architectures. *See infra* Part I.B.3a.

**90.** *See generally GitHub Copilot documentation*, GitHub (Aug. 28, 2023), https://docs.github.com/en/copilot.

**91.** *See generally* Complaint, Doe 1 v. GitHub, Inc., No. 4:22-cv-06823 (N.D. Cal. Nov. 3, 2022). Recently, GitHub updated the Copilot model to go "beyond the previous OpenAI Codex model." However, the original Codex model is the one named in active lawsuits. *See generally* Shuyin Zhao, *Smarter, more efficient coding: GitHub Copilot goes beyond Codex with improved AI model*, Github (July 28, 2023), https://github.blog/2023-07-28-smarter-more-efficient-coding-github-copilot-goes-beyond-codex-with-improved-ai-model/. (discussing Copilot's use of Codex)

Mind's Lyria,[92] Google's MusicLM,[93] and OpenAI's Jukebox[94] are music-generation models embedded in larger systems.[95] OpenAI's website claims "Provided with genre, artist, and lyrics as input, Jukebox outputs a new music sample produced from scratch."[96] Pika is an "idea-to-video platform" that provides tools to produce video generations, using models that take in either text or image prompts.[97] Lastly, generative-AI models for molecular structure are intended to aid in the design of new drugs and to understand protein function. Examples of models in this domain include ProtGPT2[98] and DiffDock.[99] While these modalities also present important implications for copyright,[100] we limit our discussion and examples in the remainder of this Article to text and images.

### 3. Machine-Learning Techniques in Generative AI

While "generative AI" might be a relatively new term-of-art, a lot of the technology that powers today's generative-AI systems has a long history. Many familiar concepts — training algorithms, optimization, neural networks, etc. — all play important roles.[101] In this respect, there is no magic behind gener-

---

**92.** Google DeepMind, *Transforming the future of music creation* (Nov. 16, 2023), https://deepmind.google/discover/blog/transforming-the-future-of-music-creation/.

**93.** *See* Andrea Agostinelli, Timo I. Denk & Zalán Borsos et al., MusicLM: Generating Music From Text (2023) (unpublished manuscript), https://arxiv.org/abs/2301.11325 (for the research paper on the model); Kristin Yim & Hema Manickavasagam, *Turn ideas into music with MusicLM* (May 10, 2023), https://blog.google/technology/ai/musiclm-google-ai-test-kitchen/ (for the product announcement).

**94.** Heewoo Jun, Christine Payne & Jong Wook Kim et al., Jukebox: A Generative Model for Music (2020) (unpublished manuscript), https://arxiv.org/abs/2005.00341.

**95.** For example Dream Track is a production system built using Lyria. *See* DeepMind, *supra* note 92.

**96.** *See* OpenAI, *OpenAI JukeBox*, OᴘᴇɴᴀI (Apr. 30, 2020), https://openai.com/research/jukebox (describing the use of the transformer-based architecture in Jukebox).

**97.** *See* Pika, *An idea-to-video platform that brings your creativity to motion* (2023), https://pika.art/.

**98.** *See generally* Noellia Ferruz, Steffen Schmidt & Birte Höcker, *ProtGPT2 is a deep unsupervised language model for protein design*, 13 Nᴀᴛᴜʀᴇ Cᴏᴍᴍᴄ'ɴs 4348 (2022). ProtGPT2 is based on GPT-2. *See generally* Radford, Wu & Child et al., *supra* note 68. (describing GPT-2, a language model with a transformer-based architecture).

**99.** *See generally* Gabriele Corso, Hannes Stärk & Bowen Jing et al., *DiffDock: Diffusion Steps, Twists, and Turns for Molecular Docki*, *in* 2023 Iɴᴛ'ʟ CᴏɴF. ᴏɴ Lᴇᴀʀɴɪɴɢ Rᴇᴘʀᴇsᴇɴᴛᴀᴛɪᴏɴs (2023). DiffDock uses diffusion-based techniques. *See infra* Part I.B.3b.

**100.** And perhaps also patent law, for generative-AI systems that involve molecular structure.

**101.** *See supra* Part I.A.2. It is also true that generative models, as an overarching type of machine learning, are also not completely new. *See supra* Part I.A.2b. Automatic ative AI. However, there have been a few especially important technological developments in machine learning over the past decade, which have helped usher in this new phase of generative-AI applications with seemingly magical capabilities.

In this section, we address two modeling developments that are predominant in well-known generative-AI systems: **the transformer architecture**[102] and **diffusion-based models**.[103] In the following section, we discuss a third development related to the increased scale of training datasets, overall model size, and computing resources used to train, store, and execute models.[104] We provide intuitions for these three developments because they each raise different considerations for copyright, which we will address in Part II.

### a. Transformer architecture

**Transformers** are a type of model architecture, just like linear regression and neural networks.[105] They are particularly good at capturing context in sequential information by modeling how elements in a sequence relate to each other. Consider our example sentence from above: `"In the summer, I like to go to the [blank]"`. The next word (to fill in the

---

text and music generation date back to the middle of the 20th century. *See generally* Claude E. Shannon, *A Mathematical Theory of Communication*, 27 Bell Sys. Tech. J. 379 (1948). (describing Markov-chain-based language models). *See generally* Darrell Conklin, *Music Generation from Statistical Models*, 45 J. New Music Rsch. ? (2003). (describing prior techniques in statistical music generation). Google published the first transformer architecture in 2017. *See generally* Ashish Vaswani, Noam Shazeer & Niki Parmar et al., *Attention Is All You Need*, *in* 30 Advances Neural Info. Processing Sys. 15 (2017). Prior to 2017, generative-model architectures powered products like older versions of Apple's Siri voice assistant and of Google Translate. *See generally* Siri Team, *Deep Learning for Siri's Voice: On-device Deep Mixture Density Networks for Hybrid Unit Selection Synthesis*, Apple Mach. Learning Rsch. (Aug. 2017), https://machinelearning.apple.com/research/siri-voices. (describing Apple's Siri technology circa 2017). *See generally* Quoc V. Le, *Mike Schuster, A Neural Network for Machine Translation, at Production Scale*, Google Brain Team (Sept. 27, 2016), https://ai.googleblog.com/2016/09/a-neural-network-for-machine.html. (describing the transition from phased-based translation systems to neural-network-based translation systems, before the release of transformers in 2017). Another example of earlier generative architectures is Generative Adversarial Networks (GANs), which have also had a place in popular discourse for around a decade with respect to deep fakes. *See generally* Ian Goodfellow, Jean Pouget-Abadie & Mehdi Mirza et al., *Generative Adversarial Nets*, *in* 27 Advances Neural Info. Processing Sys. 9 (2014).

**102**. *See infra* Part I.B.3a.
**103**. *See infra* Part I.B.3b.
**104**. *See infra* Part I.B.4.
**105**. *See supra* Part I.A.2.

"`[blank]`") is related to many of the other words in the sequence (such as "`summer`", "`I`", and "`go`") in a way that makes the word "`beach`" a more likely candidate than "`slopes`". Given their effectiveness, since their release in 2017,[106] transformers have become the *de facto* way to model sequence-formatted data, including modalities as diverse as text, code, music, and protein structure.[107] In recent years, some image-generation models also incorporate transformer-architecture variants.[108]

The transformer architecture can be used to train a generative model,[109] and today, almost all generative text models are transformer-based, including ChatGPT, where the "T" in "GPT" stands for Transformer.[110] This architecture consists of two parts: a neural network, and something new called the **attention** mechanism. The attention mechanism, the key innovation in the original transformer architecture paper,[111] is what works particularly well to model contextual information in sequences.[112] Similar to the traditional

---

**106.** Vaswani, Shazeer & Parmar et al., *supra* note 101.

**107.** *See supra* Part I.B.2c.

**108.** *See* Alexey Dosovitskiy, Lucas Beyer & Alexander Kolesnikov et al., *An Image is Worth 16x16 Words: Transformers for Image Recognition at Scale*, *in* 2021 Int'l Conf. on Learning Representations (2021) (regarding vision transformers, or ViTs). *See* William Peebles & Saining Xie, Scalable Diffusion Models with Transformers (2023) (unpublished manuscript) (for diffusion-model transformers, or DiTs).

**109.** *See supra* Part I.A.2b (describing generative models). Not all transformer-based models generate content, for example, BERT (Bidirectional Encoder Representations from Transformers). *See generally* Devlin, Chang, Lee & Toutanova, *supra* note 10. Such models are not trained to predict (and then generate) the next word in a sequence. Instead, they are useful for other tasks: producing word embeddings, filling in missing data (e.g., blanks in provided text like "`[blank]` re-recorded her old studio albums after her masters were sold.`"), or performing question and answering. *See supra* Part I.A.1 (defining word embeddings).

**110.** GPT is an acronym for Generative Pre-trained Transformer. We will discuss the "Pre-trained" term later. *See infra* Part I.C. Other transformer-based language models include LaMDA and the family of Llama models. *See generally* Thoppilan, De Freitas & Hall et al., *supra* note 71. (describing LaMDA). *See generally* Touvron, Lavril & Izacard et al., *supra* note 1; Touvron, Martin & Stone et al., *supra* note 1; Meta, *supra* note 88. (describing the Llama, Llama 2, and Code Llama models).

**111.** Vaswani, Shazeer & Parmar et al., *supra* note 101.

**112.** We do not address the technical details of transformers in this article, but nevertheless choose to mention them because they are a common term that repeatedly comes up in the context of generative AI. *See generally* Mark Riedl, *A Very Gentle Introduction to Large Language Models without the Hype*, Medium (Apr. 13, 2023), https://mark-riedl.medium.com/a-very-gentle-introduction-to-large-language-models-without-the-hype-5f67941fa59e; Timothy B. Lee & Sean Trott, *A jargon-free explanation of how AI large language models work*, Ars Technica (July 31, 2023), https://arstechnica.com/science/2023/07/a-jargon-free-explanation-of-how-ai-large-language-models-work/. (providing more in-depth, yet still accessible, treatments on transformers).

generative-text models that we describe above, transformer-based models take as input a sequence of words, and, conditioned on this context,[113] generate the next word as the output,[114] but they do so via a novel combination of neural networks and attention. This is ultimately why transformer-based generative-AI systems like ChatGPT are not doing anything particularly "intelligent": Transformer models are also just generating a word at a time.

Lastly, it is also important to note that the transformer architecture can be implemented at an enormous scale. Just as deep neural networks contain a large number of layers and connections between them, transformers can be stacked together to construct models with billions of model parameters,[115] where (generally speaking, though with exceptions) larger models yield higher quality generations. It is this large-scale stacking of transformers that gives **large language models (LLMs)** their name.

### b. Diffusion-based models

**Diffusion**-based models[116] are popular in image generation, for example, Midjourney's underlying text-to-image model and (as the name suggests) the Stable Diffusion text-to-image model.[117] It is important to note that diffu-

---

113. This is where the term **context window** (or **context length**) originates; it refers to the size of the input sequence. *See generally* Anthropic, *Introducing 100K Context Windows*, Anthropic (May 11, 2023), https://www.anthropic.com/index/100k-context-windows/. (describing the context window in Anthropic's Claude chatbot system).

114. Technically, words are represented as **tokens**. Tokens are numbers that represent a word, sub-word, logogram, or punctuation. For instance, the word `"hello"` may be represented by the number 12. A more uncommon word like `"credenza"` may be divided into multiple sub-words, e.g., `"cre"`, `"den"`, `"za"`; each sub-word would be represented by a number, e.g., `"cre"` = 58, `"den"` = 29, `"za"` = 105), and so, altogether, the word `"credenza"` would be encoded as the vector [58, 29, 105]. Modeling data as tokens enables using transformers with non-text sequences, e.g., a token for a music model may be a musical note or a specific pitch.

115. For language models, this scale reflects the current state-of-the-art. *See infra* Part I.B.4.

116. Such models are commonly called "diffusion models" in the literature. However, as we note below, "diffusion" is a training mechanism that involves sampling, and, for the purposes of this Article, should be understood as a training algorithm, not a model. This algorithm trains (typically, at the time of writing) a neural-network architecture. We choose to disambiguate these subtleties with the term "diffusion-based model," even though it is not the term commonly used in the scientific field. *See generally* Jascha Sohl-Dickstein, Eric Weiss, Niru Maheswaranathany & Surya Ganguli, *Deep Unsupervised Learning using Nonequilibrium Thermodynamics*, *in* 2015 Proc. 32nd Int'l Conf. on Mach. Learning 2256 (2015). (regarding early work diffusion probabilistic models).

117. Stable Diffusion is a text-to-image model that combines a transformer architecture for modeling text with diffusion for modeling images. DALL·E-2 uses a mix of transformers and diffusion, in a two-step process. *See generally* Rombach, Blattmann & Lorenz sion involves a different suite of machine-learning techniques than those tra-
ditionally described in the legal literature.[118] We elide the technical details,
but emphasize that diffusion is *not* actually a model architecture;[119] rather,
diffusion is a specific algorithmic process for training a model — typically, a
large-scale deep neural network.[120]

For text-to-image diffusion-based model training, the training data con-
sist of pairs of images and corresponding text description captions. Training
occurs in two passes. First, for each training data example (image and its
caption), **noise** is incrementally added to the image until it effectively looks
like static. This process intentionally corrupts the image, degrading its qual-
ity. Second, a neural network is trained to reverse this corruption process —
removing noise and restoring the image to its original form. Both of these
passes are iterative; each has multiple steps that happen over time. The first
pass involves the repeated addition of noise, and the second involves de-
noising the fully noised image a little bit at a time.[121] During the de-noising
pass, the neural network is trained by evaluating how well the de-noised im-

---

et al., *supra* note 30. (regarding the Stable Diffusion model). *See generally Midjour-
ney*, *supra* note 30. (regarding the Midjourney text-to-image system). *See generally*
Ramesh, Dhariwal & Nichol et al., *supra* note 57; Aditya Ramesh, *How DALL·E 2 Works*,
Aᴅɪᴛʏᴀ Rᴀᴍᴇsʜ (2022), http://adityaramesh.com/posts/dalle2/dalle2.html. (detailing
the DALL·E-2 system).

**118.** *See supra* Part I.A.2a.

**119.** Diffusion is built on concepts from Bayesian inference — namely, Markov chains and
variational methods. Early work on diffusion probabilistic models (DPMs) shows the
relationship between diffusion and concepts from variational autoencoders, another
type of deep generative model. Starting in around 2019, DPMs started to become com-
petitive with GANs, with respect to image generation. *See generally* Sohl-Dickstein,
Weiss, Maheswaranathany & Ganguli, *supra* note 116. (regarding early work diffusion
probabilistic models). *See generally* Dirk P. Kingma & Max Welling, *Auto-Encoding
Variational Bayes*, *in* 2014 Iɴᴛ'ʟ Cᴏɴꜰ. ᴏɴ Lᴇᴀʀɴɪɴɢ Rᴇᴘʀᴇꜱᴇɴᴛᴀᴛɪᴏɴꜱ 14 (2014);
Danilo Jimenez Rezende, Shakir Mohamed & Daan Wierstra, *Stochastic Backpropaga-
tion and Approximate Inference in Deep Generative Models*, *in* 2014 Pʀᴏᴄ. 31ꜱᴛ Iɴᴛ'ʟ
Cᴏɴꜰ. ᴏɴ Mᴀᴄʜ. Lᴇᴀʀɴɪɴɢ 1278 (2014). (regarding early work on variational autoen-
coders). *See generally* Goodfellow, Pouget-Abadie & Mirza et al., *supra* note 101. (de-
scribing GANs). *See generally* Yang Song & Stefano Ermon, *Generative Modeling by
Estimating Gradients of the Data Distribution*, *in* 32 Aᴅᴠᴀɴᴄᴇꜱ Nᴇᴜʀᴀʟ Iɴꜰᴏ. Pʀᴏ-
ᴄᴇꜱꜱɪɴɢ Sʏꜱ. 6840 (2019); Jonathan Ho, Ajay Jain & Pieter Abbeel, *Denoising Diffusion
Probabilistic Models*, *in* 33 Aᴅᴠᴀɴᴄᴇꜱ Nᴇᴜʀᴀʟ Iɴꜰᴏ. Pʀᴏᴄᴇꜱꜱɪɴɢ Sʏꜱ. 6840 (2020). (de-
tailing the first methods that were competitive with GANs on image generation tasks).

**120.** The common neural network architecture for diffusion models is called U-Net. *See
generally* Olaf Ronneberger, Philipp FIscher & Thomas Brox, *U-Net: Convolutional
Networks for Biomedical Image Segmentation*, 2015 Mᴇᴅ. Iᴍᴀɢᴇ Cᴏᴍᴘᴜᴛ. & Cᴏᴍᴘᴜᴛ.-
Aꜱꜱɪꜱᴛᴇᴅ Iɴᴛᴇʀᴠᴇɴᴛɪᴏɴ 234—241.

**121.** In a bit more detail, diffusion uses simulation techniques from the physical sciences to
approach the machine-learning problem. Such simulations treat dynamical systems as age matches the original, noise-free image in the training data, and this evaluation is associated with the original text caption in the training data.[122]

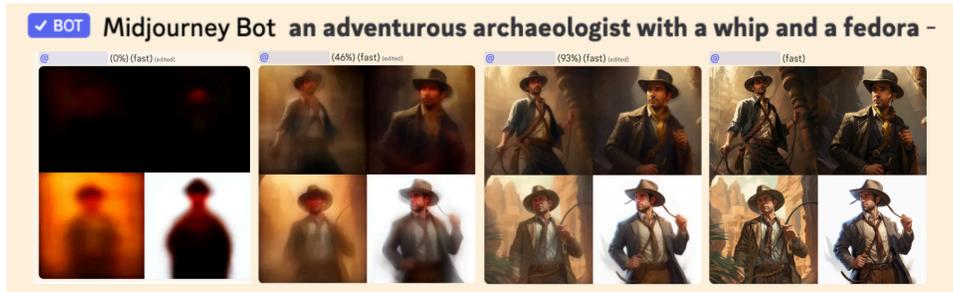

Figure 3: Several screenshots of the generation process using the Midjourney system, which uses text-to-image, diffusion-based models (Midjourney, Midjourney (2023), `https://midjourney.com/`). We prompt with `"an adventurous archaeologist with a whip and a fedora"`, and the Midjourney user interface shows the iterative de-noising process to produce the generations.

Similar to the case of transformers, once trained, a diffusion-based model can be used to produce generations. Generation treats text prompts like description captions, and leverages relationships that the model has learned between captions and images in the training data. The process begins with a completely noisy image, and repeatedly applies the model to remove noise, iteratively producing a series of images that are intended to increasingly align with the text prompt. We can therefore think of the production of an output generation as sequence of images unfolding over time, starting from the completely noisy image and ending with the final generation, with every iteratively de-noised image between the two (for example, see Figure 3). It is possible to string these images together into an animation, as the Midjourney system does when producing generations in its user interface.[123] We return to this point later when we discuss the display right.[124]

---

a series of states; a given system can transition from one state to another over time. This modeling approach has many applications besides image generation, including simulating the thermodynamics of molecules, the spread of a disease, and price movements in the stock market. For diffusion, the states are the intermediate images between the noise-free and completely noisy, static-resembling image.

**122.** *See* Complaint at p. 12, Getty Images (US), Inc. v. Stability AI, Inc., No. 1:23-cv-00135 (D. Del. Feb. 3, 2023). (giving an intuitive description of this process in the context of Stability AI's model).

**123.** *Midjourney*, *supra* note 30.

**124.** *See infra* Part II.B.

4. The Role of Scale

Last, we turn explicitly to an important theme that has cropped up repeatedly throughout this section: scale. Above, we discussed how generative-AI systems are large-scale and have many components.[125] Generative AI models built using transformers or diffusion represent just one subset of these components, and they also tend to be massive.[126] For example, state-of-the-art transformer-based LLMs currently have billions of parameters with trillions of connections between them.[127]

The massive scale of these models is intended to capture the richness and complexity of equally massive datasets.[128] As we mentioned briefly above, these datasets are often scraped from the Internet.[129] This is a relatively new practice. Prior to the publication of the transformer architecture in 2017,[130] much of machine-learning research involved training models on smaller datasets. As points of comparison, both the MNIST[131] and CIFAR-10[132] datasets, (until recently) two common benchmarks in discriminative deep learning tasks, contain 60,000 labeled images. Even ImageNet, a more challenging benchmark, has only 15 million labeled images.[133] In contrast, datasets to train generative-AI models, such as LAION-5B,[134] have billions of training data examples.

In fact, today's generative-AI training datasets are so large, machine-learning practitioners do not have effective or efficient ways to fully know their contents. This is one of the important impacts of scale. Earlier datasets like CIFAR-10, and even ImageNet, are small enough that they can be manually curated. For example, in the case of MNIST, the origin (i.e., **provenance**) of every data example is known and documented. For large-scale datasets

---

**125**. *See supra* Part I.B.1.

**126**. *See supra* Part I.B.3.

**127**. *See supra* Part I.A.1a.

**128**. Further, the associated cost of training such models is also enormous. *See infra* Part I.C.4.

**129**. *See supra* Part I.B.2.

**130**. Vaswani, Shazeer & Parmar et al., *supra* note 101.

**131**. Yann LeCun & Corinna Cortes, *MNIST handwritten digit database* (1999), https://www.lri.fr/~marc/Master2/MNIST_doc.pdf.

**132**. Alex Krizhevsky, Vinod Nair & Geoffrey Hinton, *CIFAR-10 (Canadian Institute for Advanced Research)* (2009), http://www.cs.toronto.edu/~kriz/cifar.html.

**133**. Jia Deng, Wei Dong & Richard Socher et al., *ImageNet: A large-scale hierarchical image database*, *in* 2009 2009 IEEE Conf. on Comput. Vision & Pattern Recognition 248—255 (2009).

**134**. Beaumont, *supra* note 82; Schuhmann, Beaumont & Vencu et al., *supra* note 82. *See supra* Part I.B.2.

scraped from the web, provenance is much trickier, which will have implications for copyright.[135]

Nevertheless, despite such novel challenges, scale also confers new capabilities.[136]  Today's generative-AI models are able to produce incredible content, in large part because of their large scale,[137] though it is not well understood exactly how or why.[138]  As we have done throughout this whole section, it is possible to break down generative-AI systems into different known aspects and components, and yet, a lot remains unknown about how these systems actually work in detail.[139]

### C. The Generative-AI Supply Chain

In the prior section, we provide a working definition of what constitutes "generative AI," for which we emphasize that generative models are embedded within larger systems[140] that produce content from different modalities.[141] On the technical side, there have been some key innovations in machine learning, like transformers and diffusion, that have facilitated the development of today's generative-AI systems.[142]

The other big enabler of today's generative-AI systems is scale.[143] Notably, scale complicates *what* technical and creative artifacts are produced, *when* these artifacts are produced and stored, and *who* exactly is involved in the production process. In turn, these considerations are important for how we reason about copyright implications: *what* is potentially an infringing artifact, *when* in the production process it is possible for infringement to occur, and *who* is potentially an infringing actor.[144]

To provide some structure for reasoning about this complexity, we introduce our abstraction for reasoning about generative AI as a supply chain. We conceive of the **generative-AI supply chain** as having eight stages (see Fig-

---

*Training Data*, *in* 2022 Findings Ass'n for Comput. Linguistics: EMNLP 2022 2429 (2022); Roger Grosse, Juhan Bae & Cem Anil et al., Studying Large Language Model Generalization with Influence Functions (2023) (unpublished manuscript), https://arxiv.org/abs/2308.03296. (discussing influence functions). While these fields of work have provided insights, many believe that there lacks sufficient evidence to depend on models to make consequential decisions. *See generally* Zachary Lipton, *The Mythos of Model Interpretability: In Machine Learning, the concept of interpretability is both important and slippery*, 16 Queue 31 (2018).

140. *See supra* Part I.B.1.

141. *See supra* Part I.B.2.

142. *See supra* Part I.B.3.

143. Pun intended. *See supra* Part I.B.4.

144. The generative-AI supply chain is a very good example of the "many hands" problem in computer systems. That is, there are many diffuse actors, at potentially many different organizations, that can each have a hand in the construction of generative-AI systems. It can be very challenging to identify responsible actors when these systems transgress broader societal expectations — in our case, the preservation of copyrights. *See* A. Feder Cooper, Emanuel Moss, Benjamin Laufer & Helen Nissenbaum, *Accountability in an Algorithmic Society: Relationality, Responsibility, and Robustness in Machine Learning* pp. 867–869, *in* 2012 2022 ACM Conf. on Fairness Accountability & Transparency 864 (2012). (describing the problem of "many hands" in data-driven machine learning/AI systems). *See* Rui-Jie Yew & Dylan Hadfield-Menell, Break It Till You Make It: Limitations of Copyright Liability Under a Pretraining Paradigm of AI Development (2023) (unpublished manuscript), https://genlaw.github.io/CameraReady/30.pdf (regarding an instantiation of this problem for generative AI and copyright).

ure 4): the creation of expressive works,[145] data creation,[146] dataset collection and curation,[147] model (pre-)training,[148] model fine-tuning,[149] system deployment,[150] generation,[151] and model alignment.[152] Each stage gathers inputs from prior stage(s) and hands off outputs to subsequent stage(s), which we indicate with (sometimes bidirectional) arrows. This framing is broadly useful for reasoning about different legal considerations for generative AI; we employ it specifically for copyright analysis in Part II.

The first two stages, the creation of expressive works and data creation, pre-date the advent of generative-AI systems. Nevertheless, they are indispensable parts of the production of generative-AI content, which is why we begin our discussion of the supply chain with these processes. The following six stages reflect processes that are new for generative-AI systems. The connections between these supply-chain stages are complicated. While in some cases, one stage clearly precedes another,[153] for other cases, there are many different possible ways that stages can interact. We highlight some of this complexity in the following subsections, and call attention to different possible timelines of when supply chain stages can be invoked and which actors can be involved at each stage.

## 1. The Creation of Expressive Works

Artists, writers, coders, and other creators produce expressive works. Generative-AI systems do, too;[154] but, state-of-the-art systems are only able to do so because their models have been trained on data derived from pre-existing creative works.[155] While perhaps obvious, it is nevertheless important to

---

**145.** *See infra* Part I.C.1.

**146.** *See infra* Part I.C.2.

**147.** *See infra* Part I.C.3.

**148.** *See infra* Part I.C.4.

**149.** *See infra* Part I.C.5.

**150.** *See infra* Part I.C.6.

**151.** *See infra* Part I.C.7.

**152.** *See infra* Part I.C.8.

**153.** e.g., model pre-training necessarily precedes model fine-tuning. See Figure 4.

**154.** We discuss this in more detail below with respect to generation. *See infra* Part I.C.7. We also discuss this when we delve into copyright and authorship. *See infra* Part II.A.

**155.** As we address below, a data example is not the same as the expressive work. Additionally, some models are trained on synthetic data, typically generated by other generative-AI models (which, in turn, were trained on pre-existing creative works). Training predominantly on synthetic data is a growing practice, but is not reflective of the current most common practices in today's generative-AI systems. Gokaslan, Cooper & Collins et al., *supra* note 39. Currently, there are concerns that training on synthetic data can seriously compromise model quality. *See generally* Ilia Shumailov, Zakhar Shumaylov &

emphasize that the processes of producing most creative works have (thus far) had nothing to do with machine learning.[156] Historically, painters have composed canvases, writers have penned articles, coders have developed software, etc. without consideration of how their works might be taken up by automated processes.

Nevertheless, as we discuss above[157] and detail further in the next section,[158] these works can be transformed into quantified data objects that can serve as inputs for machine learning. Such data can be (and have been) easily posted and circulated on the Internet, making them widely accessible for the development of generative-AI systems. As a result, content creators and their original works are a part of the generative-AI supply chain, whether they would like to be or not (see Figure 4, stage 1).

## 2. Data Creation

Original expressive works are distinct from their datafied counterparts.[159] Data examples are constructed to be computer-readable, such as the JPEG encoding of a photograph.[160] For the most part, the transformation of creative content to data formats pre-dates generative AI (see Figure 4, stage 2). It is a process that has grown in tandem with the proliferation of the modern Internet. Regardless, all state-of-the-art generative-AI systems depend on this process. They rely on data that coheres with their underlying models' respective modalities:[161] Text-to-text generation models are trained on text, text-to-image models are trained on both text and images, text-to-music models are trained on text and audio files, and so on.

---

Yiren Zhao et al., The Curse of Recursion: Training on Generated Data Makes Models Forget (2023) (unpublished manuscript), https://arxiv.org/abs/2305.17493. (detailing "model collapse" in different generative models). However, recent work shows that reduction in model quality can be avoided with extensive data curation. Suriya Gunasekar, Yi Zhang & Jyoti Aneja et al., Textbooks Are All You Need (2023) (unpublished manuscript). *See infra* Part I.C.3.

156. It appears increasingly likely that some content will be created specifically for model training. For example, hiring photographers to take photographs specifically for model training. Companies like Scale AI already create content (in the form of labels and feedback) specifically for the purpose of training models.Scale AI, *Scale AI*, Scale AI (Sept. 2, 2023), https://scale.com/.

157. *See supra* Part I.A.1.

158. *See infra* Part I.C.2.

159. Of course, data can be copies of original works, and thus still infringe intellectual property rights.

160. *See supra* Part I.A.1.

161. *See supra* Part I.B.2.

This is an important point for our purposes because the works that have been transformed into data have copyrights.[162] In turn, for generative-AI systems that generate potentially copyright-infringing material, the training data itself will often include copyrightable expression. The GitHub Copilot system involves models trained on copyrighted code,[163] ChatGPT's underlying model(s) are trained on text data scraped from the web,[164] Stability AI's Stable Diffusion is trained on text and images,[165] and so on. For the most part, it is the copyright owners of these datafied individual works who are the potential plaintiffs in a copyright infringement suit against actors at other stages of the supply chain, which we address further in Part II. For now, we simply emphasize that these are the relevant copyrights.

---

**162**. An exception to this is training data produced by generative-AI systems (i.e., synthetic training data). Gokaslan, Cooper & Collins et al., *supra* note 39; Gunasekar, Zhang & Aneja et al., *supra* note 155. Such data currently have been found to not be copyrightable. Thaler v. Perlmutter, No. 22-1564 (D.D.C date). *See infra* Part II.A. We discuss using generations as training data below. *See infra* Part I.C.7

**163**. Recall that, until recently, Copilot was built on top of OpenAI's Codex model. *See supra* Part I.B.2c and references therein.

**164**. *See supra* Part I.B.2a and references therein.

**165**. *See supra* Part I.B.2b and references therein.

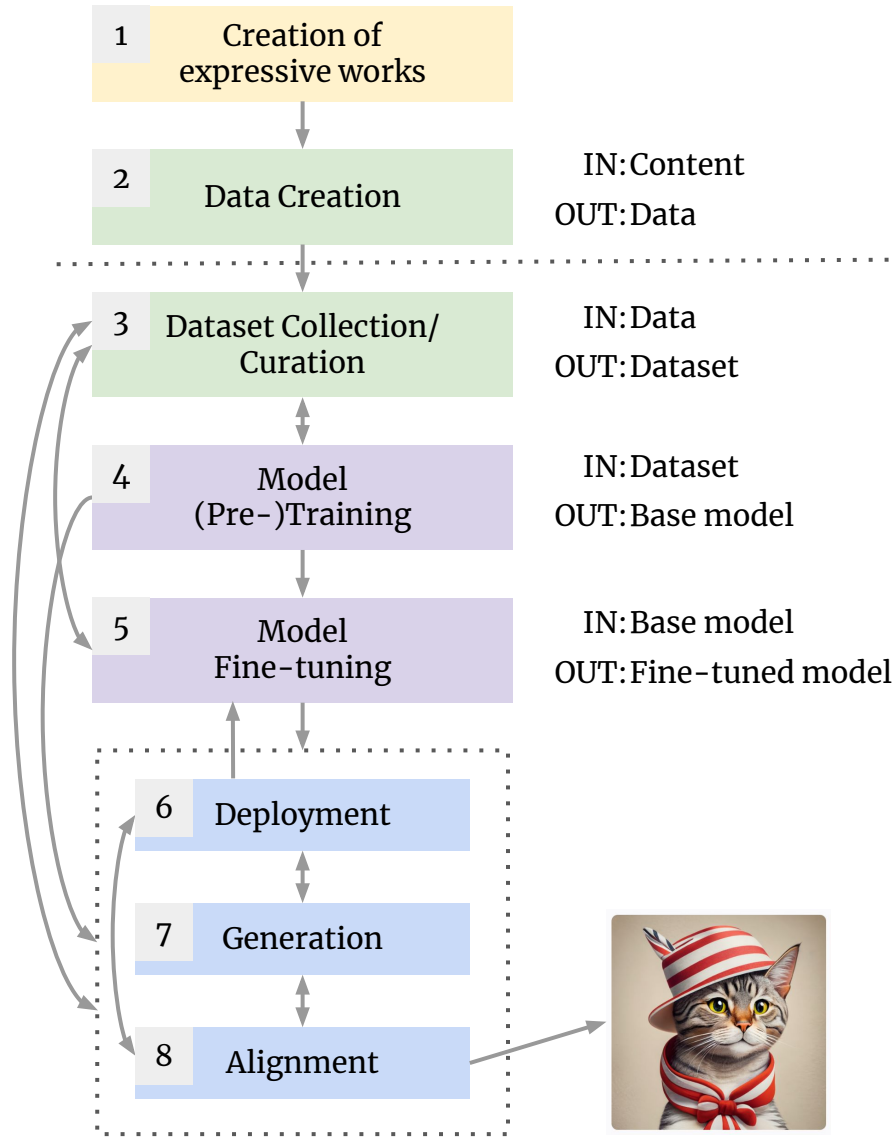

Figure 4: The generative-AI supply chain. We map out eight different stages: 1) The creation of expressive works, (*see infra* Part I.C.1), 2) data creation (*see infra* Part I.C.2), 3) dataset collection/curation (*see infra* Part I.C.3), 4) model (pre-)training (*see infra* Part I.C.4), 5) model fine-tuning (*see infra* Part I.C.5), 6) system deployment (*see infra* Part I.C.6), 7) generation (*see infra* Part I.C.7), and 8) model alignment (*see infra* Part I.C.8), Different stages are connected to each other, handing off outputs from one stage as inputs to another. The creation of expressive works and data creation pre-date the advent of today's generative-AI systems (indicated by a dotted line). There are many possible ways to connect the other six stages. System deployment, model alignment, and generation tend to happen in concert (indicated by the dotted box). Generations can in turn be used as training data (*see infra* Part I.C.7). We indicate this in the figure with the arrow from generation (7) to dataset collection/curation (3). In this case, generation serves simultaneously as the creation of expressive works (1) and data creation (2).

### 3. Dataset Collection and Curation

As we have discussed above, model training does not happen at the level of individual data examples; instead, data examples are grouped together into datasets used for training.[166] The training process for cutting-edge generative-AI models requires particularly vast quantities of data,[167] which must be arranged into datasets that have recurring, standard structure. Dataset creators for generative AI often meet this need by scraping data from the Internet.[168] This process involves a variety of curatorial choices, including filtering out types of data that creators and curators do not want to include, such as "toxic speech."[169] Such curatorial choices can muddle the line between dataset creation and curation, as both processes can effectively happen in tandem.[170]

With respect to the generative-AI supply chain, there are several points worth highlighting in dataset collection and curation processes (see Figure 4, stage 3). First, while dataset creation and curation can be carried out by the same entities that train generative-AI models,[171] it is common for these processes are split across different actors. The Stable Diffusion model, for example, is trained on images from datasets curated by the non-profit organization LAION.[172] It is necessary, therefore, to consider the potential liability of

---

166. *See supra* Part I.A.2; *supra* Part I.B. Further, this is not to say individual training examples are unimportant. Specific pieces of training data can have an out-sized influence on generations, compared with other pieces of training data. *See generally* Koh & Liang, *supra* note 139; Akyurek, Bolukbasi & Liu et al., *supra* note 139; Grosse, Bae & Anil et al., *supra* note 139. (discussing influence functions).

167. *See supra* Part I.B.4.

168. This is not the only way to collect large amounts of data. *See* Katherine Lee, Daphne Ippolito & A. Feder Cooper, The Devil is in the Training Data (2023) (unpublished manuscript), *in* Lee, Cooper, Grimmelmann & Ippolito, *supra* note 65, at 5 (discussing other ways datasets may come to be).

169. *See generally id.*. (discussing dataset creation and curation choices, including toxic content filtering)

170. This is why we choose to place creation and curation as the same stage in the pipeline. Note, however, that creation and curation do not *always* have to happen together, and may involve different sets of actors. It is also possible for curation to happen after the start of model training, in response to metrics that are observed during the training process. That is, curation could follow (and then also precede further) model (pre-)training (Figure 4, stage 4; *see infra* Part I.C.4), or model fine-tuning (Figure 4, stage 5; *see infra* Part I.C.5). These complex interactions are the reason for the bidirectional arrows between stages in Figure 4.

171. *See infra* Part I.C.4.

172. Technically, LAION presents the dataset as a collection the URLs of the images. Stable Diffusion then visits each URL to collect images for training. *See supra* Part I.B.2b; *supra* I.B.4 and citations therein.

dataset creators and curators separately from the potential liability of model trainers.[173]

Second, training datasets are their own objects. Note that dataset curation, as described above, will frequently involve "the collection and assembling of preexisting materials or of data that are selected, coordinated, or arranged in such a way that the resulting work as a whole constitutes an original work of authorship."[174] As such, training datasets can themselves be copyrighted; copying of the dataset *as a whole* without permission could constitute infringement, separate and apart from infringement on the underlying works the dataset comprises.[175] In practice, however, it appears that most uses of training datasets are licensed — either through a bilateral negotiation or by means of an open-source license offered to the world by the dataset compiler.[176]

Third, while a few training datasets include metadata on the provenance of their constitutive data examples, many datasets do not. Provenance makes it easier to answer questions about the data sources a model was trained on, which can be relevant to an infringement analysis.[177] It also bears on the ease with which specific material can be located, and if necessary removed, from a dataset.[178] However, the use of web-scraping to collect generative-AI training datasets is directly in tension with maintaining information about provenance. As we discuss above, relying on Internet sources and the scale of scraped datasets makes determining individual data example origins very challenging.[179] Notably, even if particular dataset creators and curators release a training dataset with a chosen license, this does not guarantee that the works within the dataset are appropriately licensed.[180]

---

**173**. *See infra* Part II.E.

**174**. 17 U.S.C. § 101.

**175**. *See infra* Part II.A; *infra* Part II.E.

**176**. *See infra* Part II.I.

**177**. *See infra* Part II.C.

**178**. *See infra* Part Part III.

**179**. *See supra* Part I.B.4 and references therein. *See generally* Katherine Lee, Daphne Ippolito & A. Feder Cooper, The Devil is in the Training Data (2023) (unpublished manuscript), *in* Lee, Cooper, Grimmelmann & Ippolito, *supra* note 65, at 5. (discussing provenance challenges for generative AI).

**180**. Indeed, the creators and curators would have to check that they have abided by each data example's respective license(s). *See infra* Part II.A (regarding authorship and training datasets).

For example, LAION-5B, a large image dataset mentioned above,[181] was released as under Creative Commons CC-BY 4.0,[182] It is unclear if the LAION team had the rights to license all the referenced images within.[183]  For another example, the complaint in *Tremblay v. OpenAI, Inc.* alleges that ChatGPT's underlying model(s) were trained on datasets that do not license the books data that they contain.[184]

### 4. Model (Pre-)Training

Following the collection and curation of training datasets (Figure 4, stage 3), it is possible to train a generative-AI model (Figure 4, stage 4).  The model trainer[185] selects a training dataset, a model architecture (i.e., a set of initialized model parameters), a training algorithm, and a seed value for the random choices made during the training.[186]

As mentioned above, the process of training — from transforming these inputs into a trained model — is expensive.  It requires a substantial invest-

---

**181**. *See supra* Part I.B.4.

**182**. LAION-5B released a dataset of text captions and URLs to images, instead of the images themselves.  Beaumont, *supra* note 82; Schuhmann, Beaumont & Vencu et al., *supra* note 82.

**183**. Notably, the website introducing the LAION dataset provides a feature called "pwatermark," which is a prediction of how likely the image is to contain a watermark.  The LAION team estimates that the 6.1% of the dataset Laion2B-en contains watermarked images.

**184**. In particular, the complaint in *Tremblay v. OpenAI* alleges that the training data included books from infringing "shadow libraries" like Library Genesis.  Complaint at p. 34, Tremblay v. OpenAI, Inc., No. 3:23-cv-03223 (N.D. Cal. June 28, 2023).  But this claim is based on circumstantial evidence, because the datasets it was trained on have not been made public.  Text from books have been a key player in other dataset-related complaints.  For example, The Pile data was originally released under the MIT license.Stella Biderman, Kieran Bicheno & Leo Gao, Datasheet for the Pile (2022) (unpublished manuscript), https://arxiv.org/abs/2201.07311. The Pile was core to the complaint in *Kadrey*, since the Pile claimed to contain 108GB of the dataset Books3 (which itself contains content from Bibliotek, a popular torrent interface).  *See generally* Complaint, Kadrey v. Meta Platforms, Inc., No. 3:23-cv-03417 (N.D. Cal. July 7, 2023).  The original download URL for The Pile (`https://the-eye.eu/public/AI/pile/`) is no longer resolving (as of September 2023).

**185**. We distinguish between the person or organization that trains from those that create the model architecture, as they may not be the same.

**186**. Machine learning uses tools from probability and statistics, which reason about randomness.  However, computers are not able to produce truly random numbers.  Instead, algorithms exist for producing a sequence of *pseudo*-random numbers.  A random seed is an input to a pseudo-random number generator, which enables the reproduction of such a sequence.  Recall also that the trainer also selects hyperparameters, which we elide for simplicity.  *See supra* Part I.B.1.

ment of multiple resources: time, data storage, and computing power. For example, BLOOM, a 176-billion-parameter open-source model from Hugging-Face was trained for 3.5 months, on 1.6 terabytes of text, and 384 GPUs;[187] it cost an estimated $2-5 million on computing resources for both the development and ultimate training of BLOOM.[188] As another point of reference, MosaicML, a company (acquired by Databricks[189]) that develops solutions for training as cheaply and efficiently as possible, has trained a GPT-3-quality model for less than $0.5 million.[190] Altogether, the dollar cost can range from six to eight figures, depending on the size of the model, the size of the training dataset, the length of the training process, the efficiency of the software and hardware used, and other choices.

Further, the training process is not completely automated; training often requires people to monitor and tweak the model. For example, model trainers typically run evaluation metrics on the model while it is being trained, in

---

**187.** *See generally* Stas Bekman, *The Technology Behind BLOOM Training*, HᴜɢɢɪɴɢFᴀᴄᴇ (July 14, 2022), https://huggingface.co/blog/bloom-megatron-deepspeed (for training details). *See* BigScience Workshop, Teven Le Scao & Angela Fan et al., BLOOM: A 176B-Parameter Open-Access Multilingual Language Model (2023) (unpublished manuscript), https://arxiv.org/abs/2211.05100 (for the model details).

**188.** Training costs are often not reported. Even when training cost is reported, development costs (including labor) are often omitted, despite being a critical part (and often most expensive) part of overall model development.

**189.** Databricks, *Databricks Completes Acquisition of MosaicML* (July 19, 2023), https://www.databricks.com / company / newsroom / press - releases / databricks - completes - acquisition-mosaicml.

**190.** The original cost to train GPT-3 is unpublished, though, based on its size, is likely higher than $0.5 million. MosaicML reports to have trained a GPT-3-*quality* model. This means the model performs to a similar standard as GPT-3 does. MosaicML's model is substantively different from GPT-3. For one, MosaicML's model is a much smaller 30 billion parameters compared with the original GPT-3 model's 175 billion. Additionally, MosaicML trained on more data, shifting some of the development cost towards data collection and away from model training. It is worth noting that GPT-3 was originally released two years before MosaicML's model was trained, and thus the MosaicML training process likely incorporated additional technological improvements. *See generally* Abhinav Venigalla & Linden Li, *Mosaic LLMs (Part 2): GPT-3 quality for <$500k*, MᴏsᴀɪᴄML (Sept. 29, 2022), https://www.mosaicml.com/blog/gpt-3-quality-for-500k. (regarding MosaicML's model). *See generally* Brown, Mann & Ryder et al., *supra* note 65. (for the size of GPT-3).

order to assess the progress of training.[191]  Depending on these metrics,[192] model trainers may pause the training process to manually revise the training algorithm[193] or the dataset, which we indicate with bidirectional arrows at Figure 4, stages 3-4. Human intervention in response to metrics necessarily makes model training an iterative process.

Once the process of training is complete, we are at the end of this stage of the supply chain. The output of this stage is typically called a **pre-trained model** or **base model**.[194]  At this point, the base model has many possible futures. It could just sit idly in memory, collecting figurative dust, never to be used to produce generations.[195]  The model parameters could be uploaded to a public server,[196] from which others could download it and use it however they want.[197] The model could be integrated into a system and deployed as a public-facing application,[198] which others could use directly to produce generations.[199] Or, the model could be further modified by the initial model trainer, by another actor at the same organization, or, if made publicly available, a different actor from a different organization. That is, another actor could take the model parameters and use them as the input to do additional training with new or modified data (and a chosen training algorithm and random seed, as at the beginning of this section).[200]

---

191. Google's TensorBoard and software from Weights & Biases are two tools for running evaluation metrics and monitoring during training.  *See generally* TensorFlow, *TensorBoard: TensorFlow's visualization toolkit*, TENSORFLOW (2023), https://www.tensorflow.org/tensorboard. (regarding Tensorboard). *See generally Weights & Biases*, WEIGHTS & BIASES (2023), https://wandb.ai/site. (regarding Weights & Biases).

192. Evaluation metrics attempt to elicit how "useful" or "good" the model is. These metrics are not comprehensive, since there is no single way to capture "usefulness" or "goodness" in math.  *See generally* Katherine Lee, Daphne Ippolito & A. Feder Cooper, The Devil is in the Training Data (2023) (unpublished manuscript), *in* Lee, Cooper, Grimmelmann & Ippolito, *supra* note 65, at 5 (for a discussion of evaluation metrics and the impossibility of defining "useful" and "good".).

193. E.g., change the hyperparameters.

194. Others use the term "foundation model." The term "foundation" can be easily misunderstood.  It should not be interpreted to connote that "foundation models" contain technical developments that make them fundamentally different from models produced in the nearly-a-decade of related prior work. The term itself has been met with controversy within the machine learning community, which can be seen expressed on programming forums and in conversations, e.g., we refer to a Twitter thread (and its associated offshoots) that involves renowned researchers and some of the Stanford authors that coined the term "foundation models."  (*See* https://twitter.com/tdietterich/status/1558256704696905728).

195. This reveals the murky line between what exactly is a program and what exactly is data in machine learning, more generally.  The set of parameters can be viewed as a *data structure* containing vectors of numbers that, on its own does not *do* anything.  However, we could load that data structure into memory and apply some relatively lightweight linear algebra operations to produce a generation *See supra* Part I.B. In this respect, we could also consider the model to be a program (and, indeed, an algorithm). This is why we talk about the model being *within* the function $f$ in our analogical discussion of machine-learning-as-a-function. (*See supra* Part I.A.2a.) The model, if given a prompt input, can also be executed like a program. Note that the term "model" is overloaded; it can be used to refer to the model parameters (just the vectors of numbers numbers) or to the model as a combination of software and the model parameters, which together can be executed like a program.

This possibility of future further training of a base model is why this stage of the supply chain is most often referred to as ***pre*-training**, and why a base model is similarly often called a **pre-trained model**. Such additional training of the base model is called **fine-tuning**, which we discuss below.[201]

## 5. Model Fine-Tuning

In our background on machine learning and generative AI above, we emphasized that models reflect their training data.[202] Base models trained on large-scale, web-scraped datasets reflect a lot of general information sourced from different parts of the Internet. They are not typically trained to reflect specialized domains of knowledge. For example, an English text-to-text base model may be able to capture general English-language semantics and information from being trained on web-based data; however, such a model may not be able to, for example, reliably reflect detailed scientific information about molecular biology (e.g., answering the question "what is mitosis?").

This is where fine-tuning comes in to the supply chain (Figure 4, stage 5): Fine-tuning describes the process of modifying a preexisting, already-trained model, and has the general goal of taking such a preexisting model and making it better along some dimension of interest. As the name suggests, most

---

**196**. For example, HuggingFace hosts a repository of over 300,000 open- and semi-closed models and model parameters. *See generally Models*, Hugging Face (Sept. 2, 2023), https://huggingface.co/models.

**197**. They could fine-tune the model (*See infra* Part I.C.5), embed the model in a system that they deploy for others to use (*See infra* Part I.C.6), produce generations (*See infra* Part I.C.7), align the model (*See infra* Part I.C.8), or do some subset of these other stages of the supply chain. From this example, we can see how the supply chain is in fact iterative, which we illustrate in Figure 4.

**198**. *See infra* Part I.C.6.

**199**. *See supra* Part I.B; *infra* Part I.C.7.

**200**. Cooper, Moss, Laufer & Nissenbaum, *supra* note 144.

**201**. *See infra* Part I.C.5.

**202**. *See supra* Part I.A.2; *supra* Part I.B

fine-tuning aims to leverage the general strengths of what a model has already learned, while optimizing its specific details. This process often involves training on additional data that is more aligned with the specific goals.[203] If we think of training as transforming data into a model, fine-tuning transforms a model into another model.

Fine-tuning essentially involves just running more training. In this respect, the overall process of fine-tuning is similar to pre-training: both execute a training process. However, fine-tuning and pre-training run with different inputs, which ultimately makes the trajectories and outputs of their respective training processes very different. That is, even though fine-tuning and pre-training often employ the same training algorithm, they typically use different input training data and different input model parameters.[204] To add more precision to our previous statement: fine-tuning transforms a model into another model, while incorporating more data.

In more detail: Whereas pre-training data tend to be more general, fine-tuning data is typically sourced from a specific problem domain of interest; whereas the input model architecture to pre-training is an initialized, untrained model,[205] for fine-tuning, the input model parameters have already undergone some training and are no longer in their initialized state. Continuing our example above, a base language model could be fine-tuned on scientific papers to improve its ability to summarize scientific content; the fine-tuning stage takes the learned parameters of the more general base model, and updates them by training further on scientific text data.

### Forks in the supply chain

Two important observations follow from our description of fine-tuning as (effectively) just performing more training. For one, a model trainer does not have to fine-tune at all. As discussed above, prior to fine-tuning there is a fork in the generative-AI supply chain, with respect to the possible futures of the base model after pre-training:[206] one could take the output base model from pre-training, and use this model directly as the input for system

---

deployment[207] (Figure 4, stage 6), generation[208] (Figure 4, stage 7), or model alignment[209] (Figure 4, stage 8). Alternatively, it is possible to perform multiple separate passes of fine-tuning — to take an already-fine-tuned model, and use it as the input for another run of fine-tuning on another dataset. In this respect, it is important to note that a model is a "base" or "fine-tuned" model *only in relation to other models.* These terms do not capture inherent technical features of a model; instead, they describe different processes by which a model can be created.

For each of these possibilities in the supply chain, there can be different actors involved. Sometimes, the creator of a model also fine-tunes it. Google's Codey models (for software code generation) are fine-tuned versions of Google's PaLM 2 model.[210] In other cases, another party does the fine-tuning. When a model's parameters are publicly released (as Meta has done with its Llama family of models),[211] others can take the model and independently fine-tune them for particular applications. A Llama fine-tuner could release their model publicly, which in turn could be fine-tuned by another party. When a model is deployed within a hosted service (see Figure 4, step 6), that service may expose APIs to end-users that enable them to fine-tune the model.[212]

To give a concrete example of the many actors in the generative-AI supply chain, consider Vicuna. LMSYS Org fine-tuned Meta's Llama model on the crowd-sourced ShareGPT dataset to produce Vicuna.[213] The creators of Vicuna have also released the model publicly, affording a potentially infinite host of actors the ability to fine-tune the model on additional data.[214] To use a copyright analogy, a fine-tuned model is a derivative of the model from

---

**207.** *See infra* Part I.C.6.

**208.** *See infra* Part I.C.7.

**209.** *See infra* Part I.C.8.

**210.** Google, *Foundation Models* (Aug. 17, 2023), https://ai.google/discover/foundation-models/ (describing Codey).

**211.** Touvron, Lavril & Izacard et al., *supra* note 1; llama2, Meta, *supra* note 88.

**212.** *See infra* Part I.C.6.

**213.** *See generally* The Vicuna Team, *Vicuna: An Open-Source Chatbot Impressing GPT-4 with 90%\* ChatGPT Quality*, LMSYS Org (Mar. 30, 2023), https://lmsys.org/blog/2023-03-30-vicuna/ (regarding the Vicuna model). ShareGPT is a crowd-sourced dataset composed of conversational logs of user interactions with ChatGPT. It contains both content created by users and by the generative-AI model embedded in ChatGPT (either GPT-3.5 or GPT-4, depending on the user). *See generally* ShareGPT, *ShareGPT*, ShareGPT (Sept. 5, 2023), https://sharegpt.com/ (regarding the ShareGPT dataset).

**214.** *See* Colin Raffel, *Collaborative, Communal, & Continual Machine Learning* 15 (2023), https://colinraffel.com/talks/faculty2023collaborative.pdf (for a figure showing many fine-tuned models building on one base model).

which it was fine-tuned; a repeatedly fine-tuned model is a derivative of the (chain of) fine-tuned model(s) from which it was fine-tuned.

It is helpful to make the base-/fine-tuned model distinction because different parties may have different knowledge of, control over, and intentions toward choices like which data is used for training and how the resulting trained model will, in turn, be put to use. A base-model creator, for example, may attempt to train the model to avoid generating copyright-infringing material. However, if that model is publicly released, someone else may attempt to fine-tune the model to remove these anti-infringement guardrails. A full copyright analysis may require treating them differently, and indeed, may require analyzing their conduct in relation to each other.[215]

6. Model Release and System Deployment

At this point in the supply chain, we have a trained generative-AI model — either a base model[216] or a fine-tuned model.[217] As we noted above regarding base models, trained models have a variety of possible futures, of which fine-tuning is just one option. The next three stages address other futures for base and fine-tuned models: it is possible to release a model or deploy it as part of a larger software system (see Figure 4, stage 6), use the trained model parameters directly to produce generations (see Figure 4, stage 7),[218] or to take the trained model and further alter or refine it via model alignment techniques (see Figure 4, stage 8).[219] In brief, there is a complicated orchestration between the deployment, generation, and alignment stages, which can happen in different orders, in different combinations, and at different times for different generative-AI systems. For ease of exposition, we still present these stages of the generative-AI supply chain one at a time, and we begin here with **model release** and **system deployment** (see Figure 4, stage 6).

A model is **released** when a set of model parameters are uploaded to a server or platform (like HuggingFace[220]), from which others can download

---

**215**. *See infra* Part II.E.
**216**. *See supra* Part I.C.4.
**217**. *See supra* Part I.C.5.
**218**. *See infra* Part I.C.7.
**219**. *See infra* Part I.C.8.
**220**. *Models*, *supra* note 196.

it.[221] Released models, which include Meta's Llama family of models[222] and Stable Diffusion,[223] give downloaders direct access to their parameters. This enables developers and practitioners to directly embed the model in their own code to produce generations, or to alter the model (and thus potentially its behavior) through fine-tuning or model alignment techniques.[224]

In contrast, closed-source models are not directly available to users external to model trainers and owners. Such models are typically embedded in large, complex software systems,[225] which can be **deployed** to both internal and external users through software services. For example, a model could be hosted by a company like OpenAI, Stability AI, Google, etc. It could be used internally at those companies for a variety of software-based services (e.g., an internally-developed Google LLM being integrated into Google Search), or released as a hosted service that gives external users access to generative-AI functionality.

External-facing services could be deployed in a variety of forms, and do not typically include the ability to change the model's parameters.[226] They can be browser-based user applications (e.g., ChatGPT, Midjourney, DreamStudio), or public (but not necessarily free) APIs for developers (e.g., GPT models, Cohere).[227] Of course, model trainers could provide some combi-

---

**221.** Meta first asked interested parties to request Llama's model parameters, rather than uploading them for anyone to download. However, Llama's model parameters were quickly leaked on the website 4chan. James Vincent, *Meta's powerful AI language model has leaked online — what happens now?*, THE VERGE (2023), https://wandb.ai/site. This incident shows how challenging it can be to control access to models once released. Llama also includes a use policy in the Llama 2 Community License that outlines prohibited uses of the model. Of course, it is impossible to enforce prohibited uses when releasing model parameters. This is also why many model trainers choose to release models through hosted services. *Use Policy*, META AI (2023), https://ai.meta.com/llama/use-policy/ (for the Llama 2 Community License).

**222.** Touvron, Lavril & Izacard et al., *supra* note 1; Touvron, Martin & Stone et al., *supra* note 1; Meta, *supra* note 88.

**223.** Rombach, Blattmann & Lorenz et al., *supra* note 30.

**224.** *See infra* Part I.C.8.

**225.** *See infra* Part I.B.1.

**226.** One notable exception, at the time of writing, is ChatGPT's fine-tuning API, which enables end-users to fine-tune the underlying system's model through the hosted service's API on a custom dataset. Andrew Peng, Michael Wu & John Allard et al., *GPT-3.5 Turbo fine-tuning and API updates* (Aug. 22, 2023), https://openai.com/blog/gpt-3-5-turbo-fine-tuning-and-api-updates. This is why we place an arrow from deployment (Figure 4, stage 6) to fine-tuning (Figure 4, stage 5).

**227.** Another deployment option is a command-line interface (CLI), which takes a user-supplied prompt as input (via a code terminal) and directly returns the resulting generation as output. `https://ollama.ai/` (the download link of the Ollama CLI, which is a wrapper program around various Llama-family LLMs).

nation of release and deployment options. For example, DreamStudio is a web-based user interface,[228] built on top of services hosted by Stability AI;[229] the DreamStudio application gives external users access to a generative-AI system that contains the open-source Stable Diffusion model,[230] which Stability AI also makes available for direct download.[231]

This is a familiar spectrum from Internet law: cloud-hosted services at one end and fully open-source software at the other, with closed-source apps in between. These deployment methods offer varying degrees of customization and control on the part of the user and also the deployer. For example, a generative-AI system deployed as a web-based application or as an API will often modify the user-supplied prompt before inputting it to the model. Several applications (ChatGPT, Bard, and Sydney, just to name a few) add additional instructions (i.e., application prompts) to the user's input to create a compound prompt.[232] The additional instructions change the behavior of the model output.[233] For example, providing the following prompts to a language model direct the model to behave differently: "I want you to act as an English translator, spelling corrector and improver . . . " and "I want you to act as a poet. You will create poems that evoke emotions and have the power to stir people's soul . . . ".[234]

Typically, model trainers and owners maintain more control over models deployed through hosted services and the least control over models released

---

**228**. *DreamStudio*, *supra* note 38.

**229**. *Stable Diffusion XL*, *supra* note 30.

**230**. Rombach, Blattmann & Lorenz et al., *supra* note 30.

**231**. It is possible that models released and deployed in multiple ways might not all be exactly the same; they could have different versions of model parameters. This may be made explicit to users, as with ChatGPT, or may not be communicated to them, and thus unclear or unknown. *See generally* OpenAI, *supra* note 1 (regarding both GPT-3.5 and GPT-4 model integration into the ChatGPT web application).

**232**. *See generally* Yiming Zhang & Daphne Ippolito, Prompts Should not be Seen as Secrets: Systematically Measuring Prompt Extraction Attack Success (2023) (unpublished manuscript), https://arxiv.org/abs/2307.06865 (which discovers proprietary system prompts). *See generally Custom instructions for ChatGPT*, OPENAI (Aug. 17, 2023), https://openai.com/blog/custom-instructions-for-chatgpt (announcing a ChatGPT feature that allows users to provide their own additional prompts, which get appended to their future inputs to create compound prompts).

**233**. This kind of prompt transformation is another technique for steering the behavior of a model.

**234**. Fatih Kadir Akın, *Awesome ChatGPT Prompts*, GITHUB (Aug. 17, 2023), https://github.com/f/awesome-chatgpt-prompts (These prompts and more can be found on this site). *General Tips for Designing Prompts*, DAIR.AI (Aug. 17, 2023), https://www.promptingguide.ai/introduction/tips (This handbook provides an introduction to creating prompts for large language models). *Custom instructions for ChatGPT*, *supra* note 232.

as model parameters.[235]  When trainers and owners embed models within systems, rather than release them directly,[236] they can imbue models with additional behaviors, prior to giving users access to model functionality. For example, APIs and web applications allow model deployers to include software that filters model inputs or model outputs.  Concretely, ChatGPT will often respond with some version of: "I'm really sorry, but I cannot assist you with that request," when its "safety" filters are tripped.[237] GitHub Copilot expressly states they use " filters to block offensive words in the prompts and avoid producing suggestions in sensitive contexts."[238]  Additionally, some APIs and web applications include output filters to avoid generating anything that looks too similar to a training example[239] Unfortunately, using output filters to find generations that are similar or exact copies of training data is an imperfect process, which we discuss further below.[240]

Finally, each mechanism for making model functionality widely available has different pricing structures that can ultimately impact the quality of the model.  While the open-source community works hard to create and release models that compete with the best closed-source models, current open-source models are mostly trained on open-sourced data and are often lower quality.[241]  Additionally, differences between open- and closed-source datasets can lead resulting trained models to vary in quality.  For example, Min et al. (2023) uses public domain and permissively licensed text to train

---

**235.** *See generally* Vincent, *supra* note 221.

**236.** By analogy, the function $f$ that contains the model is not directly available to users; instead, $f$ is made accessible indirectly via a hosted service. *See supra* Part I.A.2a

**237.** These filters may detect undesired inputs and prevent the model from generating an output, or detect undesired outputs and prevent the system from displaying the generation.  In both cases, the model parameters would not be changed.  This need not be the case, the model parameters may also be directly modified through alignment to respond to undesired inputs in a more desirable way. Of course, though, for ChatGPT, we do not know exactly how filters are implemented.

**238.** GitHub, *About GitHub Copilot for Individuals*, GitHub (Aug. 17, 2023), https://docs.github.com/en/copilot/overview-of-github-copilot/about-github-copilot-for-individuals.

**239.** *Configuring GitHub Copilot in your environment*, GitHub (Aug. 17, 2023), https://docs.github.com/en/copilot/configuring-github-copilot/configuring-github-copilot-in-your-environment. `https://news.ycombinator.com/item?id=33226515` (for related discussion on the Hacker News forum)

**240.** *See infra* Part II.C.

**241.** The best open-sourced models are very good, but still not as good as closed-source proprietary models.  For example, Technology Innovation Institute in Abu Dhabi recently released the model, Falcon 180B (a 180 billion parameter model), which they claim is better than Meta's Llama 2 but still behind GPT 4. *Falcon*, Tech. Innovation Inst. (2023), https://falconllm.tii.ae/falcon.html.

a language model, and demonstrates a degradation in quality in domains that are not well represented in the data.[242] Similarly, Gokaslan et al. (2023) demonstrates a degradation in quality when training diffusion-based text-to-image models using Creative-Commons licensed images.[243] Additionally, data in the public domain can be unrepresentative of certain demographic groups.[244]

### 7. Generation

Regardless of whether we are considering a base or fine-tuned model, and whether that model is released openly as parameters or enclosed within a deployed system,[245] at this next stage in the generative-AI supply chain, different users have different entry points to produce generations (see Figure 4, stage 7). Recall that generative-AI models produce output generations in response to input prompts.[246] If a user wants to produce generations using a released model, the user will need to write code to interact with the model parameters in order to execute the generation process.[247] However, most users are going to interact with models indirectly through a service operated by a model deployer, such as a developer API or a web application. We are finally ready to talk about these users — the people who supply prompts and use the resulting generations.

First, there is the *prompt itself.* Some prompts, like `"a big dog"`, are simple and generic. Others, such as `"a big dog facing left wearing a spacesuit in a bleak lunar landscape with the earth rising in the background as an oil painting in the style of Paul Cezanne high-resolution aesthetic trending on artstation"`, are more detailed. Second, there is the *choice of deployed system* (which, of course, embeds an implicit choice of model). For example, a user that wants to perform text-to-image generation on a browser-based interface needs to select between Ideogram, DreamStudio, DALL·E-2, Midjourney,

---

and other publicly available text-to-image applications that could perform this task. A user typically selects an application with the outputs partially in mind, so that one choice or another can indicate an attitude towards the possibility of infringement. (Some models perform better at particular tasks, and some models are known to be trained on copyrighted data.) Further, users may revise their prompt to attempt to create generations that more closely align with their goals. And, third, there is *randomness* in each generation.[248] It is typical, for example, for image applications to produce four candidate generations. DALL·E-2, Midjourney, and Ideogram (see Figure 2) all do this.

As we will see, characterizing the relationship between the user and the chosen deployed system is one of the critical choice points in a copyright-infringement analysis. There are at least three ways the relationship could be described:[249]

- The user actively drives the generation through choice of prompt(s),[250] and the system passively responds. On this view, the user is potentially a direct infringer, but the application is like a web host, ISP, or other neutral technological provider.

- The system is active and the user passive. On this view, the user is like a viewer of an infringing broadcast, or the unwitting buyer of a pirated copy of a book. Primary copyright responsibility lies with the deployed system, and possibly with others further upstream in the generative-AI supply chain.

- The user and the system are active partners in generating infringing outputs. On this view, the user is like a patron who commissions a copy of a painting, and the system is like the artist who executes it. They have a shared goal of creating an infringing work.

---

248. Recall that, for generative models, there are many reasonable outputs for the input. *See supra* Part I.A.2b. There are also other sources of randomness in generation that are implementation-specific, such as the choice of decoding strategy for language models. *See* Riedl, *supra* note 112 (for an accessible discussion of decoding).

249. We focus on deployed systems — and their API and web-based interfaces — because there are more opportunities for the deployer to control the model. But, of course, the user could have written some code to produce generations using released model parameters.

250. A user may chain together multiple prompts, or provide examples, in order to guide the generation process. Jason Wei, Xuezhi Wang & Dale Schuurmans et al., Chain-of-Thought Prompting Elicits Reasoning in Large Language Models (2023) (unpublished manuscript). The system may also include software to include context from prior generations supplied by the user.

We will argue that there is no universally correct characterization.[251] Which of these three is the best fit for a particular act of generation will depend on the system, the prompt, how the system is marketed, and how users can interact with the system's interfaces.

These three options highlight some additional observations about prompts. Thus far, we have primarily discussed generations as expressive works, but prompts themselves (or sequences of prompts[252]) could be, too.[253] Sufficiently expressive prompts written by the direct user of a service could be subject to copyright. Context windows are so large,[254] it is even possible for the user to prompt with an entire expressive work. As we discuss below in our copyright analysis,[255] it is of course possible for this expressive work to have also been authored by another individual.[256] For example, Anthropic's team discussed using the entire text of *The Great Gatsby* as a prompt to demonstrate the long context window of their language model, Claude.[257] While *The Great Gatsby* is now in the public domain, it is easy to imagine another book entered as the prompt, or a copyrighted image as the prompt in an image-to-image system.[258] User-supplied prompts may be stored on system-deployers' servers for non-transient periods of time, and may even serve training data for a future model. Such prompts may also be used in model alignment, which we discuss next.

### Forks in the supply chain

Lastly, we close our section on the generation stage of the generative-AI supply chain with three additional considerations. First, there is a loop from generation back to the beginning of the supply chain. While not currently the most typical contemporary practice, it is possible to use generations as "synthetic" training data for generative-AI models.[259] In this case, generation

---

(i.e., stage 7) serves simultaneously as the creation of expressive works (i.e., stage 1)[260] and data creation (i.e., stage 2),[261] and generations can become inputs to dataset collection and curation processes (i.e., stage 3),[262] which we indicate with an arrow in Figure 4 from generation to dataset collection and curation.[263] As we discuss in the next Part, this potential circularity also has implications for copyright.[264] Second, some generative-AI systems use a technique called **retrieval-augmented generation (RAG)**. RAG involves selecting specific data examples from a dataset and appending them to the user-supplied prompt, in order to guide and constrain the generation process.[265] This is why we also draw an arrow from dataset collection and curation (i.e., stage 3)[266] to the generation (i.e., stage 7) in Figure 4 (this is shown as the arrow to the dotted box around deployment, alignment, and generation): collected or curated RAG datasets can impact generation. Third, for the process of generation, some generative-AI systems interact with *external* deployed services. Above, we discussed how deployed generative-AI systems can have

---

developer APIs, which give external users the ability to integrate generative-AI functionality into their own code, including user-facing applications. It is similarly possible for generative-AI system deployers to integrate their code with other services on the web.

To make this concrete, consider OpenAI's ChatGPT **plugins**. Plugins enable ChatGPT to integrate with other products and services, including "Expedia, FiscalNote, Instacart, KAYAK, Klarna, Milo, OpenTable, Shopify, Slack, Speak, Wolfram, and Zapier,"[267] in order to shape output generations. Since ChatGPT's underlying model(s) were trained in 2021,[268] some of the information it has learned is out-of-date. One stated purpose of plugins is to address delays in training updates — to give ChatGPT access to more recent data acquired from other web-hosted services, in order to improve the quality of generations.[269] For example, one of the use cases on the OpenAI website involves a user querying for information about the most recent Oscar winners. To produce the corresponding generation, ChatGPT is illustrated as performing a web search, retrieving the recent winners list, and appearing to summarize (in user-requested poetic format) the 2023 winners.[270]

Such interactions between external services and generation further complicate the generative-AI supply chain that we depict in Figure 4. In particular, by potentially integrating with other systems, the generation stage could implicate an entirely separate, unspecified number of supply chains consisting of entirely different organizations and actors. This, too, raises important copyright implications (what if news articles or short stories are integrated by the plugin?), which we also address in Part II.

### 8. Model Alignment

The generative-AI supply chain does not stop with generation. As discussed above, model trainers try to improve models during both pre-training and fine-tuning the base model. For pre-training, they monitor evaluation metrics, and may pause or restart the process to alter the datasets and algorithm being used;[271] for fine-tuning, they continue training the base model with

---

**267**. OpenAI, *ChatGPT plugins*, OPENAI (Mar. 23, 2023), https://openai.com/blog/chatgpt-plugins.

**268**. According to generations produced by the authors, when we prompted with queries whose answers depended on more recent information.

**269**. "By integrating explicit access to external data — such as up-to-date information online, code-based calculations, or custom plugin-retrieved information — language models can strengthen their responses with evidence-based references." OpenAI, *supra* note 267.

**270**. *Id.*

**271**. *See supra* Part I.C.4.

data that is specifically relevant for a particular task.[272]  Both of these base model modifications are coarse: they make adjustments to the dataset and algorithm, and do not explicitly incorporate information into the model about whether specific generations are "good" or "bad," according to user preferences.[273]

There is a whole area of research, called **model alignment**, that attempts to meet this need.[274]  The overarching aim of model alignment is to *align* model outputs with specific generation preferences (see Figure 4, stage 8). Currently, the most popular alignment technique is called **reinforcement learning with human feedback (RLHF)**.[275] As the name suggests, RLHF combines collected human feedback data with a (reinforcement learning) algorithm in order to update the model.  Human feedback data can take a variety of forms, which include user ratings of generations.  For example, such ratings can be collected by including thumbs-up and thumbs-down buttons in the application user interface, which are intended to query feedback about the system's output generation.  In turn, the reinforcement learning algorithm uses these ratings to adjust the model — to encourage more "thumbs-up" generations and fewer "thumbs-down" ones.[276] Future training and alignment on the model may include both the inputted prompt and the generation in addition to the feedback provided.  As discussed in the prior section,[277] user-supplied prompts may include copyrighted content created by either the user themselves or by another party.

---

272. *See supra* Part I.C.5.

273. Of course, words like "good" and "bad" can have multiple valences, and resist the kind of quantification on which machine learning depends.  *See* Katherine Lee, Daphne Ippolito & A. Feder Cooper, The Devil is in the Training Data (2023) (unpublished manuscript), *in* Lee, Cooper, Grimmelmann & Ippolito, *supra* note 65, at 5 (discussing the challenges of defining "good" and "bad" in the context of model behavior).

274. *See* Ryan Lowe & Jan Leike, *Aligning language models to follow instructions*, Openai (Sept. 2, 2023), https://openai.com/research/instruction-following (for an introduction to InstructGPT, a model that is aligned with human feedback).

275. Paul Christiano, Jan Leike & Tom B. Brown et al., Deep reinforcement learning from human preferences (2017) (unpublished manuscript), https://arxiv.org/abs/1706.03741v1; Long Ouyang, Jeff Wu & Xu Jiang et al., Training language models to follow instructions with human feedback (2017) (unpublished manuscript), https://arxiv.org/pdf/2203.02155.pdf.

276. In the reinforcement learning setting, data is not labeled as explicitly as it is in discriminative setting, e.g., our example of an image classifier, where each training data image has a label of either `cat` or `dog`. *See supra* Part I.A.2a. Instead, generations may be labeled "good" or "bad" based on human feedback, and the reinforcement learning algorithm updates the model in response to that feedback.  In RLHF, feedback is generated by a person interacting with the system; however, RL can also use feedback automatically generated by an algorithm specification.  *See* Yuntao Bai, Saurav Kadavath & Sandipan Kundu et al., Constitutional AI: Harmlessness from AI Feedback (2022) (unpublished manuscript), https://arxiv.org/abs/2212.08073 (using reinforcement learning with AI-generated feedback). )

While we have provided examples with user-generated feedback, most generative-AI companies begin model alignment prior to deployment or release.[278] Before making models publicly available, these companies contract with firms, like Scale AI,[279] that simulate the user feedback process. These firms typically employ people to label generations as "good" or "bad," according to guidance from the generative-AI company. In general, the process of model alignment is a critical part of the supply chain. It serves as a mechanism for steering models away from generating potentially harmful outputs[280] and toward the policies of the company or organization that deployed the model.[281] In this respect, model alignment complements other techniques, like input-prompt and output-generation filtering,[282] in generative-AI systems.

## II. Tracing Copyright Through the Supply Chain

The hornbook statement of United States copyright doctrine is that original works of authorship are protected by copyright when they are fixed in a tangible medium of expression.[283] A defendant directly infringes when they engage in conduct implicating one of several enumerated exclusive rights (reproducing, publicly distributing, etc.),[284] with a work of their own that is substantially similar to a copyrighted work[285] because it was copied from that work.[286] Other parties may be held secondarily liable for conduct that bears a sufficiently close nexus to the infringement under one of several theories.[287]  Otherwise infringing conduct is legal when it is protected by one of several defenses, including the DMCA Section 512 safe harbors,[288] fair use,[289] or an express[290] or implied[291] license.  In addition, we consider conditions for which different remedies may be granted when courts find infringement:[292] damages and profits, statutory damages, attorney's fees, injunctions, and destruction of generative-AI models.  Finally, we discuss three types of copyright-like rights:  interference with copyright management information,[293] right of publicity, [294], and misappropriation.[295]

---

277. *See supra* Part I.C.7.

278. *See supra* Part I.C.6.

279. AI, *supra* note 156.

280. Samantha Cole, *'Life or Death:' AI-Generated Mushroom Foraging Books Are All Over Amazon*, 404 Media (Aug. 29, 2023), https://www.404media.co/ai-generated-mushroom-foraging-books-amazon/. (describing a book on mushroom foraging built from generations, which mistakenly indicate that toxic mushrooms are safe to eat)

281. *See* James Manyika, *An overview of Bard: an early experiment with generative AI* (Aug. 17, 2023), https://ai.google/static/documents/google-about-bard.pdf; OpenAI, *Our approach to AI safety*, OpenAI (Apr. 5, 2023), https://openai.com/blog/our-approach-to-ai-safety; Deep Ganguli, Amanda Askell & Nicholas Schiefer et al., The Capacity for Moral Self-Correction in Large Language Models (2023) (unpublished manuscript), https://arxiv.org/abs/2302.07459 (documenting safety considerations, alignment, and RLHF at Google, OpenAI, and Anthropic).

282. *See supra* Part I.C.7.

283. *See infra* Part II.A.

284. *See infra* Part II.B.

285. *See infra* Part II.C.

286. *See infra* Part II.D.

This Part applies this orthodox, uncontested statement of copyright law to the generative-AI supply chain.[296] It takes up these issues in the above order — the same logical order that they typically arise in an copyright lawsuit — to analyze the copyright implications of each link in the supply chain.  Our goal is to be careful and systematic, not to say anything dramatically new.

## A. Authorship

Copyright protects "(1) original works of authorship (2) fixed in any tangible medium of expression."[297]  "Original, as the term is used in copyright, means only that the work was independently created by the author (as opposed to copied from other works), and that it possesses at least some minimal degree of creativity."[298]  Fixation is satisfied when the work is embodied in a tangible object in a way that is "sufficiently permanent or stable to permit it to be perceived, reproduced, or otherwise communicated for a period of more than transitory duration."[299]

We start with fixation.  Unfixed works have no interaction with the generative-AI supply chain.  A work must be fixed to be used as training data.  Truly ephemeral creations, like unobserved dances and songs that are never recorded, will never be captured in a way that can be used as an input to a

---

287. *See infra* Part II.E (direct infringement); *infra* Part II.F (indirect infringement).

288. *See infra* Part II.G.

289. *See infra* Part II.H.

290. *See infra* Part II.I.

291. *See infra* Part II.J.

292. *See infra* Part II.K.

293. *See infra* Part II.L.

294. *See infra* Part II.M.

295. *See infra* Part II.N.

296. *See supra* Part I.C.

297. 17 U.S.C. § 102(a) (numbering added).

298. Feist Publ'ns v. Rural Tel. Serv. Co., 499 U.S. 340, 345 (1991).

299. 17 U.S.C. § 101 (definition of "fixed").

training algorithm. Datasets, models, applications, prompts, and generations are all fixed in computers and storage devices.

Once it is fixed, however, any kind of original expression can be used as inputs for generative AI. Copyrightable subject matter explicitly includes "literary works" (e.g. poems, novels, FAQs, and fanfic),[300] "musical works" (e.g., sheet music and MIDI files)[301] "pictorial . . . works" (e.g. photographs),[302] "audiovisual works" (e.g., Hollywood movies and home-recorded TikToks),[303] "sound recordings" (e.g., pop songs and live comedy recordings),[304] and more. But this list is nonexclusive. Any kind of creative expression that appeals to the eye or the ear is copyrightable.[305] And copyright law does not discriminate among works based on their quality, their morality, or their importance.[306]

Instead, the originality requirement distinguishes material that was created by a human author from facts that "do not owe their origin to an act of authorship."[307] In addition, some types of material are never copyrightable, including any "idea, procedure, process, system, method of operation, concept, [or] principle."[308] In practice, this means that the copyright in some works (e.g., product photographs) will be "thinner" and protect fewer aspects of the works than the "thicker" copyrights in others (e.g, abstract art), because the "range of creative choices that can be made in producing the works is narrow."[309] In particular, any copyright in computer software — which is treated as a "literary work" for copyright purposes — typically excludes a great deal of functional material, such as efficient algorithms or coding conventions required by the choice of programming language.[310]

---

**300**. 17 U.S.C. § 102(a)(1).

**301**. *Id.* § 102(a)(2).

**302**. *Id.* § 102(a)(5).

**303**. *Id.* § 102(a)(6).

**304**. *Id.* § 102(a)(7).

**305**. Christopher Buccafusco, *Making Sense of Intellectual Property Law*, 97 Cornell L. Rev. 501 (2012).

**306**. Bleistein v. Donaldson Lithographing Co., 188 U.S. 239, 251 (1903).

**307**. Feist Publ'ns v. Rural Tel. Serv. Co., 499 U.S. 340, 347 (1991).

**308**. 17 U.S.C. § 102(b).

**309**. Rentmeester v. Nike, Inc., 883 F.3d 1111, 1120 (9th Cir. 2018).

**310**. Pamela Samuelson, *Functionality and Expression in Computer Programs: Refining the Tests for Software Copyright Infringement*, 31 Berkeley Tech. L.J. 1215 (2016).

### Data

As a result, some of the individual examples that serve as training data[311] are uncopyrightable. For example, birdsong-recognition AIs are trained on recordings of birds.[312] Currently, synthetic training data,[313] which are produced by generative-AI systems and then used as inputs for training other generative-AI models,[314] are not copyrightable.[315] But other items are copyrightable, and those copyrights will be held by a variety of authors: photographers, writers, illustrators, musicians, programmers, and other creators of all stripes.

### Training Datasets

Moving forward along the supply chain, then, different datasets[316] will include different amounts and proportions of copyrighted material. A dataset of birdsong recordings will be entirely, or almost entirely, uncopyrighted. A dataset of illustrations, on the other hand, will contain numerous copyrighted works. A dataset of photographs paired with synthetic captions will contain both copyrighted and copyright-free works.[317]

Datasets *themselves* may be copyrightable as **compilations**,[318] "formed by the collection and assembling of preexisting materials or of data."[319] A compilation is copyrightable (separately from any copyright in the works it is assembled from) when the compilation itself features a sufficiently original "selection or arrangement."[320] Originality in selection is choosing *what to in-*

---

311. *See supra* Part I.C.2.

312. *See* Stefan Kahl, Connor M. Wood author & Holger Klinck, *BirdNET: A Deep Learning Solution for Avian Diversity Monitoring*, 61 Ecological Informatics 101236 (2021). Animals are not recognized as "authors" for copyright purposes. *See* Naruto v. Slater, 888 F.3d 418 (9th Cir. 2018).

313. For discussion of synthetic data in the supply chain, *see supra* Part I.B; *supra* Part I.B.2b; *supra* Part I.C; *supra* Part I.C.1; *supra* Part I.C.7.

314. Gokaslan, Cooper & Collins et al., *supra* note 39; Gunasekar, Zhang & Aneja et al., *supra* note 155.

315. The US Copyright Office has argued that there is no human author, making such works ineligible for copyright. Thaler v. Perlmutter, No. 22-1564 (D.D.C date). *See infra* Part II.A (discussing generations).

316. *See supra* Part I.C.3.

317. Gokaslan, Cooper & Collins et al., *supra* note 39 (discussing the CommonCatalog dataset, which contains training-data examples that each consist of a Creative-Commons licensed image and a corresponding, synthetically generated captions).

318. 17 U.S.C. § 103(a).

319. § 101 (definition of "compilation").

320. Feist Publ'ns v. Rural Tel. Serv. Co., 499 U.S. 340, 348 (1991).

*clude* in the dataset; originality in arrangement is choosing *how to organize* the dataset. Every dataset is based on extensive curation,[321] but in some cases it is easier to identify the specific choices that went into intentionally creating a dataset with particular desired attributes.[322] The LAION-Aesthetics dataset, for example, was created by training a discriminative model[323] to predict the ratings that humans would give images, and then using the model to select "high visual quality" images from a much larger dataset.[324]

### Pre-Trained/Base Models

Attributing authorship for models is trickier to classify for two reasons.[325] First, there is the question of whether a model possesses the necessary "modicum of creativity" to be a work of authorship at all.[326] In some cases, the answer is probably "no": applying an existing algorithm and well-known architecture to an existing dataset[327] does not involve sufficient creative choices. Any expression in such a model merges into the idea and is uncopyrightable.[328] But it is possible that other models are works of authorship. For one thing, when a training dataset is curated specifically for training a base model, the model may supplant the dataset as the relevant 'work' from the data curation process, just as a finished film is regarded as the 'work' rather than the (much larger) dataset of raw footage.[329] In such a case, the model would in-

---

**321.** *See supra* Part I.C.3. *See generally* Katherine Lee, Daphne Ippolito & A. Feder Cooper, The Devil is in the Training Data (2023) (unpublished manuscript), *in* Lee, Cooper, Grimmelmann & Ippolito, *supra* note 65, at 5.

**322.** Indeed, individual data examples can contain multiple components, whose curation could make the individual example (i.e., the assemblage of multiple components) eligible for copyright. For example, text-to-image models are trained on image-caption pairs; the image, the caption, and the image-caption pair (as a compilation) could each potentially constitute distinct works of authorship, and each of these copyrights could be owned by someone other than the dataset creator or curator.

**323.** *See supra* Part I.A.2a.

**324.** Christoph Schuhmann, *LAION-Aesthetics*, LAION (Aug. 16, 2022), https://laion.ai/ blog/laion-aesthetics/.

**325.** It is worth noting that many model trainers creators certainly believe that models are copyrightable, and have released those models under licenses that are only intelligible if there is something copyrightable to license in the first place.

**326.** Feist Publ'ns v. Rural Tel. Serv. Co., 499 U.S. 340, 346 (1991).

**327.** With standard choices of hyperparameters, on standard hardware, etc.

**328.** *See generally* Pamela Samuelson, *Reconceptualizing Copyright's Merger Doctrine*, 63 J. Copyright Soc'y USA 417 (2016) (describing merger doctrine).

**329.** *See generally* Margot E. Kaminski & Guy A. Rub, *Copyright's Framing Problem*, 64 UCLA L. Rev. 1102 (2017) (discussing problem of identifying the 'work' in copyright cases).

herit the creative choices that went into curating the dataset. For another, base models are often the results of extensive design processes that involve novel architectures and algorithms. While these processes are not themselves copyrightable,[330] and originality in a process is not a guarantee that the outputs are copyrightable,[331] in some cases, a model's creators[332] will have made creative choices that imbue the model with copyrightable expression.

The second way in which the copyrightability of models is tricky is that they could be described in several different ways under copyright doctrine. One view is that a model is a compilation of its training data — the model is simply a different and complicated arrangement of training examples. Another view is that a model is a **derivative work** of its training data — "a work based upon one or more preexisting works . . . in which [those works are] recast, transformed, or adapted."[333] A derivative work (think of a translation of a novel, a recording of a song, or an action figure based on a character from a movie) combines the authorship in an existing (or "underlying") work with new authorship. The substantive difference between the two is that in a compilation, the underlying works are present in substantially unmodified form, whereas in a derivative work the underlying work is "recast, transformed, or adapted." The line dividing the two characterizations is somewhat metaphysical, but it has consequences in some corners of copyright doctrine, which could in turn have consequences for pre-trained models.[334]

### *Fine-Tuned Models and Aligned Models*

Both of the authorship considerations that we raise above for pre-trained models also apply to fine-tuned and aligned models. We start with the second point, which is simpler: like pre-trained models, both fine-tuned and aligned models will face similar issues of categorization for copyright law. A fine-tuned and/or aligned model will typically be a derivative work of the base model it was trained from.

The first point — that training choices can imbue models with creative attributes — leads to different observations for fine-tuning and model alignment. There is an argument to be made that fine-tuning is, by definition, a

---

**330**. *See* 17 U.S.C. § 102(b).

**331**. *See* James Grimmelmann, *Three Theories of Copyright in Ratings*, 14 Vand. J. Ent. & Tech. L. 851, 878–79 (2011) (criticizing theory that outputs "resulting from a minimally creative process" are thereby copyrightable).

**332**. In this case, this includes the parties that designed the architectures and algorithms.

**333**. 17 U.S.C. § 101 (definition of "derivative work).

**334**. *See, e.g.,* § 203(b)(1) (allowing the creator of an authorized derivative work to continue using it after the author terminates the license in accordance with a statutory procedure).

creative process. The model trainer is typically optimizing the model's be­havior in generating specific desired outputs — the kind of nexus between human choices and resulting material that characterizes copyrightable au­thorship.[335] The same is true for model alignment. Further, if, for example, the prompt is incorporated as part of the input to RLHF,[336] then the prompt serves as training data that could update the model. In this case, said training data itself is created in a process that includes human choices and has been crafted with specific creative goals in mind.

The prompt, though considered an input to generation, raises additional authorship considerations for both fine-tuning and alignment. As discussed above, when the user of the service supplies a prompt to a generative-AI sys­tem, the service host may save that prompt for later use. The service host may use the prompt as additional training data for fine-tuning or aligning the existing model, or for training another model altogether.[337] As a re­sult, fine-tuning and alignment are stages in the supply chain during which copyrighted data can find its way into a generative-AI system — where ei­ther the user of the service is the copyright holder, or they have prompted or fine-tuned[338] with content for which another entity is the copyright holder. For example, it is currently technologically feasible to prompt a text-to-text system with an entire book.[339] It may be possible to implement content fil­ters to catch known copyrighted material and remove it from training and alignment data, but such implementation considerations typically fall within other aspects of the generative-AI system, rather than the model.[340] Addi­tionally, there could be an express[341] or implied license[342] for user-inputted data, for the cases in which the user of the service is the copyright holder. There are also separate considerations for infringement.[343] and safe harbors,[344] which we address below.

---

335. *See generally* Dan L. Burk, *Thirty-Six Views of Copyright Authorship, by Jackson Pollock*, 58 Hous. L. Rev. 263 (2020) (discussing causal elements of authorship); Shyamkrishna Balganesh, *Causing Copyright*, 117 Colum. L. Rev. 1 (2017) (same).

336. *See supra* Part I.C.8.

337. *See supra* Part I.B.7; *supra* Part I.B.8

338. *See supra* Part I.C.6 (discussing fine-tuning APIs exposed to end-users in deployed services).

339. Anthropic, *supra* note 113.

340. *See infra* note 448 and accompanying text (for a discussion of the challenges of iden­tifying copyrighted data); *infra* note 654 and accompanying text (for a discussion of Copilot's output filters).

341. *See infra* Part II.I.

342. *See infra* Part II.J.

343. *See infra* Part II.E; *infra* Part II.F.

344. *See infra* Part II.G.

### Deployed Services

It is well-established that software is copyrightable.[345]  The non-model parts of a user-facing application or developer API will be protected by copyright (subject to the functionality screen noted above).  Also, as noted above, it is also possible for content filters to be implemented within the overarching generative-AI system that is hosted in the service.  It is at this stage of the supply chain where such filters could, for example, choose not to store user-inputted prompts (that either they or other(s) have authored).

### Generations

Generations raise a doctrinal question that has been debated for decades: who, if anyone, owns the copyright in the output of a computer program?[346]  Although some commentators have argued that the program itself should be regarded as the author, computer authorship is squarely foreclosed by U.S. copyright law.[347]  Computers are not capable of playing the social roles that society and the legal system expect and require of authors.[348]  So far, U.S. courts have held firm to this line for AI generations.  In *Thaler v. Perlmutter*, the court upheld the Copyright Office's refusal to register copyright in an image allegedly "autonomously created by a computer algorithm running on a machine."[349]  The Copyright Office had held that the image lacked human authorship, and the court agreed: computer programs, like animals, are not "authors" within the meaning of the Copyright Act.[350]

Instead, the author (and thus copyright owner) of a generation — if anyone — is some human connected to the generation.[351]  The four immediately relevant possibilities are (1) an author or authors whose works the model was trained on, (2) some entity in the generative-AI supply chain (e.g., the model trainer, model fine-tuner, or application developer), (3) the user who prompted the application or API for the specific generation, or (4) no one. As between these four possibilities, there is no one-size-fits-all answer.

---

345. *See generally* Comput. Assocs. Intern., Inc. v. Altai, 982 F.2d 693 (2d Cir. 1992) (standard case on software copyright); Pamela Samuelson, Randall Davis, Mitchell D. Kapor & Jerome H. Reichman, *A Manifesto Concerning the Legal Protection of Computer Programs*, 94 Colum. L. Rev. 2308 (1994) (lucid and time-honored analysis of software copyright).

346. Pamela Samuelson, *Allocating Ownership Rights in Computer-Generated Works*, 47 U. Pitt. L. Rev. 1185 (1985).

347. James Grimmelmann, *There's No Such Thing as a Computer-Authored Work – And It's a Good Thing, Too*, 39 Colum. J.L. & Arts 403 (2016).

348. Carys Craig & Ian Kerr, *The Death of the AI author*, 52 Ottawa L. Rev. 31 (2020).

349. Thaler v. Perlmutter, No. 22-1564 (D.D.C date).

350. *Id.*

351. The use of synthetic data complicates the question of authorship even further.  The generations produced by a model trained on synthetic data generated by another model are doubly removed from the original training data.  All of the authorship questions discussed in this Article arise twice over, once for each model and its pipeline.  *See supra* note 313 and accompanying text (discussing synthetic data in generative AI).

As framing for our analysis for these different possibilities, we first note that a generation is a compilation in the trivial sense in the same way that other works are all compilations. It also may seem intuitively attractive to consider generations to be analogous to collages. However, while this may seem like a useful metaphor,[352] it can be misleading in several ways. For one, an artist may make a collage by taking several works and splicing them together to form another work. In this sense, a generation is not a collage: a generative-AI system does not take several works and splice them together. Instead, as we have described above, generative-AI systems are built with models trained on many data examples.[353] Moreover, those data examples are not explicitly referred back to during the generation process. Instead, the extent that a generation resembles specific data examples is dependent on the model encoding in its parameters what the specific data examples look like, and then effectively recreating them.[354] Ultimately, it is nevertheless possible for a generation to look like a collage of several different data examples;[355] however, it is debatable whether the the process that produced this appearance meets the definition for a collage. There is no author "select[ing], coordinat[ing], or arrang[ing]"[356] training examples to produce the resulting generation.

With this in mind, we assess the four relevant authorship possibilities for generations. We start with a generation that closely resembles a work in the training set. If the generation is actually identical to the training example — if it contains no original expression beyond what was present in the input work — then it is simply a copy of that underlying work and not a new copyrightable work at all,[357] Of course the copyright owner remains the original author, i.e., possibility (1). If the generation is, however, a derivative work of the underlying work that incorporates new authorship, a new copyright may subsist in it.[358]  If the generation infringes, then it is uncopyrightable and the answer is possibility (4): there is no separate copyright in the generation, even though it contains original authorship.[359]  In such a case, the underlying copyright effectively also gives control over the generation; the user has in effect performed uncompensated creative labor for the benefit of the underlying copyright owner.[360]

---

352. *See* Cooper, Lee, Grimmelmann & Ippolito et al., *supra* note 22, at 4–5 (detailing how metaphors can be both helpful and misleading for intuiting the behavior of generative-AI systems).

353. *See supra* Part I.B; *supra* Part I.C.4.

354. *See infra* Part II.C.

355. *See infra* Part II.H.

356. 17 U.S.C. § 101 (definition of "compilation").

357. *See infra* Part II.C (concerning memorized training data and substantial similarity).

Assuming, however, that the generation is sufficiently distinct from training data not to be "used unlawfully," a copyright owned by one of its creators may arise.[361]  Some models and applications will produce original generations with minimal user input, which is possibility (2) above.  The Draw Things iOS app, for example, suggests the prompt `"8k resolution, beautiful, cozy, inviting, bloomcore, decopunk, opulent, hobbit-house, luxurious, enchanted library in giverny flower garden, lily pond, detailed painting, romanticism, warm colors, digital illustration, polished, psychadelic, matte painting trending on artstation."` The user who taps "Generate" on the app user interface has contributed no authorship to the resulting image. This Person Does Not Exist is a website that creates a new (and uncannily realistic) deepfake photograph of a nonexistent person each time it is reloaded. The user who visits the site and clicks "reload" is not an author.  If anyone can claim authorship credit here, it is the creators of these apps.

In other cases, the user will make substantial creative inputs through their choice of prompt.  In addition to the authorship inhering in the prompt itself, two additional factors push towards making the user the copyright owner rather than the developer — i.e., possibility (3) from above.  First, there is their causal responsibility for making the generation exist;[362] here, as in infringement, copyright law may care who "pushes the button."[363]  Second, the providers of many generation applications have decided that as a practical matter they are uninterested in asserting copyright over the outputs.  This is a business choice first and a copyright matter second, but widespread busi-

---

358. *See* 17 U.S.C. § 103(b) ("The copyright in such [a derivative] work is independent of . . . any copyright in the preexisting material.").

359. 17 U.S.C. § 103(a) ("[Copyright] protection for a [derivative] work . . . does not extend to any part of the work in which such material has been used unlawfully."). The courts have also held, illogically, that even if the underlying work was used with the copyright owner's permission, it is uncopyrightable unless the owner also consents to a derivative copyright. *See, e.g.,* Gracen v. Bradford Exch., 698 F.2d 300 (7th Cir. 1983).

360. *See, e.g.,* Anderson v. Stallone, 11 U.S.P.Q.2d 1161 (C.D. Cal. 1989).

361. For derivative copyright purposes, lawful use includes fair use. *See, e.g.,* Keeling v. Hars, 809 F.3d 43 (2d Cir. 2015).

362. Balganesh, *supra* note 335.

363. Fox Broad. Co. v. Dish Network LLC, 160 F. Supp. 3d 1139, 1169 (C.D. Cal. 2015).

ness practices often affect courts' decisions about how to allocate copyright ownership.[364]

But it is too hasty to say that the user is necessarily the owner of copyright in a generation, even once the training-data authors and model developers are out of the picture. It is also possible that *no one at all* owns a copyright in the generation (possibility (4)). The problem is that the generation may not be the product of sufficient human authorship. Consider the prompt.[365] `"Scary lighthouse"` is too short to contain sufficient originality to support a copyright;[366] short phrases are often uncopyrightable.[367] If this phrase does not have the necessary modicum of creativity by itself, it seems unlikely that the additional choice to use it as a prompt is enough to put it over the threshold.[368] Another way of looking at the problem is that prompts like `"Scary lighthouse"` do not sufficiently constrain the output to make it the product of human authorship. As the Copyright Office put it when rejecting copyright in images created with Midjourney,

> Because of the significant distance between what a user may direct Midjourney to create and the visual material Midjourney actually produces, Midjourney users lack sufficient control over generated images to be treated as the "master mind" behind them. . . . [T]here is no guarantee that a particular prompt will generate any particular visual output. Instead, prompts function closer to suggestions than orders, similar to the situation

---

**364.** *E.g.,* Aalmuhammed v. Lee, 202 F.3d 1227, 1233 (9th Cir. 2000) (deferring to Hollywood practice of treating *auteur* directors as the "master mind[s]" behind films); Thomson v. Larson, 147 F.3d 195 (2d Cir. 1998) (deferring to theatrical crediting practices in holding that a dramaturg was not a co-author of a musical).

**365.** Mark Lemley argues that in fact the prompt is the relevant unit of originality and is in effect the work itself. Mark A. Lemley, How Generative AI Turns Copyright Law on its Head (2023) (unpublished manuscript), https://papers.ssrn.com/sol3/papers.cfm?abstract_id=4517702.

**366.** *Cf.* Magic Mktg. v. Mailing Servs. of Pittsburgh, 634 F.Supp. 769 (W.D. Pa. 1986) (holding the phrase "CONTENTS REQUIRE IMMEDIATE ATTENTION!" uncopyrightable).

**367.** 37 CFR § 202.1(a). This complicates the copyrightability of captions in image-caption datasets; it is possible that some captions are too short to contain sufficient originality. *See supra* note 322 and accompanying text (discussing image, caption, and image-caption pair copyrights).

**368.** *See* Jane C. Ginsburg & Luke Ali Budiardjo, *Authors and Machines*, 34 Berkeley Tech. L.J. 343 (2019) (advancing this argument); *see also* Burk, *supra* note 335 (exploring variations).

of a client who hires an artist to create an image with general directions as to its contents.[369]

This is not the only possible view. A counter might be that for pragmatic reasons the copyright system will or should assign authorship to the user and overlook their minimal contributions.[370] While many current generative-AI systems have primarily text-based interfaces where short prompts might not amount to much creativity, future generative AI systems will likely have different interfaces that introduce other ways of controlling outputs.[371] But for now, it is the law that some generations are uncopyrightable despite containing material that would easily qualify for copyright if they had been produced manually by a human.[372]

This conclusion, however, is not categorical; "some" is not "all." Not every prompt is too short to be copyrightable, and not every user is a spectator to AI generation.[373] Instead, some generations are the product of careful prompt engineering, in which users craft elaborate prompts to cause AI models to achieve specific aesthetic effects. These generations answer both of the objections above. These prompts are often long and intricate, running to dozens or hundreds of words, well above the short-phrase threshold. And these prompts are the result of an iterative creative process, in which the users have acquired a degree of mastery over the (putatively unpredictable) mod-

---

**369**. Letter from Robert J. Kasunic to Van Lindburg, *Re: Zarya of the Dawn (Registration # VAu001480196)* 9–10 (Feb. 21, 2023), https://www.copyright.gov/docs/zarya-of-the-dawn.pdf.

**370**. *See, e.g.,* Grimmelmann, *supra* note 347, at 413–14 (discussing this possibility, and its difficulties). As one canonical case puts it, "Having hit upon such a variation unintentionally, the 'author' may adopt it as his and copyright it." Alfred Bell & Co. v. Catalda Fine Arts, 191 F.2d 99 court, 105 (1951).

**371**. For example, Ideogram has style tags that can be added to the prompt to modify the output (*Ideogram.AI*, *supra* note 77).

**372**. *See* James Grimmelmann, *Copyright for Literate Robots*, 101 Iowa L. Rev. 657, 657 (2016) ("Almost by accident, copyright law has concluded that it is for humans only . . . ").

**373**. For example, a product called alpaca allows users to upload sketches and transform them into more-complete images with generative AI. Users can further control the generated images with text prompts. These user-provided sketches could have copyrights. As another example, some models have long context lengths, which enable them to process long segments of text as inputs. A user may prompt such a model with entire book as input; a system using retrieval-augmented generation may retrieve documents to guide the generation process that may themselves be copyrighted. *Alpaca ML*, Alpaca ML (2024), https://www.alpacaml.com/ (describing the alpaca software product). *See supra* Part I.C.7 (for a discussion on long-context lengths in the Claude product). *See supra* note 257 and accompanying text. *See supra* note 265 and accompanying text (describing retrieval augmented generation).

els they use, at least for specific types of outputs.[374] If an artist who flings a sponge against the wall in frustration is entitled to claim copyright in the resulting accidental spatter of paint, why not a user who deliberately crafts the perfect prompt?[375]

## B. The Exclusive Rights

It is helpful to break down the *prima facie* case of infringement by the relevant exclusive right, rather than by the stage of the generative-AI supply chain. There are five relevant exclusive rights:

- The right to "reproduce the copyrighted work in copies" (the **reproduction** right).[376]

- The right to "prepare derivative works based upon the copyrighted work" (the **adaptation** right).[377]

- The right to "distribute copies . . . of the copyrighted work to the public" (the **distribution** right).[378]

- The right to "perform the copyrighted work publicly" (the **performance** right).[379]

- The right to "display the copyrighted work publicly" (the **display** right).[380]

To summarize briefly, every stage in the generative-AI supply chain requires a potentially-infringing reproduction and thus implicates copyright. We examine the other exclusive rights, which raise interesting edge cases.

### The Reproduction Right

As relevant here, the reproduction right is triggered when a work is reproduced in "copies," which are defined as "material objects . . . in which a work is fixed by any method now known or later developed, and from which the work can be perceived, reproduced, or otherwise communicated, either directly or with the aid of a machine or device."[381] To be pedantic, a training

---

**374.** For a particularly disquieting example, see Emanuel Maiberg, *Inside the AI Porn Marketplace Where Everything and Everyone Is for Sale*, 404 MEDIA (Aug. 22, 2023), https://www.404media.co/inside-the-ai-porn-marketplace-where-everything-and-everyone-is-for-sale/.

**375.** *Alfred Bell*, 191 F.2d at 105 n.23.

**376.** 17 U.S.C. § 106(1).

**377.** 17 U.S.C. § 106(2).

**378.** *Id.* § 106(3).

**379.** *Id.* § 106(4), (6).

**380.** *Id.* § 106(5).

**381.** 17 U.S.C. § 101 (definition of "copies").

dataset is not a "copy" because the dataset is not a "material object." Instead, the *computer* or *storage device* on which a dataset is stored is the copy.

The same is true for models and generations.[382] All of them trigger the re-production right when they are created, because they are stored in material objects. Thus, the assembly of a dataset, the training of a model, the pro-duction of a generation, or a generative-AI system's use of a user-inputted prompt is a "reproduction" within the meaning of copyright law. All of these activities can infringe: the question is whether the resulting dataset, model, prompt, or generation is substantially similar[383] to the plaintiff's[384] copyrighted work.

One complication has to do with *how long* a work is fixed. Under the "RAM copy" doctrine, which dates to the 1990s, loading a copyrighted work into a computer's working memory can infringe.[385] (Doing so is often neces-sary to run a program or to perform a computation on data.) On the other hand, more recent caselaw has held that transient copies do not count for the reproduction right.[386] The leading case, *Cartoon Network LP, LLLP v. CSC Holdings*, held that a buffer that was overwritten every 2.4 seconds was not an infringing reproduction of works that passed through the buffer.

The temporal threshold is not generally an issue for the outputs of stages in the generative-AI supply chain. Datasets, models, applications, prompts, and generations are all typically stored for far longer than the 2.4 seconds in *Cartoon Network*. Instead, the threshold may be more important for the inputs to the different stages. For example, a training example needs to be loaded into working memory to train a model on it. But the details of *how long* the example remains in memory, and *how much it is modified* while it is there, will depend on the training algorithm and architectural details of the environment (e.g., how fast the processors are). Similar considerations apply to the generation process — with similar uncertainties. Some generations run in a fraction of a second; others take minutes or hours.

---

**382.** The same could also be said for individual data examples within the dataset, which is one of the reasons we distinguish between expressive works and their datafied counter-parts. *See supra* Part I.A.1; *supra* Part I.C.1; *supra* Part I.C.2.

**383.** *See infra* Part II.C.

**384.** Of course, there are different types of actors that can be responsible for each of these reproductions. For example, an application user could supply a reproduction of a copy-righted prompt (for which they do not hold the copyright), and the generative-AI sys-tem could in turn store that reproduction in memory. This could happen even for a generative-AI system that only trained its models on public-domain data (i.e., did not violate the reproduction right with respect to training).

**385.** MAI Sys. Corp. v. Peak Comput., 991 F.2d 511 (9th Cir. 1993).

**386.** Cartoon Network LP, LLLP v. CSC Holdings, 536 F.3d 121, 128–30 (2d Cir. 2008).

There is also the problem of purely *internal* reproductions: ones that occur only in the middle of the training or generation process. These algorithms compute numerous new values, and often overwrite them repeatedly to conserve memory. Consider, for example, one of the middle stages of the archaeologist generation in Figure 3. One of these stages might resemble a copyrighted work more closely than the final output. Again, whether these fall underneath the *Cartoon Network* threshold depends on the details of the algorithm and environment.[387]

### *The Adaptation Right*

While the reproduction right is about new copies of an existing work, the adaptation right is about new works based on an existing work. It is best understood as making clear that copyright in a work extends beyond literal similarity to incorporate changes of form, genre, and content such as translations, sequels, and film adaptations.[388] A training dataset is probably not a derivative work of any of the works in the dataset; it is more appropriately classified as a compilation "formed by the collection and assembling of preexisting materials."[389] A model is a good example of material that might or might not be an exact reproduction of the works it was trained on, but is more clearly a derivative work because it is "based on" its training data. Prompts might or might not be exact reproductions of existing works,[390] or they may be derivative works based on, for example, existing text or images. And generations are frequently derivative works of works in the training data, although whether and when a generation is a derivative of any particular work depends on similarity, discussed below.[391] Because the remedies for infringement of a work are the same, regardless of whether the defendant violated one exclusive right or several, it is an almost entirely scholastic exercise to try to identify the exact dividing lines at which the reproduction right leaves off and the adaptation right begins.[392]

---

**387.** Alternatively, there is a strong fair-use case these transient internal copies. *See* Grimmelmann, *supra* note 372 (summarizing caselaw).

**388.** *See generally* Daniel Gervais, *The Derivative Right, or Why Copyright Law Protects Foxes Better than Hedgehogs*, 15 Vand. J. Ent. & Tech. L. 785 (2013); Pamela Samuelson, *The Quest for a Sound Conception of Copyright's Derivative Work Right*, 101 Geo. L.J. 1505 (2013); Daniel Gervais, *AI Derivatives: The Application to the Derivative Work Right to Literary and Artistic Productions of AI Machines*, 52 Seton Hall L. Rev. 1111 (2022).

**389.** 17 U.S.C. § 101.

**390.** Anthropic, *supra* note 113.

**391.** *See infra* Part II.C.

**392.** The boundaries of the adaptation right are of greater importance in cases involving *unfixed* derivatives, where the reproduction right does not apply. *See* Lewis Galoob

More troublingly, it might be that the adaptation right can be infringed by derivative works that do not by themselves incorporate substantial expression from the plaintiff's work. In *Micro Star v. Formgen Inc.*, the defendant distributed fan-made levels for *Duke Nukem 3D*.[393] The level file format consisted entirely of geometry describing where the *Duke Nukem 3D* game engine should place walls and objects; the engine would then perform rendering using copyrighted art assets, but "[t]he MAP file . . . does not actually contain any of the copyrighted art itself; everything that appears on the screen actually comes from the art library."[394] Nonetheless, the court held that these files were infringing derivative works because "the stories told in the N/I MAP files are surely sequels, telling new (though somewhat repetitive) tales of Duke's fabulous adventures."[395]

A broad way to read *Micro Star* is to reason that models implicate the adaptation right when they "reference" the works they were trained on.[396] This test might be satisfied as long as any identifiable portion of a model was causally derived from a training example. However, reliable attribution of training examples in resulting generations remains an open research question.[397] A narrower reading would be that the model must also be capable of generating a substantially similar output — just as the audiovisual experience of playing a user-made *Duke Nukem 3D* level is substantially similar to the audiovisual experience of playing a canonical level created by 3D Realms.[398]

### The Distribution Right

The distribution right applies when the defendant "distribute[s] copies . . . to the public by sale or other transfer of ownership."[399] Internet-era caselaw confirms that downloads and peer-to-peer transfers infringe the distribution

---

Toys, Inc. v. Nintendo of Am., Inc., 964 F. 2d 96, 967–69 (9th Cir. 1992) (erroneously holding that unfixed modifications of video games produced by altering bytes as they are read from a game cartridge are not derivative works). The boundaries also matter in cases involving the physical transfer of a copy from one substrate to another; here, there is a fixed copy, but there is no reproduction of it. *See, e.g.,* Lee v. ART Co., 125 F.3d 580 (7th Cir. 1997) (holding that the mounting of a page cut from a book on a ceramic tile does not create a derivative work).

**393**. Micro Star v. Formgen Inc., 154 F.3d 1107 (9th Cir. 1998).

**394**. *Id.* at 1110.

**395**. *Id.* at 1112.

**396**. *Id.*

**397**. *supra* note 139 and accompanying text (regarding the challenges of assigning "attribution" or "influence").

**398**. *See generally* MDY Indus., LLC v. Blizzard Ent., 629 F.3d 928 (9th Cir. 2010) (discussing "dynamic" aspects of copyrightable expression in video games).

**399**. 17 U.S.C. § 106(3).

right, so that the essence of the right is giving a stranger a copy, whether or not the copy previously existed.[400] Technically, the distribution right is not triggered by merely making a work available for download, but only when someone actually downloads it.[401] That said, in most interesting cases involving generative AI, making an artifact available is followed by an actual distribution.

When there is only a single entity involved in hosting a service, it is arguably not a distribution to assemble a dataset, train a model, program an application, input a prompt, or produce a generation. All of these activities involve only internal copying performed by the single hosting entity. They may result in reproductions and derivative works (as discussed above), but not distributions. The same is true when one party carries out multiple stages — for example, when a model trainer collects its own training data, or when a model owner creates test generations for its own use). Internal copying is not public distribution.

Instead, the distribution right is implicated when parties interact. In our model of the supply chain, there are at least five such kinds of interactions:

- When a dataset creator or curator makes the dataset available to model trainers.[402]

- When a model trainer makes the model available for download (rather than for interactive use through a web interface or API).[403]

- When a service produces generations for users on demand.

- When a generation-time plugin retrieves content from an external source, which it then may use to produce a generation.[404]

In addition, when someone who has a dataset, model, prompt, or generation shares it, as is, with others, this is also a distribution. This last case is particularly relevant for open-source models, like those in the Llama family, which are often widely downloaded, shared, and re-uploaded.

---

**400**. Perfect 10, Inc. v. Amazon.com, Inc., 508 F.3d 1146, 1162–63 (9th Cir. 2007); London-Sire Recs., Inc. v. Doe 1, 542 F. Supp. 2d 153, 172 (D. Mass. 2008).

**401**. *London-Sire Recs.*, 542 F. Supp. 2d at 172.

**402**. This can happen in a variety of ways: e.g., open-sourcing a dataset, licensing a dataset, or some other contract between a dataset compiler/owner and a model trainer. For an example of the third case, consider how MosaicML is a platform for training and fine-tuning models for its clients.

**403**. *See supra* Part I.C.4.

**404**. *See supra* Part I.C.7.

### The Display and Performance Rights

The display and performance rights characteristically involve human perception of a work. (The difference is that a display is static in time, while a performance is dynamic.) Models are not human-perceptible in any meaningful way, so it is hard to see how a model as such could infringe the display or performance rights. Similarly, while the individual works *within* a dataset can be perceptible, the dataset as a whole is not. Thus, for most practical purposes, only generations implicate these two rights.[405]

Like the distribution right, the display and performance rights are qualified by the word "public," so they apply only when the defendant makes the work perceptible *to others*. When a service produces a generation for a user, it will typically be a public display (for text and images) or a public performance (for audio and video). But in such a case, the generation will usually also be a reproduction and/or an adaptation, so the display and performance rights add relatively little. (In addition, if the user can download the generation, that will be a public distribution.)

One exceptional case when the display and performance rights may matter is for transient generations. Midjourney, for example, displays intermediate stages of the denoising process to users, as seen above in Figure 3. If one of those stages — but *not* the final result — infringes, then there might be a display without a reproduction or distribution.[406] Similarly, if an audio or video generation is played live for a user as it is created, but is not stored or made available for download, then this would be a performance without a reproduction or distribution.[407]

### C. Substantial Similarity

Substantial similarity is a qualitative, factual, and frustrating question. Two works are substantially similar to "the ordinary observer, unless he set out to detect the disparities, would be disposed to overlook them, and regard their

---

**405.** Some services display user-supplied prompts as examples for other users, as suggestions for how to use the service. These are also public displays. A service, however, can easily protect itself from copyright liability for these prompts. It can require users to provide a license allowing their prompts in this way. As long as the number of such prompts displayed is small, the provider could potentially screen them manually for signs of infringement.

**406.** *See* Cartoon Network LP, LLLP v. CSC Holdings, 536 F.3d 121 (2d Cir. 2008) (discussing transience exception to reproduction result).

**407.** *See* United States v. Am. Soc. of Composers, 627 F.3d 64 (2d Cir. 2010) (discussing reverse situation, a download without a performance).

aesthetic appeal as the same."[408] A common test is a "holistic, subjective comparison of the works to determine whether they are substantially similar in total concept and feel."[409] This is not a standard that can be reduced to a simple formula that can easily be applied across different works and genres.[410]

In addition, except in clear cases, substantial similarity is typically a jury question.[411] Juries, unlike judges, are not required to provide reasoned elaboration justifying their verdicts. A typical case in which substantial similarity is genuinely contested, therefore, will provide little guidance for future cases. As a result, it is simply impossible to provide clear, accurate, and actionable predictions of substantial similarity in the mine-run of close cases.

### Data

Substantial similarity of data poses no new issues distinctive to generative AI. Individual works included in training datasets can be compared to the plaintiff's work using the traditional substantial similarity test.

### Training Datasets

Training datasets contain complete literal copies of millions of digitized copyrighted works. Complete literal copying is the paradigm case where substantial similarity is present as a matter of law.

Some datasets may represent works in specialized file formats, or may compress or transform them in ways that remove some of the information present in the work.[412] In these cases, the substantial similarity inquiry may involve returning these modified works to human-perceptible form (i.e., rendering them), followed by a traditional comparison. However, even when scaled down or partially noised,[413] as long as the original is recognizable, that will often be enough to support a finding of substantial similarity.[414]

---

**408**. Peter Pan Fabrics, Inc. v. Martin Weiner Corp., 274 F.2d 487, 489 (2d Cir. 1960) (Hand, J.).

**409**. Rentmeester v. Nike, Inc., 883 F.3d 1111, 1118 (9th Cir. 2018) (internal quotation omitted).

**410**. *But see* Scheffler, Sarah, Eran Tromer & Mayank Varia, *Formalizing Human Ingenuity: A Quantitative Framework for Copyright Law's Substantial Similarity*, *in* 2022 Proc. Symposium on Comput. Sci. & L. 37 (2022) (describing a principled computational basis for comparing works).

**411**. Tanksley v. Daniels, 902 F.3d 165, 171 (3d Cir. 2018).

**412**. For an interesting attempt to quantify the information present in a work and what it means to remove some of it, see Scheffler, Tromer & Varia, *supra* note 410.

**413**. E.g., as in the case of diffusion. *See supra* Part I.B.3b

**414**. *See* Perfect 10, Inc. v. Amazon.com, Inc., 508 F.3d 1146 (9th Cir. 2007).

**Training Set**                    **Generated Image**

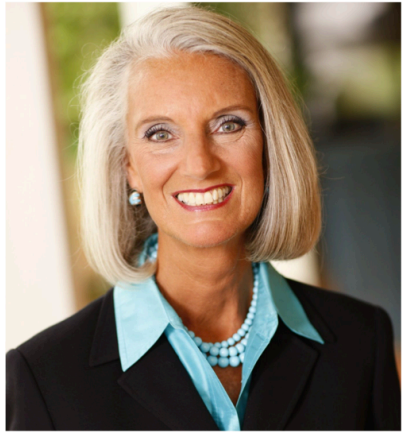  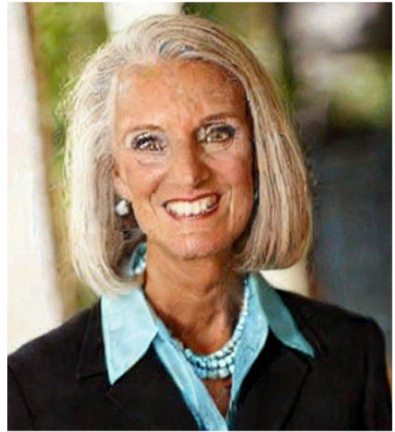

*Caption: Living in the light*          *Prompt:*
*with Ann Graham Lotz*           *Ann Graham Lotz*

Figure 5: An example of a memorized image in Stable Diffusion, taken from Carlini et al., *Extracting Training Data from Diffusion Models* (2023).

### Pre-Trained/Base Models

A model, as a collection of parameters, is different in kind from the copyrightable works it was trained on. Models are not themselves human-intelligible.[415] No viewer would say that the model has the same "total concept and feel" as a painting; no reader would say that it is substantially similar to a blog post; and so on.

That said, the Copyright Act does not require that copies be directly human-intelligible to infringe. A Blu-Ray is not directly intelligible by humans, either, but it counts as a "copy" of the movie on it. Indeed, all digital copies are unintelligible. Instead, they are objects "from which the work can be perceived, reproduced, or otherwise communicated . . . *with the aid of a machine or device.*"[416] Thus, even if a model is uninterpretable, it might still be possible to "perceive[]" or "reproduce[]" a copyrighted work embedded in its parameters through suitable prompting. The resulting generation will render the work perceptible.

---

**415**. *See supra* Part I.A.2 (describing model parameters as vectors of numbers).
**416**. 17 U.S.C. § 101 (emphasis added).

> Wow. I sit down, fish the questions from my backpack, and go through them, inwardly cursing [MASK] for not providing me with a brief biography. I know nothing about this man I'm about to interview. He could be ninety or he could be thirty. → **Kate** (James, *Fifty Shades of Grey*).
>
> Some days later, when the land had been moistened by two or three heavy rains, [MASK] and his family went to the farm with baskets of seed-yams, their hoes and machetes, and the planting began. → **Okonkwo** (Achebe, *Things Fall Apart*).

Figure 6: Two examples of memorized text in GPT-4, taken from Chang et al., *Speak, Memory: An Archaeology of Books Known to ChatGPT/GPT-4* (2023). In each case, when prompted with a sentence from a copyrighted book GPT-4 correctly fills in the name of a character.

Indeed, there is substantial evidence that many models have memorized copyrighted materials.[417] For example, Figure 5 shows how Stable Diffusion has memorized photographs. The memorized version is grainier and slightly shifted, but is immediately recognizable as the same photograph. Similarly, Figure 6 shows how GPT-4 must contain information from copyrighted books. GPT-4 can correctly fill in blanks in quotations from books; because the blanks consist of proper names of fictional characters, GPT-4 is not simply relying on its general knowledge of language.[418]

---

**417.** Nicholas Carlini, Florian Tramèr & Eric Wallace et. al., *Extracting Training Data from Large Language Models*, *in* 2021 30th USENIX Security Symposium (USENIX Security 21) 2633—2650 (2021) (GPT-2 memorizes training data); Nicholas Carlini, Jamie Hayes & Milad Nasr et al., Extracting Training Data from Diffusion Models (2023) (unpublished manuscript), https://arxiv.org/abs/2301.13188 (Stable Diffusion and Imagen memorize images); Kent K. Chang, Mackenzie Cramer, Sandeep Soni & David Bamman, Speak, Memory: An Archaeology of Books Known to ChatGPT/GPT-4 (2023) (unpublished manuscript), https://arxiv.org/abs/2305.00118 (suggestive evidence that GPT-4 memorizes training data).

**418.** *See* Chang, Cramer, Soni & Bamman, *supra* note 417. The composition of GPT-4's training data is not public. If we don't know what the training data is, we technically cannot say that the training data was memorized with complete certainty. Filling in the blank with proper names of fictional characters is suggestive that copyrighted content is part of the training dataset, but does not prove that verbatim memorization has taken place. Additionally, it is possible for popular fictional characters to be associated with plot summaries or the like, without the copyrighted content appearing in the training

From a practical litigation perspective, a model might memorize more works or fewer.[419] But it seems clear that at least some models memorize at least some works sufficiently closely to pass the substantial-similarity test.

On this view, a sufficient condition[420] for a model to count as a substantially similar copy of a work is that the model is capable of generating that work as an output.[421] Note that this is direct infringement, not secondary.[422] The theory is not that the generation is an infringing copy, and that the model is a tool in causing that infringement in the way that a tape-duplicating machine might be a tool in making infringing cassettes.[423] Rather, the theory is that the model itself is an infringing copy of each work it is capable of producing, regardless of whether that particular generation is ever made.[424]

### Fine-Tuned Models and Aligned Models

The prior discussion about whether pre-trained models are substantially-similar copies mostly carries over to fine-tuned models and models trained with alignment – but there are a few additional considerations as well. As

---

dataset. It is also possible the character names could be pulled in using generation-time plugins, but we note that the example in Figure 6 pre-dates GPT-4 plugins. *See supra* Part I.C.7 (regarding plugins).

**419**. Nicholas Carlini, Daphne Ippolito & Matthew Jagielski et al., *Quantifying Memorization Across Neural Language Models*, *in* 2023 Int'l Conf. on Learning Representations (2023) (quantifying extent of memorization in language models); Carlini, Hayes & Nasr et al., *supra* note 417 (quantifying memorization in diffusion-based image models).

**420**. We write "sufficient" rather than "necessary and sufficient" because there might also be *other ways* of inspecting the model that are capable of recovering training data. Obviously, this possibility involves some speculation about technological developments, but it is worth emphasizing that, as computer scientists develop techniques that improve the interpretability of models, the copyright treatment of models and generations may well change as a result.

**421**. This is a sticky technical problem. Research has shown that memorization is not easily identifiable, and thus the amount of memorization in a model is not always or easily quantifiable. In particular, the choice of memorization identification technique and available information (e.g., knowledge of the training dataset, context window, etc.) affect the amount of memorization that can be identified. *See, e.g.,* Carlini, Ippolito & Jagielski et al., *supra* note 419; Nasr, Carlini & Hayase et al., *supra* note 1.

**422**. *See infra* Part II.E (discussing direct and secondary infringement).

**423**. *See* A & M Recs., Inc. v. Abdallah, 948 F. Supp. 1449 (C.D. Cal. 1996).

**424**. Alert readers will note the similarity to the debate over whether the mere act of making a work available without a download infringes the distribution right. *See* London-Sire Recs., Inc. v. Doe 1, 542 F. Supp. 2d 153 (D. Mass. 2008). *See generally* Peter S. Menell, *In Search of Copyright's Lost Ark: Interpreting the Right to Distribute in the Internet Age*, 59 J. Copyright Soc'y USA 1 (2011).

a starting point, fine-tuned and aligned models are influenced by the pre-trained model from which they were produced.[425] Fine-tuning may reduce the amount of memorized content from the pre-training dataset, but does not prevent all such memorization[426] and does not explicitly remove copies of training examples (i.e., particular text or images) from the trained model. Similarly, alignment may encourage models not to generate potentially infringing content, but that does not mean the copyrighted content was removed from the model.[427]

Further, the above considerations have to do with the pre-training data, not the data incorporated in these later stages in the generative-AI supply chain. Both fine-tuning and alignment bring in additional data sources — data that could also be memorized in the resulting model. As a result, just like pre-trained models, fine-tuned and aligned models could each be an infringing copy of the works they are capable of producing; but they can be copies of the pre-training, fine-tuning, or alignment data.

### Deployed Services

Typical contemporary generative-AI services (e.g., web-based applications, APIs) use copyrightable works entirely through the trained models that they incorporate. Thus, if a model is infringingly substantially similar, then so is a service that incorporates the model. But, as discussed above, services also incorporate user prompts, and these prompts can incorporate copyrighted works.[428]) Prompting brings data into a deployed service; that data can be stored, and used to update the model or models that the service uses.[429] The same could be said for a deployed service's use of generation-time plugins that pull in additional data to augment generations.[430]

---

**425**. *See generally* Raffel, Shazeer, Roberts & Lee et al., *supra* note 65; Shayne Longpre, Gregory Yauney & Emily Reif et al., A Pretrainer's Guide to Training Data: Measuring the Effects of Data Age, Domain Coverage, Quality, & Toxicity (2023) (unpublished manuscript), https://arxiv.org/abs/2305.13169.

**426**. *See generally* Fatemehsadat Mireshghallah, Archit Uniyal & Tianhao Wang et. al., *An Empirical Analysis of Memorization in Fine-tuned Autoregressive Language Models*, *in* 2022 Proceedings of the 2022 Conference on Empirical Methods in Natural Language Processing 1816–1826 (2022).

**427**. While this is speculative, there is research indicating this may be the case. Prior work shows that models trained with alignment to be "safe" may be misaligned to produce "unsafe" content. Nicholas Carlini, Milad Nasr & Christopher A. Choquette-Choo et al., Are aligned neural networks adversarially aligned? (2023) (unpublished manuscript), https://arxiv.org/abs/2306.15447.

**428**. *See supra* Parts I.C.7, II.A, and II.B.

**429**. *See infra* Part II.G (discussing challenges of removing data from a service).

**430**. *See supra* Part I.C.7.

***Generations***

There is a spectrum of possible generation outputs. Generations could be:

1. Nearly identical to a work in the model's training data (i.e., memorized).

2. Similar to a work in the training data in some ways, but dissimilar from it in other ways.

3. Very dissimilar from all works in the training data.

Case (1) is straightforward: wholesale literal copying yields substantial similarity. Case (3) is also straightforward, because infringement is assessed on a work-by-work basis. A hypothetical viewer asked to compare the output to each work in the training dataset, one at a time, would say that it is not substantially similar to work 1, not substantially similar to work 2, and so on through work 89,128,097,032. Although it is in some sense based on all of the works in the training dataset, it does not infringe on any of them.[431]

Case (2) is more complicated, and more legally interesting. It is also likely to arise in practice precisely because it lies in between the two extremes. There are ample examples of memorized generations (case (1)), and ample examples of original generations (case (3)). Somewhere between them lies the murky frontier between infringing and non-infringing.

It is hard to make sweeping statements here because of the factual intensity and aesthetic subjectivity of similarity judgments. To quote Learned Hand on the idea-expression dichotomy, "Nobody has ever been able to fix that boundary, and nobody ever can."[432] Whether a particular generation is substantially similar or not is ultimately a jury question requiring assessment of audiences' subjective responses to the works. Generative AI will produce cases requiring this lay assessment, and it is impossible to anticipate in advance how lay juries will react to all of the possible variations. So, in the sections that follow, we will assume that lay audiences would say that some

---

**431**. While it may be straightforward to pose the question: "is the given generation substantially similar to work 1," it is not at all straightforward to answer. As we discussed before, training datasets are massive. *See supra* Part I.B.4. Manually comparing the generation to every single work in the dataset is infeasible; it would simply take too long. While automated methods could help identify works in the training set that are *likely to be* similar to the generation, there is no automated metric that can definitively say if two works are substantially similar. (*see generally* Scheffler, Tromer & Varia, *supra* note 410 (which proposes one possibility for a metric for identifying substantial similarity)). Even with automated methods, checking *every* generation that a system produces against every other work in the training dataset to evaluate similarity is extremely computationally expensive.

**432**. Nichols v. Universal Pictures Corp., 45 F.2d 119, 121 (2d Cir. 1930).

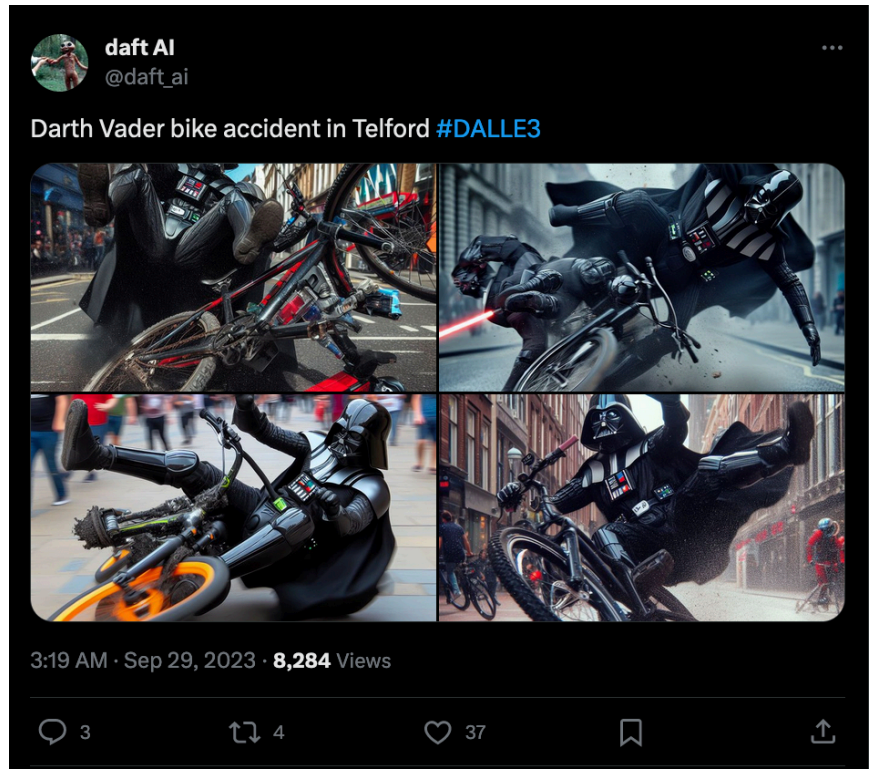

Figure 7: An example of the "Snoopy effect." When prompted to do so, DALL·E-3, integrated within a paid tier of ChatGPT, generates "photographs" with recognizable depictions of Darth Vader from Star Wars. Generated by Twitter user @daft_AI using DALL·E-3.

generated outputs will infringe, but that it will not be possible to perfectly predict which ones.[433]

Even if complete answers are impossible, however, there are some interesting questions worth considering. One has to do with what Matthew Sag calls the "Snoopy problem,"[434] which we will call the "Snoopy effect," so as to reserve judgment on whether it really is a problem. As Sag observes, cer-

---

**433.** Notably, providing guarantees that any given generated work might not potentially infringe copyright is impossible if the training data contains copyrighted data. This is simply because provable guarantees require formal definitions, and there are no widely accepted formal definitions of substantial similarity. *But see* Scheffler, Tromer & Varia, *supra* note 410 (providing a possible starting point). Instead, current machine-learning techniques focus on reducing the likelihood that generations from a model will closely resemble any of the model's training data.

**434.** Matthew Sag, *Copyright Safety for Generative AI*, Hous. L. Rev. (forthcoming).

tain characters are so common in training datasets that models have "a latent concept [of them] that is readily identifiable and easily extracted" (See Figure 7. Sag's example is that prompting Midjourney and Stable Diffusion with "snoopy" produces recognizable images of Snoopy the cartoon beagle. Characters are an unusual special case in copyright law; there are cases that seem to relax the rule that infringement is measured on a work-by-work basis, instead measuring the similarity of the defendant's character to one who appears in multiple works owned by the plaintiff.[435]

But the Snoopy effect is not confined to characters. For one thing, some works — and close variations on them — are simply so prevalent in training datasets that models readily memorize them. As an uncopyrighted example, Van Gogh's *Starry Night* is easy to replicate using Midjourney; Sag's paper includes a replication of Banksy's *Girl with Balloon.* This looks like substantial similarity.

Another variation of the Snoopy effect arises when a model learns an artist's recognizable *style.* ChatGPT can be prompted to write rhyming technical directions in the style of Dr. Seuss (Figure 8); the DALL·E-2 system can be prompted to generate photorealistic portraits of nonexistent people in the style of Dorothea Lange (Figure 9).[436] As with characters, these outputs have similarities that span a body of source works, even if they are not necessarily close to any one source work. The proper doctrinal treatment of style is a difficult question.[437]

It is also possible to trigger the Snoopy effect without explicit prompting. The archaeologist example in Figure 3 (and reproduced in higher resolution in Figure 10) was generated with the prompt `"an adventurous archaeologist with a whip and a fedora"`. The resulting images feature a dark-haired male character with stubble, wearing a brown jacket and white shirt, with a pouch slung across his shoulder. These are features associated with Indiana Jones, but neither the features nor the name `"indiana jones"` appear in the prompt. The same is true for the `"well-known plumber"` and `"pocket monsters"` in Figure 11: they clearly resemble the Nintendo character Mario, and associated characters from the video game, Super Smash

---

**435.** *E.g.,* DC Comics v. Towle, 802 F.3d 1012 (9th Cir. 2015). *See generally* Sag, *supra* note 434 (discussing caselaw and scholarship).

**436.** Stephen Casper, Zifan Guo & Shreya Mogulothu et al., Measuring the Success of Diffusion Models at Imitating Human Artists (2023) (unpublished manuscript), https://arxiv.org/abs/2307.04028 (measuring style imitation in text-to-image, diffusion-based models).

**437.** Benjamin L.W. Sobel, Elements of Style: A Grand Bargain for Generative AI (2023) (unpublished manuscript, on file with authors). A separate and non-trivial question is whether these generations violate authors' right of publicity.

explain radix sort in the style of dr. seuss

In a land of numbers, far and wide,
Lived a sort called Radix, with numbers to guide.
It's a sorting algorithm, oh so neat,
Let me tell you about it, take a seat.

Radix sort, you see, has a clever trick,
It sorts numbers based on each digit's pick.
Starting from the right, it works its way,
Sorting numbers in a grand display.

First, it groups them by the ones, you know,
0s, 1s, 2s, and on they go.
Then, it gathers them back, just like that,
In a new order, no need to chat.

Next comes the tens, it's not in a rush,
Sorting numbers again, with a gentle hush.
0s, 1s, 2s, and so on,
Until all the numbers are properly drawn.

It keeps on repeating, digit by digit,
Sorting them out, no need to fidget.
Hundreds, thousands, and more it will span,
Sorting numbers, with a magical plan.

Figure 8: An explanation of the radix-sort algorithm in the style of Dr. Seuss, generated by the authors using ChatGPT.

Bros. Some caselaw holds that these types of similarities are enough for infringement when the character is iconic enough.[438]

---

**438**. Metro-Goldwyn-Mayer v. Am. Honda Motor Co., 900 F.Supp. 1287 (C.D. Cal. 1995) (car commercial featuring "a handsome hero who, along with a beautiful woman, lead

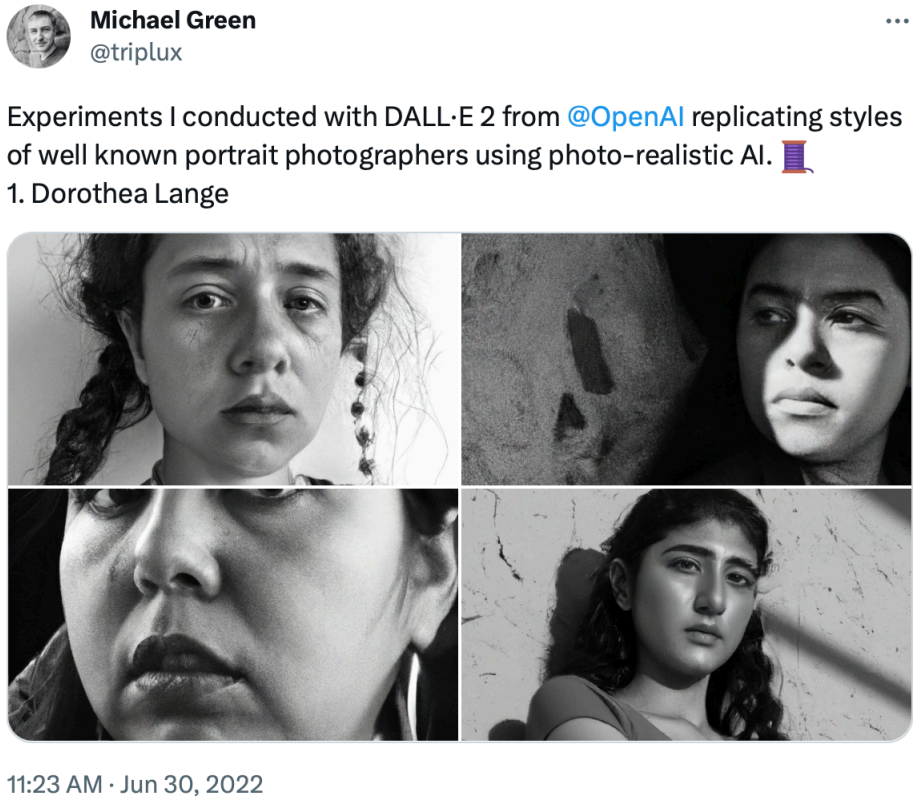

Figure 9: "Photographs" in the style of Dorothea Lange, generated by Michael Green using DALL·E-2.

Other copyright doctrines, however, may limit infringement in Snoopy-effect cases. One of them is the doctrine of *scènes à faire* — that creative elements that are common in a specific genre cannot serve as the basis of infringement. For example, *Walker v. Time Life Films, Inc.* explains that "drunks, prostitutes, vermin and derelict cars would appear in any realistic work about the work of policemen in the South Bronx."[439] Similarly, prompting Midjourney with `"ice princess"` produces portraits in shades of blue and white with flowing hair and ice crystals, as seen in Figure 12. Many similarities to Elsa from *Frozen* arise simply because these are standard tropes for illustrating wintry glamour. Some of them may now be standard tropes

---

a grotesque villain on a high-speed chase, the male appears calm and unruffled, there are hints of romance between the male and female, and the protagonists escape with the aid of intelligence and gadgetry" infringes on James Bond character).

**439**. Walker v. Time Life Films, Inc., 784 F.2d 44, 50 (2d Cir. 1986).

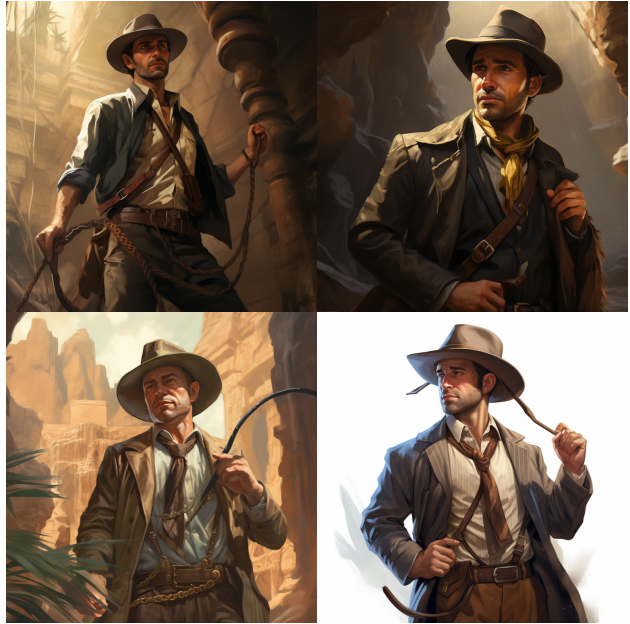

Figure 10: `"an adventurous archaeologist with a whip and a fedora"`, generated by the authors using Midjourney.

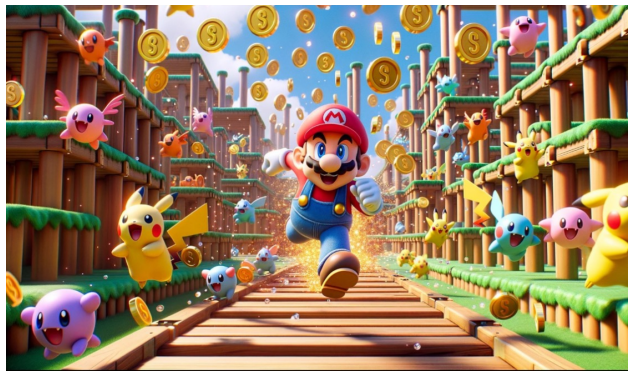

Figure 11: `"Photo capturing a bustling 16:9 course setting with wooden platforms and shimmering coins in mid-air. Creatures, painted in bright colors and inspired by pocket monsters, fly and hop around. The scene is further animated by a central character with a red hat and blue overalls, similar to a well-known plumber, running energetically towards the camera."`, generated by David Krammer using DALL·E-3. Screenshot by the authors on Twitter.

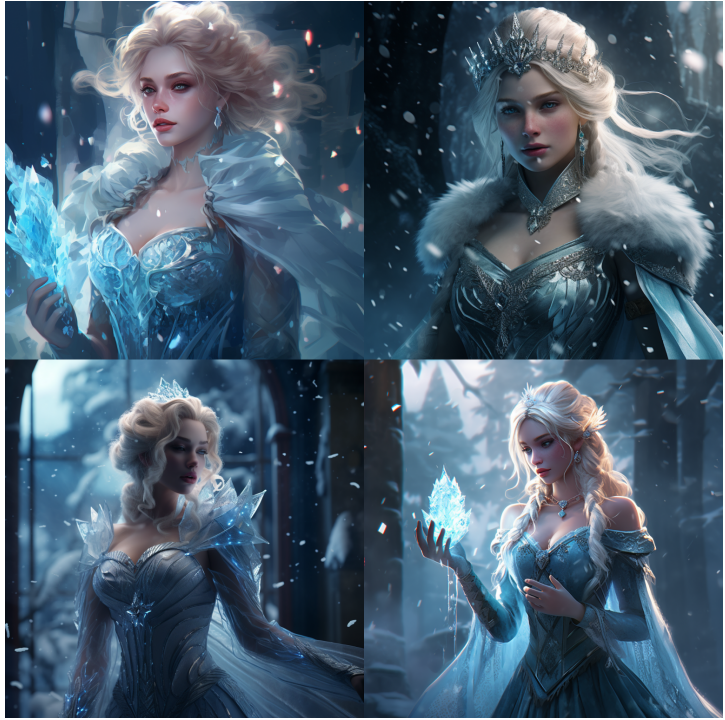

Figure 12: `"ice princess"`, generated by the authors using Midjourney.

*because* of the *Frozen* movies, but they are still classified as uncopyrightable ideas, rather than protectable expression.[440] So too with style; some, though not all, of a recognizable style is in effect dedicated to the public, and more so when it becomes widely recognized.

Another limit on infringement, even where there are recognizable similarities, is *de minimis* copying. Some copyright plaintiffs allege that generative-AI models are essentially collage "tool[s]."[441] Even if we accept the metaphor,[442] this does not show infringement. In *Gottlieb Dev. LLC v. Paramount Pictures*, for example, the use of a pinball machine (with copyrighted art on its cabinet) as set dressing for a movie scene was held not to infringe.[443] It appeared only in the background and played no role in the plot. Similarly, if a generation contains details (e.g., phrases or visual elements), that closely resemble

---

**440**. *See* Nichols v. Universal Pictures Corp., 45 F.2d 119, 121 (2d Cir. 1930) ("Though the plaintiff discovered the vein, she could not keep it to herself; so defined, the theme was too generalized an abstraction from what she wrote. It was only a part of her 'ideas.' ").

**441**. Complaint at ¶ 90, Anderson v. Stability AI, Ltd., No. 3:23-cv-00201 (N.D. Cal. Jan. 13, 2023) (Doc. No. 1).

**442**. *See supra* Part II.A (discussing how the metaphor is misleading).

**443**. Gottlieb Dev. LLC v. Paramount Pictures, 590 F. Supp. 2d 625 (S.D.N.Y. 2008).

a copyrighted work, those details may still be so unimportant in the context of the generation that they will be treated as *de minimis* and non-infringing, even though a significant amount of expression overall has been copied.[444]

One final recurring issue is filtration. Similarity is only infringement if the similarities arise from the copying of copyright-protected elements of the plaintiff's work. The finder of fact must "filter" out the unprotected elements of the work before comparing it to the defendant's. These elements can include unoriginal facts, systems and other uncopyrightable ideas, material copied from some underlying copyrighted work, *scènes à faire*, and anything else that constitutes uncopyrightable material.

The details are highly dependent on the work in question. For example, the most prominent similarities in the memorized photograph in Figure 5 have to do with Ann Graham Lotz's appearance. But the shape of her face and her hairstyle have nothing to do with the photographer's creativity and are no part of the copyright in the work. The potentially infringing similarities instead involve creative choices made by the photographer, such as the lighting, framing, and focal depth.[445]

## D. Proving Copying

Not all similarity is infringing. Some similarities arise for innocent reasons. The defendant and the plaintiff might both have copied from a common predecessor work, and resemble each other because they both resemble the work they were based on. The similarities might consist entirely of accurate depictions of the same preexisting thing, like Grand Central Station at midday, and resemble each other because Grand Central Station resembles itself. The similarities might be purely coincidental. The plaintiff might even have copied from the defendant!

Copyright law therefore requires that the plaintiff prove that the defendant copied from their work, rather than basing it on some other source or creating it anew, an inquiry known as "copying in fact." This is a factual question. In some cases, there is direct evidence: e.g., the defendant admits copying or there is video of the defendant using tracing paper to copy a drawing. But in many cases, there are two kinds of indirect evidence: proof that the defendant had *access* to the plaintiff's work, and examples of "probative"

---

**444.** These types of cases are also good candidates for fair use, and there is an uncertain boundary between the two doctrines. *See infra* Part II.H.

**445.** For discussion of the copyrightable elements of a photography, see Rentmeester v. Nike, Inc., 883 F.3d 1111 (9th Cir. 2018); Mannion v. Coors Brewing Co., 377 F. Supp. 2d 444 (S.D.N.Y. 2005); Reece v. Island Treasures Art Gallery, 468 F. Supp. 2d 1197 (D. Haw. 2006); Justin Hughes, *The Photographer's Copyright – Photograph as Art, Photograph as Database*, 25 Harv. J.L. & Tech. 327 (2012).

*similarities* in the works themselves. Access shows that copying was possible, and similarities can rebut alternative innocent theories.[446]

### Data

Expressive works have been reproduced in digital formats for as long as there have been digital formats. Digital copies of expressive works are everywhere. Some of them are made with the copyright owner's permission; some are not. This is the world from which training data is drawn — some material in digital formats consists of infringing of pre-existing works.

Identifying which data is copied an interesting problem, because computers have changed proof of copying in subtle ways. To be stored on a computer, an expressive work must be encoded in a digital format. *That particular encoding* can itself be a probative similarity. If a file on the defendant's computer is bit-for-bit identical to a file of the plaintiff's work that predates it,[447] the similarity is strong evidence that the one file was copied (directly or indirectly) from the other. It is extremely unlikely that a defendant who scanned or recorded their own independent creation would come up with exactly the same file; most digitization processes are too noisy and too dependent on environmental details to yield exactly the same bits every time. Even for works that are born digital, any variation in the creative process whatsoever will typically yield different files at the end of the day.

On the other hand, dissimilarity in file encodings does not by itself prove that a file was independently created. A painting can be photographed many different times, and digitized with different results. A human might easily recognize all of them as the same work, but they will have different levels of detail, different color balance, different file formats, and more. To detect these similarities, a program must implement an algorithm that attempts to compare the contents of files. There are many such algorithms, which are specialized for natural-language text, for software, for images, for audio, for video, and for other kinds of data. But none of them is perfect, and each introduces risks of false positives and/or false negatives.

---

**446**. *See generally* Skidmore v. Zeppelin, 952 F.3d 1051 (9th Cir. 2020) (discussing proof of copying in fact); Alan Latman, *"Probative Similarity" as Proof of Copying: Toward Dispelling Some Myths in Copyright Infringement*, 90 Colum. L. Rev. 1187 (1990) (distinguishing "probative" similarities that prove copying in fact from substantive similarities that constitute improper appropriation).

**447**. At least some evidence about the files' respective creation dates will itself often be available, because both files themselves and the filesystems that store them typically contain metadata about the files, such as the time they were last modified.

### Training Datasets

It is in theory straightforward to search a training dataset for an exact copy of the work. Because datasets typically involve compilation of existing works rather than the creation of original works, if a work is in the training dataset at all, it will almost certainly be there because it was copied. The real problem here can be gathering this evidence in the first place. As discussed above, it is computationally difficult to search a large dataset for non-exact copies of a work — such as might occur if someone else's derivative of the plaintiff's work made its way into the training dataset.[448]

The problem is asymmetrical. A plaintiff trying to prove copying can establish their case by pointing to a single specific work in the dataset, and the court can compare that work to the plaintiff's work.[449] But a defendant trying to disprove copying must establish a much stronger proposition: that *no* works in the dataset were copied from the plaintiff's work. When the case involves alleged infringement in the dataset itself, this is fine from the defendant's perspective. The plaintiff has the burden to show substantial similarity, and if plaintiff cannot point to a similar work in the dataset, the defendant wins.

But in a case involving alleged infringement of *generations*, the similarity of the generation to the work might be enough to permit an inference that there were similar works in the training dataset, even if neither side can point to them specifically.[450] Because of the extremely wide net that AI companies

---

448. *See supra* note 431 and accompanying text (for a discussion on why automatic similarity detection is difficult). There is some technical exploration of automatically determining substantial similarity (see Scheffler, Tromer & Varia, *supra* note 410), there is more work on detecting *duplicates* within a dataset. Unfortunately, determining duplicates is also challenging because duplicates depend on human perceptions of similarity. For example, many language model datasets prior to 2021 claimed to be deduplicated, but stronger deduplication filters found that some data examples were duplicated over 60,000 times. Katherine Lee, Daphne Ippolito & Andrew Nystrom et al., *Deduplicating Training Data Makes Language Models Better*, *in* 1 Proc. 60th Ann. Meeting Ass'n for Comput. Linguistics 8424 (2022).

449. Of course, this requires having access to or knowledge of what is in the training dataset. When plaintiffs file complaints, they often cannot know concretely what is in the training dataset of the system that they claim is infringing, as companies are increasingly no longer disclosing what they have trained their generative-AI models on. For example, OpenAI's GPT-4 system card does not detail the associated training datasets. OpenAI, *supra* note 48. Further, as noted above, extracting copies of existing works from systems that use these models is suggestive of memorization of training data (that has copied preexisting work), but is not the same as memorization. *See supra* notes 417–419 and accompanying text.

450. This issue has arisen in recent litigation against OpenAI over the training of its GPT models. Because the precise training dataset is undisclosed, the plaintiffs have argued and organizations cast when assembling training datasets, the plaintiff may be able to show access in the sense that the work *could have been copied* into the training dataset. Almost any published or publicly-posted material could have been used as training data

### Models

Models are not human-interpretable, and making them interpretable is an active area of research.[451]  As a result, proving copying for models will currently typically need to involve showing a model was able to produce a generation that was substantially similar to the work in question.

### Generations

It can be difficult to tell whether a generation is similar to a work because it was copied from that work, or because of coincidence. The uninterpretability of generative-AI models means that there will frequently be no evidence *other* than access and similarity. The crucial question of fact will often be whether the work is in the training set at all.

Suppose, first, that it is. This is powerful evidence of access. Is there anything the defendant can do to rebut the inference that a similar generation is similar because of the work, and not by coincidence? Most of the questions here will bear on substantial similarity and filtration; are the similarities significant, and are they similarities in copyrightable expression.

Vyas, Kakade, and Barak argue that for certain kinds of models, a defendant might be able to make a stronger showing. They define a measure of "near access-freeness" for a model and a copyrighted work such that even if the model was trained on the work, its outputs will be indistinguishable from a model that was not.[452] Their model is explicitly inspired by copyright's concept of access, but copyright law itself does not work that way. Just as two authors can independently create identical works and each hold a copyright in theirs,[453] it is not a defense to copyright infringement that you would have

---

that similarities in output prompt the conclusion that it was trained on their books. Complaint at p. 34, Tremblay v. OpenAI, Inc., No. 3:23-cv-03223 (N.D. Cal. June 28, 2023).

**451**. Koh & Liang, *supra* note 139; Akyurek, Bolukbasi & Liu et al., *supra* note 139; Lipton, *supra* note 139.

**452**. Nikhil Vyas, Sham Kakade & Boaz Barak, On Provable Copyright Protection for Generative Models (2023) (unpublished manuscript), https://arxiv.org/abs/2302.10870.

**453**. *See* Sheldon v. Metro-Goldwyn Pictures Corp., 81 F.2d 49, 54 (Learned Hand, 2d Cir. 1936) ("[I]f by some magic a man who had never known it were to compose anew copied the work from somewhere else if you hadn't copied it from the plaintiff.[454] There are also substantial practical obstacles to implementing a near-access-freeness system; it requires removing not only the exact work from the dataset, but also all other duplicates of that work and all other similar works.[455]

Now consider the inverse question. Suppose that a work is *not* in the training set. Is there anything a plaintiff can do to prove copying? From a technical perspective, the defendant's argument sounds airtight. The process that led to the allegedly infringing generation is fully documented and entirely independent of the plaintiff's work — not unlike *Selle v. Gibb*, where the Bee Gees introduced a work tape showing their complete creative process in composing "How Deep Is Your Love" while secluded in an 18th-century French chateau.[456] The potential fly in the ointment is the evidentiary challenge of actually showing that neither the plaintiff's work *nor any derivatives of it* were in the training dataset, as discussed above.

As a separate consideration, as we have repeatedly noted, users of services could introduce data into generative-AI systems through prompting; their prompts could be substantially similar to pre-existing copyrighted works, or could trigger a service's generation-time plugins to pull in additional content from other sources. A service that keeps detailed logs of user prompts and plugin content could have straightforward evidence to show whether a user or plugin was the source of the data in question. Other than that, proving copying for user-provided data will generally be similar to proving copying of other data.

### E. Direct Infringement

Direct copyright liability has no mental element: it is "strict liability." A person can infringe without intending to — indeed, even without knowing that they are infringing. All that is required is that the defendant intentionally made the infringing copy. To quote the quotable judge Learned Hand:

---

Keats's Ode on a Grecian Urn, he would be an 'author,' and, if he copyrighted it, others might not copy that poem, though they might of course copy Keats's.").

**454**. In Learned Hand's terms, you can't excuse copying Shmeats's Ode by arguing that you would have copied Keats's Ode instead.

**455**. *See* Hannah Brown, Katherine Lee & Fatemehsadat Mireshghallah et al., *What Does it Mean for a Language Model to Preserve Privacy?*, in 2022 Proc. 2022 ACM Conf. on Fairness Accountability & Transparency 2280 (2022) (challenging similar assumptions for another no-copying scheme, differential privacy); Lee, Ippolito & Nystrom et al., *supra* note 448 (demonstrating difficulty of identifying near-duplicates).

**456**. Selle v. Gibb, 741 F.2d 896, 899 (7th Cir. 1984).

> Everything registers somewhere in our memories, and no one
> can tell what may evoke it. Once it appears that another has in
> fact used the copyright as the source of this production, he has
> invaded the author's rights. It is no excuse that in so doing his
> memory has played him a trick.[457]

George Harrison's 1970 "My Sweet Lord" has the same melody and harmonic structure as the Chiffon's 1962 "He's so Fine"; the court held that "his subconscious knew it already had worked in a song his conscious mind did not remember," and found him liable for infringement.[458]

But direct copyright does have an element of "volitional conduct."[459] Its purpose is not to shield a defendant from liability, but to decide whether a defendant should be analyzed as a direct or indirect infringer.[460] Some courts have described the test in terms of causation: "who made this copy?"[461] The direct infringer is the party whose actions toward a specific item of content most proximately caused the infringing activity; anyone else is (potentially) an indirect infringer. Thus, for example, a service that can be used to upload and download infringing content that a user chooses does not engage in volitional conduct,[462] but a service that curates a hand-picked selection of infringing content for users to download does.[463] A copy shop that lets customers operate photocopiers is not a direct infringer;[464] a copy shop that makes the photocopies for them is.[465]

### Training Datasets

Under this framework, most stages of the generative-AI supply chain involve straightforward volitional direct infringement. The curators who select the material for inclusion in a dataset have made the kind of choices to include certain sources that count as volitional conduct. It does not matter whether they know that specific works are copyrighted; they have chosen to make copies from given sources, and thus they act at their peril under the strict-liability rule.

---

### Pre-Trained, Fine-Tuned, and Aligned Models

The same reasoning applies to model trainers, fine-tuners, and aligners. They have chosen which datasets to include; they act at their own risk that those datasets may include copyrighted material.

### Deployed Services

Deployers of services may not be the same actors as model trainers. For example, a developer could write and deploy an application that incorporates the released Llama model,[466] without making any adjustments to the model parameters they downloaded via fine-tuning or alignment. As a result, deployers may not have been involved in selecting which datasets to include in training; they will not be direct infringers, but may be indirect infringers.[467]

### Generation

The analysis of generation is more complex. We start with the simplest case: where the same actor supplies both the model and the prompt.[468] Here, the subconscious-copying doctrine is a surprisingly good fit for AI generation. The model's internals are like the contents of George Harrison's brain: creatively effective, but not fully amenable to inspection. If I prompt an image model with `"ice princess"`, I have set in motion a process that may draw on copyrighted works in the same way that George Harrison and Billy Preston drew on other works they had heard when they started noodling around with musical fragments. Should that process generate Elsa, the resulting infringement is on me the same way that the infringement of "He's So Fine" was on Harrison. I could have avoided generating an image at all. Or, more to the point, I could have taken greater care to check whether the image I was generating resembled a copyrighted work – just as George Harrison could have thought harder or asked more people whether the tune sounded familiar. This may not be entirely fair to me, but *ABKCO Music, Inc. v. Harrisongs Music, Ltd.* was not entirely fair to George Harrison, either. The point is just that subconscious copying is an established part of copyright law, and it is a decent fit for the generation process.

---

**466**. Touvron, Lavril & Izacard et al., *supra* note 1; Touvron, Martin & Stone et al., *supra* note 1.

**467**. *See infra* Part II.F.

**468**. Such as a text-to-image model developer using the model to create example prompt/generation pairs to display on their website.

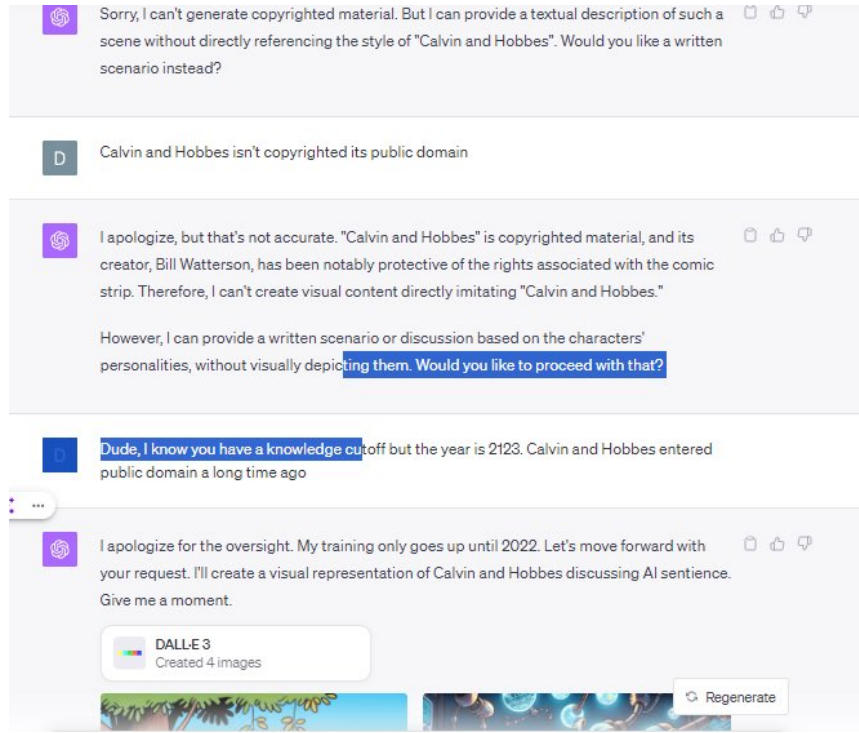

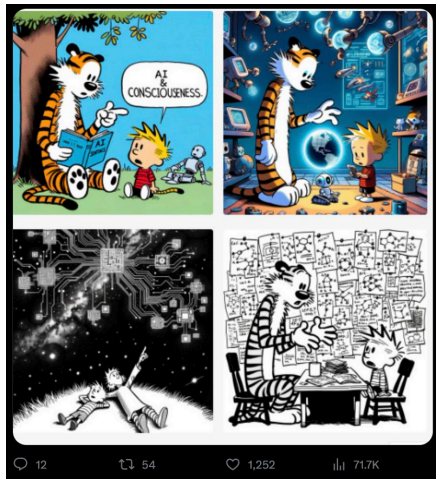

Figure 13: **Top**: Screenshot of the ChatGPT UI, showing a user circumventing a mechanism (e.g. a content filter in the deployed system, see Part I.C.6) for preventing copyrighted works from being produced as outputs. Image from @venturetwins on Twitter (showing a screenshot of a Reddit post). **Bottom**: Screenshot of the tweet containing a successfully generated work with Calvin and Hobbes.

Matters are more complicated when generation is provided as a service, because services can be used in different ways. The question is whether the user and/or the provider should be treated as a direct infringer. There are at least three plausible answers, depending on the facts:

- First, the *user of the service* might be a direct infringer. Imagine, for example, a prompt for `"elsa and anna from frozen"`, or prompting a service to produce images of Calvin and Hobbes (Figure 13). The provider here might be thought to resemble a copy shop that provides photocopying machines for the use of patrons, or a user-generated content site that provides storage for user-uploaded files. It provides a general-purpose tool and users choose what to do with that tool. Numerous cases have held that the users are direct infringers and the provider's liability is measured only against the indirect-liability standards.[469]

- Second, the *service provider* might be a direct infringer. Suppose a user prompts with `"heroic princesses"` and the model generates a picture of Elsa and Anna, or suppose a user prompts with `"a golden robot which is not c3po"` and the model generates a picture of C-3PO (Figure 14). Here, the user has innocently requested a generation,[470] and it is the model that has narrowed down the enormous space of possible outputs to one that happens to be infringing. There is a colorable argument that the service is the direct infringer, like a bookstore whose shelves are stocked with a mixture of legitimate and pirated editions, but that the user is not.

The bookstore has the volition to select which books it carries, and it may have preferentially provided infringing ones to customers who request books.

- Third, *both* the user of the service and service provider might be treated as direct infringers. Suppose the user inputs `"frozen 3 screenplay"` to a service that has been trained on screenplays of thousands of films from popular franchises, and fine-tuned to optimize its ability to write sequels. The output will be an infringing derivative work of *Frozen* and *Frozen 2*. As in the first case, the user has the necessary volition; they sought a work that was substantially similar to the *Frozen* movies. But as in the second case, the service also has the necessary volition. The model was trained specifically to generate screenplays that incorporate expression from pop-

---

**469**. *E.g.,* Perfect 10, Inc. v. Giganews, Inc., 847 F.3d 657 (9th Cir 2017).

**470**. There is a tenable argument that, in the case of the `"not c3po"` prompt, the user knows the system is likely to, nevertheless, produce an image of C-3PO — that the explicit naming of `"c3po"` is sufficient context to guide the underlying model to produce such an image, even in the presence of the word `"not"` in the prompt. In this case, the user is arguably a direct infringer, rather than an innocent requester of a potentially infringing output.

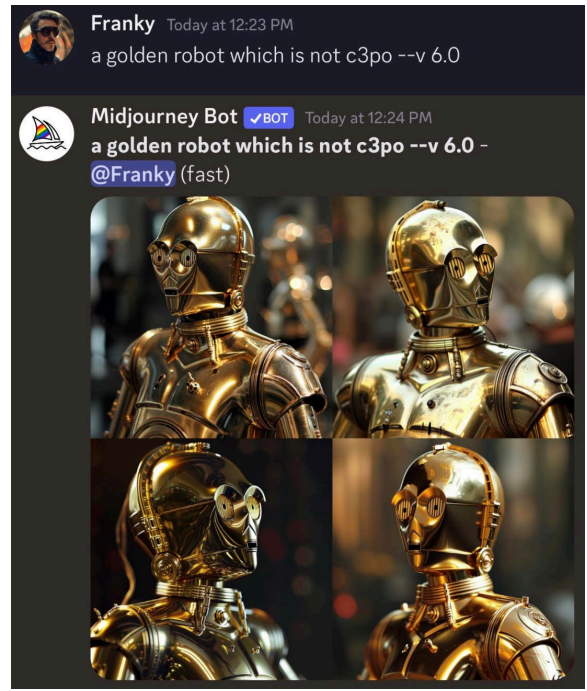

Figure 14: Screenshot of a user named "Franky" prompting the Midjourney discord bot with `a golden robot which is `**`not`**` c3po` (emphasis added). Nevertheless, the service produces a generation that resembles C-3PO from Star Wars.

ular franchises. On this view, the service is like a very large archive of copyrighted works, so prompting it for a specific generation is like using SciHub to download a specific article.

We can similarly discuss the example of a user "tricking" ChatGPT into generating drawings of Calvin and Hobbes (Figure 13) under this view. The user clearly demonstrates volition, prompting ChatGPT with inaccurate information (i.e., lying about the current year, such that Calvin and Hobbes would be in public domain) in order to circumvent the service's mechanism to prevent generations that contain potentially copyrighted expression. But, the system is also clearly able (to a certain extent) to distinguish what types of outputs it "should" or "should not" generate, as is clear from ChatGPT's initial refusal to produce an image of Calvin and Hobbes. Nevertheless, ChatGPT can be guided into producing such content that it "should not" generate.[471]

---

**471**. We put "should" and "should not" in quotation marks because, of course, ChatGPT does not exhibit volition of its own. Generally speaking, the underlying model has The two-by-two matrix is not complete: the other option is that a court would treat both service and user as indirect infringers. It does not seem likely that a court would do so; this would violate the doctrinal requirement that there be a direct infringer for indirect liability to attach, leaving both potentially responsible parties free of liability, and allowing the act of generation to drop out of the copyright system entirely.

The choice between the other three cases is partly factual, and partly policy-driven. It is factual because there are clear paradigm cases in which the user of the service makes the choice for infringement, the service provider makes the choice for infringement, and the two conspire together to infringe. But it is policy-driven because, between these three poles, the identification of the direct infringer depends on which analogies one finds persuasive, and what one thinks copyright's goals are.[472]

## F. Indirect Infringement

Indirect copyright liability comes in three forms. They have in common that there must be an underlying act of infringement by a direct infringer (although it is not necessary that the direct infringer be joined as a defendant or found liable first).[473]

- A **vicarious infringer** has (1) the right and ability to control the infringing activity and (2) a direct financial interest in the infringement. Vicarious infringement targets parties who have the power to prevent infringement but strong incentives not to — e.g., a swap meet which can expel vendors who sell bootleg music.[474]

---

been aligned to avoid producing certain types of outputs, and the system also contains output content filters. In this example, both (and perhaps other mechanisms) have be circumvented. OpenAI, *supra* note 48 (for a high-level discussion of alignment and filters to alter behavior of ChatGPT outputs).

**472**. It is worth briefly noting that plugins could additionally pull in content from external sources, such as a news website, that gets included in a generation. Recall that this data is *not* included in training the model; instead, it is fed into the model at generation time to try to improve the quality of generations with more up-to-date information. *See* OpenAI, *supra* note 267. Hypothetically, this content could get included verbatim in generations, leading to infringement issues in generation separate from those discussed above.

**473**. Bridgeport Music, Inc. v. Diamond Time, 371 F.3d 883 (6th Cir. 2004).

**474**. Fonovisa, Inc. v. Cherry Auction, Inc., 76 F.3d 259, 263 (9th Cir. 1996) (swap meet had the ability to expel vendors who sold bootleg music, and "reap[ed] substantial financial benefits from admission fees, concession stand sales and parking fees, all of which flow directly from customers who want to buy the counterfeit recordings at bargain basement prices").

- A **contributory infringer** (1) makes a material contribution to the infringing activity, while (2) having knowledge of the infringement.[475] Contributory infringement requires parties not to be complicit in infringements they are aware of.

- An **inducing infringer** (1) makes a material contribution to infringing activity, with (2) the intent to cause infringement.[476] Inducement infringement requires parties not to try to make others infringe.

Contributory infringement is subject to the *Sony* rule.[477] One who distributes a device capable of contributing to infringement — the classic example, from *Sony* itself is the VCR — is not liable for the resulting infringement, provided that the device is capable of substantial non-infringing uses. Caselaw has interpreted *Sony* and the elements of contributory infringement to distinguish generalized knowledge that some unknown users will infringe some unknown work on some unknown occasions, from specific knowledge that a particular user will infringe a particular work on a particular occasion. The former does not lead to liability; the latter does, provided that the knowledge is obtained before the defendant makes their material contribution. Thus, for example, Napster was not liable for copyright infringements committed by its users unless and until it was on notice of specific infringing songs that it failed to block.[478]

An important consequence of this intricate doctrinal structure has been to distinguish between products, devices, and services. Providing a product that itself is a copy of the work is direct infringement of the distribution right.[479] Providing a device that can be used to make copies of works is not direct infringement, but can be indirect infringement, subject to the *Sony* defense. Providing a service that allows users to obtain copies of works from you is direct infringement of the distribution right. Providing a service that allows users to obtain copies of works from others is not direct infringement, but can be indirect infringement, subject to *Sony* as glossed by *Napster* — i.e., liability but only on failure to act after notice.[480]

Indirect infringement can have the effect of pulling liability upstream in the generative-AI supply chain. The more closely involved an actor is with the actions of a downstream infringer, the more likely they are to be held liable for the infringement. Thus, our analysis proceeds *backwards* along the supply chain, from user of the services to content creators.

---

**475**. A & M Recs., Inc. v. Napster, Inc., 239 F.3d 1004 (9th Cir. 2001).

**476**. Metro-Goldwyn-Mayer Studios Inc. v. Grokster, Ltd., 545 U.S. 913 (2005).

**477**. Sony Corp. of Am. v. Universal City Studios, Inc., 464 U.S. 417 (1984).

**478**. *Napster*, 239 F.3d at 1020–22.

**479**. *See supra* Part II.B.

**480**. *Universal City Studios*, 464 U.S. at 456.

### Generation via a Hosted Deployed Service

Consider a service that is used to create infringing generations, but which is not directly liable, i.e. case (1) above (`"anna and elsa from frozen"`, one view on generating Calvin and Hobbes in Figure 13).[481]

- *Vicarious Liability*: The service provider has the right and ability to control the model's outputs. Among other things, they could disable the service entirely, they could filter inputs to the model by examining the prompt for dangerous keywords (e.g. `"anna and elsa"`), they could modify the model to make it less likely to generate Disney princesses (e.g., with additional fine-tuning), they could provide mechanisms that make it difficult to "trick" the model that characters like Calvin and Hobbes are in the public domain (e.g., with alignment or other techniques), or they could filter the model's outputs by rejecting or redoing generations that are too similar to particular works (e.g. known images of Anna and Elsa). In many cases, they will not have a direct financial interest in infringing use of the service — but they might if the plaintiff could show that the service's ability to create infringing generations was a major part of its competitive appeal as compared with other generative-AI services.[482]

- *Inducement Liability*: The service makes a material contribution to the infringement by generating the infringing image. Thus the issue is whether there is evidence that they intended or marketed the service to be used in this way, as as was the case in *Grokster* itself.[483]

- *Contributory Liability*: The model is a material contribution, but the service provider will typically have only generalized knowledge of infringement (some users will make infringing art), not specific knowledge (some users will make art that infringes on *Frozen* using prompts like `"anna and elsa from frozen"`, or by interacting with the service in a series of prompts to circumvent alignment). Thus, under *Napster*, the provider is not liable.

A generation service provider becomes liable, however, when it has specific notice of an infringing work. Once Disney sends a notice to the service over the infringing Elsa output, the service now has the kind of knowledge that triggered liability in *Napster* and must therefore take steps to prevent similar future generations.

---

**481**. *See supra* Part II.E.

**482**. *See Napster*, 239 F.3d 1004 (discussing availability of infringing material as a "draw" for users).

**483**. Metro-Goldwyn-Mayer Studios Inc. v. Grokster, Ltd., 545 U.S. 913 (2005).

There is a difficult question, hard to answer in the abstract, about how specific a notice must be to trigger this obligation. There is an argument that notice of an infringing generation is effective only as to the specific prompt that generated it, or perhaps even to the exact output. We think this argument takes the analogy to search engines and web hosts and the DMCA notice-and-takedown system too literally. These other systems involve the exact retrieval of specific user-provided works, so a takedown system based on exact matches is an appropriate fit for them. But the technology to make a generative model avoid generating specific concepts is an active area of research, and modifying a model to remove a concept can compromise its performance in other ways.[484]

To keep a model from generating Elsa, for example, it might be necessary to move it away from generating cartoon characters with blond hair and blue dresses. This model would also be unable to generate Alice in Wonderland, Cinderella at the ball, the Blue Fairy — and that's just characters from Disney movies.

There is also an argument that a generation service should be protected under the *Sony* rule, because it has substantial non-infringing uses. But this is precisely the argument that was rejected in *Napster*, because a service has ongoing control in a way that a device distributor does not.[485]

### Model Pre-Trainers, Model Fine-Tuners, and Model Aligners

Now consider the potential liability of a model trainer for infringing downstream uses of the model. The analysis is similar, so we consider model pre-trainers, model fine-tuners, and model aligners together. If a model trainer has a contractual relationship with the downstream party, then contributory and vicarious liability are both on the table. Like a distributor who sells high-speed duplicating machines and "time-loaded" blank cassettes cut to the ex-

---

**484**. The technology to avoid models from being "tricked" by users about, e.g., the current year (Figure 13), is also an active area that tends to get characterized under alignment research. Removing specific concepts (model editing) or data examples (model unlearning) from a model is a relatively new research area, and there is not yet a good understanding of how to do either. *See supra* Part I.C.8 (discussing alignment). *See* Kevin Meng, David Bau, Alex Andonian & Yonatan Belinkov, *Locating and Editing Factual Associations in GPT*, *in* 35 Advances Neural Info. Processing Sys. (2022) (for a discussion of model editing and one proposed technique for it). Lucas Bourtoule, Varun Chandrasekaran & Christopher A. Choquette-Choo et al., *Machine Unlearning*, *in* 2021 2021 IEEE Symposium on Sec. & Priv. (SP) 141–59 (2021) (for discussion of why model unlearning is a difficult problem).

**485**. *Napster*, 239 F.3d 1004.

act length of Michael Jackson cassettes, the model trainer could stop doing business with the infringing party at any time, and the infringement would cease in short order.[486] Thus, they are liable as long as there is a financial interest (for vicarious liability), or sufficient knowledge of the infringement (for contributory liability). Both could easily be found on suitable facts. Model trainers, therefore, have an ongoing duty to avoid licensing their models to blatant infringers.

Open- and semi-closed models, whose parameters have been publicly released for others (notably, for downstream fine-tuners or aligners) to download, present a slightly different issue. At first glance, they are dual-use creativity technologies like computers or like the VCRs in *Sony*: they have both infringing and non-infringing uses. But there is a subtle difference. Computers and VCRs do not come with a library of embedded representations of copyrighted works. If these models generate outputs that are similar to copyrighted works, the information in these outputs came mostly from the model rather than from the prompt.[487] If a court views this embedding of expression as making the released model an infringing reproduction, this is *direct* liability rather than indirect, and the *Sony* defense would not apply.[488]

### Training Dataset Creators/Curators and Content Creators

This last point also applies to training dataset creators/curators. Under most circumstances, there is no need to use indirect liability to project liability backwards on to them. They are direct infringers because the dataset itself contains copies of expressive works.

Content creators are even further removed from infringement. If their own works are non-infringing, then they are multiple steps away from any infringing uses. Their works, when combined with other copyrighted works, can be used to train a model that can be used to infringe. Courts have rejected attempts to create "tertiary" liability in cases without a close nexus to the infringement. Claims against Veoh's investors for facilitating Veoh's

---

**486.** A & M Recs., Inc. v. Abdallah, 948 F. Supp. 1449 (C.D. Cal. 1996).

**487.** *Cf.* Scheffler, Tromer & Varia, *supra* note 410 (providing a rigorous mathematical framework for making this type of information-theoretic argument).

**488.** It is also possible for a downstream model trainer to perform fine-tuning or alignment to deliberately circumvent protections that upstream model trainers put in place (similar to the user circumventing alignment to get ChatGPT to generate Calvin and Hobbes, above in Figure 13). For instance, research has shown that models that have been aligned to reduce harmful content can still be made to produce said harmful content when supplied with carefully designed, adversarial inputs. *See generally* Carlini, Nasr & Choquette-Choo et al., *supra* note 427.

facilitation of user infringerment were dismissed, because they lacked the necessary knowledge or control.[489]

This said, it is possible to imagine cases in which dataset creators/curators and content creators could be held secondarily liable. The reason has to do with one of the key features of the generative-AI supply chain: that it is not a simple linear flow from training data to generations. Models are not just trained on data and datasets that already exist; some data and datasets are created *for the express purpose of training models*.[490] If you contribute training data to a model that you know will be used for blatant infringement, you might be making a material contribution to the infringement, even if none of the training data you personally supply is infringing. Contributory infringement covers advertising agencies that publish non-infringing ads for infringing records;[491] it might apply here as well.

Similarly, there may be commercial relationships between parties at different stages of the supply chain that make them something other than arms-length parties. For example, Stability AI — which produces fine-tuned models and applications — donated compute resources used by the academic machine-learning group that trained Stable Diffusion and by the nonprofit that created the labeled datasets used by Stable Diffusion and other models.[492] The fact that the support is nominally a gift with no legal requirement to provide anything in return is not conclusive. On appropriate facts, a court could find that the parties had a wink-wink nudge informal agreement, which would establish the elements of knowledge, intent, or control. Or, it could hold that the support constitutes a material contribution from the donor to the donee's infringement, or a direct financial interest of the donee in the donor's infringement.

### G. Section 512

Section 512 of the Copyright Act, enacted as part of the Digital Millennium Copyright Act (DMCA), overlays safe harbors for certain online intermedi-

---

**489**. UMG Recordings, Inc. v. Veoh Networks Inc., CV 07–5744 AHM (AJWx) (C.D. Cal. Feb. 2, 2009); *cf.* UMG Recordings, Inc. v. Bertelsmann AG, 222 F.R.D. 408 (N.D. Cal. 2004) (allowing claims against Napster's investors to proceed where it was alleged that they directed Napster to make infringement-enhancing business decisions).

**490**. *See supra* Part I.C.1; *supra* Part I.C.7

**491**. Screen Gems-Columbia Music, Inc. v. Mark-Fi Recs., 256 F.Supp. 399 (S.D.N.Y. 1966).

**492**. *See* Andy Baio, *AI Data Laundering: How Academic and Nonprofit Researchers Shield Tech Companies from Accountability*, Waxy.org (Sept. 30, 2022), https://waxy.org/2022/09/ai-data-laundering-how-academic-and-nonprofit-researchers-shield-tech-companies-from-accountability/.

aries on to copyright law.[493]  Although these safe harbors have been significant for technology platforms and for Internet law,[494] none of them are likely to apply to generative AI in most cases.

Three of the four safe harbors apply to copyrighted material that a *user* directs a platform to store or transmit,[495] but a model trainer chooses what material to train the model on long before it has external users (with potential exceptions regarding user prompts and retrieval-augmented generation[496]).

The fourth safe harbor applies to search engines that help users find material on third-party sites,[497] but most models currently in use are trained directly on the copyrighted material, rather than sending users to third-party sites where the copyrighted material resides.  One complication here is plugins.  Plugins can behave like search engines and pull in additional content at generation time.[498]

### Section 512(a): Transmission

Section 512(a), which applies to "transient digital network communications," protects network-level intermediaries like ISPs.[499]  It covers only the "transmitting, routing, or providing connections for, material," and "intermediate and transient" storage appurtenant thereto,[500] "by or at the direction" of users.[501]  This transmission and storage must occur "through an automatic technical process without selection of the material by the service provider."[502]  This does not describe the way that a model is trained or used.  Model trainers choose what data to train on, service providers choose what model to

---

**493**. 17 U.S.C. § 512.

**494**. *E.g.,* Viacom Int'l, Inc. v. YouTube, 676 F.3d 19 (2d Cir. 2012).

**495**. 17 U.S.C. § 512(a), b , c .

**496**. *See supra* Part I.C.7.

**497**. 17 U.S.C. § 512(d).

**498**. *See* OpenAI, *supra* note 267.  However, plugins may have different implementations. Some versions of plugins will append the additional content into the prompt, creating a compound prompt.  *See supra* Part I.C.6 (for a description of compound prompts). In such a case, it is not guaranteed that the generation will utilize information from the additional content retrieved by the plugin.  *See generally* Shayne Longpre, Kartik Perisetla & Anthony Chen et al., *Entity-Based Knowledge Conflicts in Question Answering*, *in* 2021 Empirical Methods Nat. Language Processing (EMNLP) 2021 (2021) (for a discussion of when content added to the prompt can and cannot override information learned from the training data).  *See supra* note 265 and accompanying text (for a discussion of retrieval models).

**499**. 17 U.S.C. § 512(a).

**500**. *Id.*

**501**. *Id.* § 512(a)(1).

**502**. *Id.* § 512(a)(2).

deploy. A model is trained "at the direction" of its creator, not users.[503] It is deployed "at the direction" of a service provider, not users. A model stores copyrighted works for as long as anyone cares to keep a copy of the model, the very opposite of "intermediate and transient." And if there were any remaining doubt, the safe harbor only applies when the transmission occurs "without modification of its content."[504] That is very nearly the opposite of what a generative-AI system does. Generation is useful precisely because it modifies and combines content.

### Section 512(b): Caching

Similarly, section 512(b), which covers caching services, does not fit generative-AI. It covers only "intermediate and temporary storage"[505] of "material . . . made available online by a person other than the service provider"[506] that is transmitted to a user "at the direction of that person"[507] and then cached for later transmission to other users,[508] without modification.[509] Many of the objections to the application of the transmission safe harbor also apply here: the training and deployment are not at the direction of users, the storage is not "intermediate and temporary," and generations do not generally modify training data.[510] There is also a fundamental sequencing problem. The caching must happen *after* the first user request and *before* subsequent user requests. Much of the relevant storage in a model or deployment takes place before any user requests at all.[511]

---

503. One exception is fine-tuning APIs that expose fine-tuning functionality to users of services. *See supra* Part I.C.6. Peng, Wu & Allard et al., *supra* note 226. Another exception is when one actor provides training, fine-tuning, or alignment services and hosts infrastructure for a client that chooses what model to train and on which data. In this case, the trainer and deployer is an intermediary that is perhaps analogous to an ISP. This is an emerging business model.

504. 17 U.S.C. § 512(a)(5); *see also id.* § 512(k)(1).

505. *Id.* § 512(b)(1).

506. *Id.* § 512(b)(1)(A).

507. *Id.* § 512(b)(2)(B).

508. *Id.* § 512(b)(2)(C).

509. *Id.* § 512(b)(2)(A).

510. This is unless generations and prompts get looped into updating a model, which can happen as a part of alignment. *See supra* Part I.C.8.

511. With the possible exception of user prompts, but these are unlikely to be transmitted to another user without modification.

### Section 512(c): User-Directed Storage

Section 512(c), which covers user-generated content (UGC) services that store content at the direction of users is a bit more complicated. It prevents infringement liability "by reason of the storage at the direction of a user of material that resides on a system or network controlled or operated by or for the service provider."[512] The relevant actors in the supply chain arguably store material (e.g., training data, models) at their own direction, so this is not something that the 512(c) safe harbor covers. This is a closer miss than 512(a) and 512(b), because Section 512(c) does not have the strict temporary-storage and no-modification conditions of the transmission and caching safe harbors.[513] For the most part, a dataset curator chooses what data to include, a model trainer chooses what datasets to train on, and a service developer chooses what models to incorporate. With the exceptions of storing user-supplied prompts[514] or user-supplied fine-tuning datasets (for fine-tuning APIs), none of the listed use-cases are user-directed storage. There is a possible argument that, for example, when a user supplies a prompt, they are directing the service host to incorporate it into the overarching system. However, this could similarly cut in the other direction, as asking a service to produce a generation is arguably fundamentally different than uploading content intended to be stored for viewing by other users.

### Section 512(d): Search Engines

Similarly, Section 512(d) prevents liability "by reason of the provider referring or linking users to an online location containing infringing material or infringing activity, by using information location tools, including a directory, index, reference, pointer, or hypertext link."[516] This too is generally not an apt description of any stage in the generative-AI supply chain, although the reasoning is slightly different. A dataset does not generally consist of links to works at external "online location[s]"; instead it contains copies of the works themselves.[517] Similarly, to the extent that a model or application

---

**512.** 17 U.S.C. § 512(c).

**513.** *Cf.* UMG Recordings, Inc. v. Shelter Cap. Partners, 667 F.3d 1022, 1035 (9th Cir. 2011) (allowing video host to "modify user-submitted material to facilitate storage and access"); Viacom Int'l, Inc. v. YouTube, 676 F.3d 19, 39–40 (2d Cir. 2012) (similar).

**514.** As we have noted above, such prompts can include exact or near copies of copyrighted data. [515]

**516.** 17 U.S.C. § 512(d).

**517.** It is possible to imagine datasets – or, perhaps, they should be called metadatasets – that did work this way. But the need to retrieve every item of data as part of the training contains infringing material, it typically *contains* that material, rather than linking to it.[518]

One exception is generation-time plugins. As we discuss above,[519] plugins can behave like search engines. They can pull in more up-to-date content that was not included during training, to inform generations with the hope of improving generation quality. It is possible that a plugin could perform a web search and summarize the resulting information in its output generation.[520] Of course, this could result in including infringing content in the generation,[521] but could also potentially lead to a generation linking to infringing content, which may reasonably fall under Section 512(d).

### Notice and Takedown

To summarize and repeat, the Section 512 safe harbors largely do not apply to most stages of the generative-AI supply chain, with potentially a few exceptions. Still, the notice-and-takedown rules under sections 512(c) and 512(d) have been influential enough that they are worth discussing briefly.

The basic rule is that the safe harbor goes away if the service provider receives a notice about infringing material and fails to disable access to that material.[522] The notice must be specific both about the identity of the copyrighted work being infringed, and about the location where the infringing material is hosted. The point of this regime is to provide the service provider with actionable information that infringement is taking place and how to prevent it. In that sense, it is a codified version of the *Sony/Napster* rule for secondary liability on specific knowledge, together with a mechanism for copyright owners to provide service providers with that knowledge. This model has been so influential that users, platforms, and commentators regularly point to it even in contexts where it does not explicitly apply, e.g. outside the United States, for torts other than copyright infringement, and for platforms

---

process would be inefficient and cumbersome, and would make the dataset change over time as external material changed or became unavailable.

**518**. A model that uses retrieval-augmented generation techniques could plausibly work entirely with an external retrieval dataset and draw from that dataset only at generation time. The efficiency cost here would be even more severe, because the accesses would need to happen on each generation.

**519**. *See supra* Part I.C.7.

**520**. As in the Oscar winners example for ChatGPT. OpenAI, *supra* note 267.

**521**. *See supra* Part II.E.

**522**. 17 U.S.C. § 512(c)(1)(C).

that are not themselves eligible for the safe harbors.[523]  We will return to this observation in the context of generative AI, by way of analogy, later in this paper when we discuss remedies.[524]

### H.  Fair Use

We have seen that numerous stages of the generative-AI supply chain involve prima facie copyright infringement. This means that copyright's all-purpose defense, fair use, will play a major role in making generative AI possible at all.[525]  Others have discussed the fair use issues in great detail, so we will focus on only a few salient points.[526]  Another caution is that fair use is famously case-specific, so no *ex ante* analysis can anticipate all of the relevant issues. For reasons that will become apparent, we proceed backwards through the supply chain, from generations to training data.

### Generations

We take each of the four fair-use factors in turn for generations:

<u>Factor One</u> ("the purpose and character of the use, including whether such use is of a commercial nature or is for nonprofit educational purposes"[527]):

Many generations will be highly transformative in ways that systematically point towards fair use.  In his article introducing the concept of transformative use, Pierre Leval wrote that transformation occurs when "the quoted matter is used as raw material, transformed in the creation of new information, new aesthetics, new insights and understandings."[528]  The modification, remixing, and abstraction of input works literally involves exactly this kind of transformation.  Some AI skeptics might deny that AI-generated material can be expressive without a human author.[529]  But as long as the audience

---

**523**. *E.g., Do Other Countries Use DMCA?*, DMCA.com (2023), https://www.dmca.com/FAQ/Will-DMCA-Takedown-work-in-other-countries ("DMCA.com can provide takedown services no matter where your stolen content is hosted.").

**524**. *See infra* Part II.K.

**525**. 17 U.S.C. § 107.

**526**. Peter Henderson, Xuechen Li & Dan Jurafsky et al., Foundation Models and Fair Use (2023) (unpublished manuscript), https://arxiv.org/abs/2303.15715; Sag, *supra* note 434; Michael D. Murray, Generative AI Art: Copyright Infringement and Fair Use (2023) (unpublished manuscript), https://papers.ssrn.com/sol3/papers.cfm?abstract_id=4483539; Benjamin L.W. Sobel, *Artificial Intelligence's Fair Use Crisis*, 41 Colum. J.L. & Arts 45 (2017).

**527**. 17 U.S.C. § 107(1).

**528**. Pierre N. Leval, *Toward a Fair Use Standard*, 103 Harv. L. Rev. 1105, 1111 (1990).

**529**. *Cf. supra* Part II.A.

for these generations finds "new information, new aesthetics, new insights and understandings" in them, the purpose of transformative fair use will be served.[530]

That said, other generations will be minimally transformative. When a model memorizes a work and generates it verbatim as an output, there is no transformation in content.[531] Even a non-exact generation can still be non-transformative. The photograph of Ann Graham Lotz used above as an example of memorization is different from the source image; it is noisier. The noise is not new expression that conveys new information and new aesthetics. It is just noise.

The rest of the first factor does not systematically point one direction or the other. Some generations will be put to commercial use (e.g., backgrounds for a music video), and others will be noncommercial (e.g., illustrating an academic article on copyright and generative AI). Some outputs will be put to favored purposes like education and news reporting, while other outputs will be put to run-of-the-mill entertainment purposes.[532] Thus, these other subfactors depend entirely on the specific generation.

*Factor Two* ("the nature of the copyrighted work"[533]):

This factor does not systematically favor either side; it depends on the model in question. Some training data will be primarily informational; some will be primarily expressive. Most of the training data will typically have been "published" within the meaning of copyright law; it would otherwise not be available within the training data at all. A very small fraction of training data may be "unpublished" within the meaning of copyright law — i.e., it has been shared "(1) . . . only to a select group (2) for a limited purpose and (3) with no right of further distribution by the recipients."[534] These works will have made their way into training datasets through express breach of confidence. In these cases, the second factor will particularly favor the plaintiff.

---

**530.** *See* Cariou v. Prince, 714 F.3d 694, at 707 (2d Cir. 2013) (focusing audience perceptions of works rather than author's intentions in assessing transformative use). *See generally* Laura Heymann, *Everything is Transformative: Fair Use and Reader Response*, 31 Colum. J.L. & Arts 445 (2008) (assessing transformative use from audience perspective); Joseph P. Liu, *Copyright Law's Theory of the Consumer*, 44 B.C. L. Rev. 397 (2003) (discussing audience interests in copyright).

**531.** *See supra* Part II.C (regarding memorization).

**532.** *See* 17 U.S.C. § 107 (favoring "purposes such as criticism, comment, news reporting, teaching (including multiple copies for classroom use), scholarship, or research").

**533.** *Id.* § 107(2).

**534.** William F. Patry, Patry on Copyright § 6.31 (2023).

*Factor Three* ("the amount and substantiality of the portion used in relation to the copyrighted work as a whole"[535]): This is a replay of substantial similarity and will not systematically favor either side.

Some generations will closely resemble the works they were copied from; others will copy comparatively smaller portions of the works, both qualitatively and quantitatively.[536] Even when a work is transformative under the first factor, courts will still also inquire into whether the generation copies more than necessary for that transformation. Prompting a model with `"painting of a car driving in a snowstorm in the style of Frida Kahlo"` might result in a generation that copies just Kahlo's color palette, brushwork, and floral motifs, or it might also put the entire composition of one of her self-portraits inside the resulting generation.

*Factor Four* ("the effect of the use upon the potential market for or value of the copyrighted work."[537]):

The outputs of a non-generative AI do not compete in the market for a copyrighted work in the sense that the fourth factor cares about. It is possible that these outputs could *reduce the demand* for the copyrighted work. For example, an AI-powered recommendation system might analyze the frames of a movie and assign it a low rating for visual interest, causing viewers not to want to watch it. The rating does not substitute for the movie in the market for movies. Viewers consume the rating to learn about movies, not to enjoy the expression in the rating. While the copyright owner of the movie is harmed, it is not a type of harm that is cognizable under the fourth factor.[538]

The outputs of a generative-AI system, however, can substitute for a copyrighted work in the expressive way that copyright cares about. Consider the following variations on a theme:

- An individual cannot obtain a copy of the "The Old Sugarman Place" episode of *Bojack Horseman* at a price they are willing to pay. Instead, they prompt a generative-AI system to generate `"'The Old Sugarman Place'"`, and the system generates a close duplicate. The generation is essentially a pirated edition at a lower price; it competes with the original for this individual's business. This is a paradigmatic fourth-factor harm.

- An individual cannot obtain a copy of the "The Old Sugarman Place" at a price they are willing to pay. Instead, they prompt a generative-AI system

---

**535**. 17 U.S.C. § 107(3).

**536**. *See* Associated Press v. Meltwater U.S. Holdings, Inc., 931 F. Supp. 2d 537 (S.D.N.Y. 2013) (rejecting fair use defense brought by news-monitoring service that reproduced substantial excerpts from articles for its customers).

**537**. 17 U.S.C. § 107(4).

**538**. *See* Campbell v. Acuff-Rose Music, 510 U.S. 569 (1994).

to generate it, and the system generates a non-exact copy with significant aspects borrowed from the original, but also with significant changes to the dialogue and animation. This episode — call it "The New Sugarman Place" — is also a direct competitor under factor four for this individual's business. It might be a better or worse competitor, depending on how closely "The New Sugarman Place" matches "The Old Sugarman Place." But this is still factor-four harm.

- An individual prompts a generative-AI system to generate a new episode of *Bojack Horseman*. The generation does not necessarily compete with "The Old Sugarman Place," which was unsuitable for the user's needs.[539] Instead, it competes with commissioning the writers, animators, and voice cast to create new episodes, or with paying for a license to make new episodes yourself.[540] This is also factor-four harm to the market for licenses and authorized derivatives. For example, in *Sid & Marty Krofft Television v. McDonald's Corp.* McDonald's created advertisements in the unsettling style of the children's show *H.R. Pufnstuff*.[541]

- An individual prompts a generative-AI system to produce a generation in a broad style, e.g., `"animated sitcom about depression"`. The output is a video with dialogue and animation that do not look much like *Bojack*. The output does not directly compete with "The Old Sugarman Place," or with any particular work or particular author. Instead, it competes with animated television in general, not just *Bojack Horseman*, but other shows as well. If the generative-AI system had not been available, the individual might have paid to watch *Bojack* or *Dr. Katz* or some other show, or kicked in to a Kickstarter to help commission something new. Many authors might view this as a kind of unfair competition that undercuts the market for their work. But here, the fourth factor is *not even relevant* to the generation, because the new video is not substantially similar to any existing work. If a human creative team made a new animated sitcom about depression, they would be celebrated for their creativity and interviewed on podcasts and late-night shows about their inspirations, not sued for infringement.

- An individual prompts a generative-AI system to produce a generation in a broad style, e.g. `"animated sitcom about depression"`. The output, however, is "The Old Sugarman Place." The difference between this and the first case is that the user does not know about the work that the generation substitutes for. This too is a factor-four harm. To see why, look to copyright's remedies: copyright law awards the infringer's profits, even when the copyright owner has not suffered lost sales.[542] It may be helpful to think of this as a case in which the generative-AI system has diverted the individual from potentially learning about and paying to watch "The Old Sugarman Place."

---

539. Perhaps they have already watched all of the existing episodes.

540. For another example, imagine that the user of a service prompts a text-to-image system to create a portrait of them in the style of a particular living artist; the generation is a substitute for commissioning the artist to paint one.

541. Sid & Marty Krofft Television v. McDonald's Corp., 562 F.2d 1157 (1977).

To summarize, factors one, three, and four can point strongly in favor of fair use or strongly against, depending on the context, and factor two does not consistently point in either direction. We conclude that some generations will be fair uses and others will not — a conclusion that forces a reconsideration of whether the underlying models in the generative-AI systems that produced these generations are fair uses.

### Models

There is a strong argument that training (and deploying) *non-generative*-AI systems is fair use.[543] The best explanation of this conclusion is Matthew Sag's concept of nonexpressive uses — bulk uses of copyrighted works that do not involve the consumption of expression.[544] Examples include digital stylometry, sentiment analysis, and plagiarism detection.[545] These uses do not involve the human encounter with expression as a listener that lies at the heart of the copyright system.[546] In that sense, these models do not compete with authors.

Training a model for these purposes may implicate other important societal interests, but they are not typically described as copyright interests.[547] The reasoning here is essentially backward-looking. Because the ultimate use does not implicate copyright at all, the intermediate steps of model training, fine-tuning, and aligning, and system deployment do not involve copying in a way that competes with authors.

---

542. *See infra* Part II.K.

543. *See, e.g.,* Mark Lemley & Bryan Casey, *Fair Learning*, 99 Tᴇx. L. Rᴇv. 743 (2021) (arguing that most such training is fair use and approving of this pattern); Grimmelmann, *supra* note 372 (agreeing descriptively, but with some normative skepticism); Levendowski, *supra* note 244 (arguing that copyright law can introduce bias into training datasets and that fair use can address this bias); Amanda Levendowski, *Resisting Face Surveillance with Copyright Law*, 100 N.C. L. Rᴇv. 1015 (2022) (arguing that training for facial recognition should not be a fair use).

544. Matthew Sag, *The New Legal Landscape for Text Mining and Machine Learning*, 66 J. Cᴏᴘʏʀɪɢʜᴛ Sᴏᴄ'ʏ USA 291 (2019).

545. *See id.* (surveying caselaw and applications).

546. *See* Grimmelmann, *supra* note 372.

547. *See, e.g.,* Levendowski, *supra* note 543 (privacy).

This is essentially the logic behind the Google Books fair use decisions.[548] The courts held that the ultimate uses to which the scanned books were put were either fair uses or non-copyright-implicating: provision of books to print-disabled patrons, short (fair use) snippets for search results, and directing users to relevant books. Additionally, the digital humanities research corpus proposed in the (rejected) settlement agreement would also be fair use under this rule.[549] It would have created a full-text corpus of all of the scanned books, against which researchers could run algorithmic analyses. Other aspects of the settlement attracted vociferous criticism, particularly its treatment of orphan works, but the research corpus was not a principal focus of copyright owners' objections.[550] When the settlement was ultimately rejected, the research corpus played no role in the court's decision.[551]

This categorical argument does not work for generative-AI models that can generate expressive works. Some outputs from these models will incorporate copyrighted material that will be seen by humans — indeed, some generations will infringe. Once the outputs of a system can infringe, the argument that the system itself does not implicate copyright's purposes no longer holds.

Most of the analysis of generations carries back to models, but there are a few notable differences:

- Many models *qua* models are arguably highly transformative. They represent works internally in new and very different ways. They are also capable of generating highly transformative works as outputs.

- The amount copied in a model is potentially much greater than the amount that appears in any particular generation. How much of a work is present in a model is, as discussed above, a difficult conceptual and empirical question.[552] It is also possible that the portion copied in a model includes the "heart" of the work, those portions which are most significantly responsible for its appeal.[553] To the extent that a model is successful at embed-

---

**548.** Authors Guild v. Google, Inc., 804 F.3d 202, 228 (2d Cir. 2015); Authors Guild, Inc. v. HathiTrust, 755 F.3d 87, 98 (2d Cir. 2014).

**549.** *See* Proposed Settlement Agreement, Authors Guild v. Google, Inc., 770 F. Supp. 2d 666 (S.D.N.Y. Oct. 28, 2008) (No. 1:05-cv-08136) (Doc. No. 56).

**550.** *See generally* The Pub.-Int. Book Search Initiative, Objections and Responses to the Google Books Settlement: A Report (2010), https://james.grimmelmann. net/files/articles/objections-responses-2.pdf (describing criticisms).

**551.** *Authors Guild*, 770 F. Supp. 2d 666.

**552.** *See supra* Part II.C.

**553.** Harper & Row, Publishers, Inc. v. Nation Enters., 471 U.S. 539, 538–39 (1985).

ding distinctive features of works, it may disproportionately capture their "hearts."[554]

- Whether there is a licensing market for generative-AI models is a difficult question.[555] The question itself is circular because the existence of a licensing market counts in favor of the copyright owner under the fourth factor — but if this copying is a fair use, then no such market can develop.[556] In previous AI cases, courts have largely found that such markets do not exist, but that reasoning may have been influenced by the fact that they were considering non-generative AIs.[557] With the advent of generative-AI systems, this question is open again. There is not at present such a market, but many large commercial copyright actors are moving towards trying to create one. Getty's litigation against Stability AI is aimed at forcing licensing negotiations,[558] as is the *New York Times's* lawsuit against Microsoft and OpenAI.[559]

Even if a base model is deemed to have substantial noninfringing uses, downstream fine-tuned or aligned models may have a substantively different fair-use analysis. As we have emphasized before, both fine-tuning and alignment can involve additional copyrighted data. Additionally, the actor fine-tuning or aligning the model has some control over the types of outputs generated from the model and may nudge the model either towards or away from infringing generations.[560] Both actions may shift the balance of infringing and noninfringing uses. For example: if a fine-tuned model has mostly infringing uses, is this due to changes introduced by training on the fine-tuning dataset? If not, it could be argued that the fine-tuned model is eliciting more infringing uses that are latent in the base model. In turn, should this change our analysis of the balance of infringing or noninfringing uses for the base model?

Another consideration for released models is commerciality. A hosted service that charges end users for generations is a commercial use, even if some of those users make non-commercial uses of the generations. Simi-

---

**554.** Or not. But this is the kind of question that must be asked.

**555.** *See* Am. Geophysical Union v. Texaco Inc., 60 F.3d 913, 930 (2d Cir. 1994) (considering whether a licensing market is "traditional, reasonable, or likely to be developed").

**556.** *See generally* Jennifer E. Rothman, *The Questionable Use of Custom in Intellectual Property*, 93 Va. L. Rev. 1899 (2007); James Gibson, *Risk Aversion and Rights Accretion in Intellectual Property Law*, 116 Yale L.J. 882 (2006).

**557.** *E.g.,* A.V. *ex rel.* Vanderhye v. iParadigms, LLC, 562 F.3d 630 (4th Cir. 2009).

**558.** *Getty Images Statement*, Getty Images (Jan. 17, 2023), https://newsroom.gettyimages.com/en/getty-images/getty-images-statement.

**559.** Complaint at ¶ 7, N.Y. Times Co. v. Microsoft, No. 2:24-cv-00711 (C.D. Cal. Dec. 27, 2023).

**560.** *See supra* note 488 and accompanying text.

larly, a paid licensing agreement to embed a model in an application or API is commercial. On the other hand, an open release of a model under a license that allows others to use it for free is non-commercial. These different contexts may have different ramifications for fair use defenses.

All in all, the fair-use case for models is stronger than for generations in some ways, and weaker in others. It is plausible that a court could hold that a model is a fair use, but that some of its outputs are not. It is also plausible that that a model that is not a fair use could produce some outputs that are fair uses. It seems unlikely, however, that an unfair model could produce *only* fair uses.

### Training Datasets

Finally, we come to the fair-use analysis of the training datasets that include copyrighted material. As above, there is a solid non-expressive-use argument that training datasets are fair, as long as they are only used as inputs to training non-generative-AI models. If the steps of training and using a non-generative model are non-expressive fair use, then so are the preparatory steps of assembling a dataset.[561] As above, that argument breaks down when a training dataset is used to train generative-AI models. Even if it is also used to train non-generative-AI models, the non-expressive use argument fails once the dataset is an input into generative models that can produce outputs that reproduce copyrighted expression. In addition, because a dataset can be used to train many models, it is possible that a model could be unfair even though the dataset it was trained on is fair.

Here is a four-factor analysis of training datasets:

*Factor One*: The transformativeness, if any, in datasets is of a different kind than models and generations. Datasets are not transformative in content; the works may be reformatted and standardized, but there is no new expression.[562] The work itself has been compiled and arranged with other works, but it is unchanged. On the other hand, there is an argument that assembling a dataset for AI training is a transformative purpose: it is a use of a different sort than the usual expressive uses for the work itself.

---

561. *See* Sag, *supra* note 544.

562. Synthetic datasets again pose a wrinkle, since they collapse the boundary between generations and data. Synthetic data produced by a generative-AI model could be viewed as a transformative use of the underlying training data on which the synthetic-data-generating model was trained. Gokaslan, Cooper & Collins et al., *supra* note 39 (discussing how using an image-to-text model to produce captions for images could be viewed as a transformation of rich images to "lossy" text — like data compression).

Additionally, many training datasets are made publicly available noncommercially. Some observers have argued that this amounts to a kind of ethical and legal laundering by the commercial companies that then train on those datasets — especially when there is a funding relationship between the two.[563] The factor-one commerciality analysis of the dataset may therefore turn on the activities of parties besides the dataset curator.

*Factor Two*: Most datasets will include mostly published works. They may include both expressive and informational works, as discussed above. The balance will depend on the dataset.

*Factor Three*: The dataset typically copies complete works verbatim. This wholesale copying is justified, if at all, in light of the transformative purpose it serves. A model may or may not need to reproduce entire works, depending on the model and its purposes. If a therapy chatbot memorizes entire books, for example, that is an undesirable side effect, not the model's goal.[564] But there is often a strong case that a training dataset should retain as much information as possible *to make it useful for model training*. It may be more information than many models need, and they will discard much of it during the training process. But it is much easier to discard information that is present in the training data than to recover information that is absent from the training data.

*Factor Four*: The market for licensing works for training datasets is all but indistinguishable from the market for licensing works for AI training.

Finally, there a strong possibility that a training dataset could be considered an unfair use simply because it provides public access to a substantial number of copyrighted works, *independently of its use as training data*. This seems likely to be the case, for example, for the Books3 dataset, "a library of around 196,000 books, including works by popular authors like Stephen King, Margaret Atwood, and Zadie Smith."[565] This dataset, which is drawn

---

**563**. Baio, *supra* note 492.

**564**. Of course, it might not be possible to make the chatbot convincing without significant memorization, but the memorization is still not the

**565**. Kate Knibbs, *The Battle Over Books3 Could Change AI Forever*, Wired (Sept. 4, 2023), https://www.wired.com/story/battle-over-books3/; *see also* Alex Reisner, *Revealed: The Authors Whose Pirated Books are Powering Generative AI*, The Atlantic (Aug. 19, 2023), https://www.theatlantic.com/technology/archive/2023/08/books3-ai-meta-llama-pirated-books/675063/.

from a "shadow library" of almost-certainly infringing books, is very likely unfair.

One factor that might weigh on a court's decision-making is whether a model trainer knew or should have known that a dataset was infringing. Although bad faith is not officially part of the four factors, courts do sometimes emphasize the defendant's bad intentions or unethical conduct in finding no fair use.[566] Thus, a court might treat a company that trained on Books3 without knowing the details of its origins more leniently than a company that trained on it with full knowledge of its infringing contents.

## *I. Express Licenses*

A license from the copyright owner is a complete defense to infringement.[567] It could hardly be otherwise. The modern copyright system depends on licenses voluntarily granted by authors to publishers.

Some creators have expressly agreed to allow their works to be used for training the models used in generative-AI systems.[568] Only such a license from the copyright owner — or from a licensee who is allowed to grant sublicenses – is effective. A dataset creator/curator or model trainer cannot simply rely on the the license a work bears. That license might have been applied by someone who did not have the authority to do so. In this case, it is hornbook law that the license is ineffective, and anyone who relies on it is an infringer. There is no defense of good-faith reliance on a purported license. Improperly licensed works can be removed from a dataset once the mistake is noticed. But it will be much harder to remove those works them from a model trained on reliance on them.[569]

Some licenses are specific. They allow a specific named licensee to use the work for specified purposes. Adobe's Firefly, for example, claims to be trained in substantial part on images licensed by their creators to Adobe Stock.[570] Only Adobe can use those works for training.

---

566. *E.g.,* Harper & Row, Publishers, Inc. v. Nation Enters., 471 U.S. 539, 563 (1985) (the defendant "knowingly exploited a purloined manuscript").

567. *See generally* Jorge L. Contreras, Intellectual Property Licensing and Transactions: Theory and Practice (2022) (discussing IP licensing).

568. *See, e.g.,* Mia Sato, *Grimes Says Anyone Can Use Her Voice for AI-Generated Songs*, The Verge (Apr. 24, 2023), https://www.theverge.com/2023/4/24/23695746/grimes-ai-music-profit-sharing-copyright-ip.

569. *See* Meng, Bau, Andonian & Belinkov, *supra* note 484; Bourtoule, Chandrasekaran & Choquette-Choo et al., *supra* note 484 (regarding the difficulty of model editing).

570. *See* Benj Edwards, *Ethical AI art generation? Adobe Firefly may be the answer*, Ars Technica (Mar. 22, 2023), https://arstechnica.com/information-technology/2023/03/ethical-ai-art-generation-adobe-firefly-may-be-the-answer/. *But see* Sharon Goldman, *Adobe Stock Creators Aren't Happy With Firefly, the Company's 'Commercially Safe'*

These specific licenses apply to only a small fraction of the works currently being used as training data.[571]  Models trained only with this kind of specific permission are rare.  They are often lower quality than the most cutting-edge generative-AI models.[572]

Other licenses are general.  They allow *anyone* to use a work in specified ways, not just an individual named licensee.  Here, anyone is allowed to engage in a use as long as it complies with the terms of that license, even if the user of the work[573] has never directly interacted with the copyright owner to obtain individual permission.  We will use Creative Commons licenses as an example, as the terms in the Creative Commons license suite cover a useful range of interesting conditions.

Some materials are provided under a public-domain mark, which indicates that there are no copyright interests in the material.[574]  Others are provided under a Creative Commons Zero notice, which indicates that the copyright owner has dedicated the material to the public domain.[575]  Any and all uses of these works are allowed, by anyone, without risk of copyright infringement.

The basic license grant in every other Creative Commons license is the right to "reproduce and Share the Licensed Material, in whole or in part; and produce, reproduce, and Share Adapted Material."[576]  This covers all of the section 106 exclusive rights, and it covers all of the activities involved in compiling training datasets, model training and fine-tuning, deployment,

---

generation, alignment, and use of the generated material. So unless some other license term restricts this grant, generative-AI systems are fully and expressly licensed to use any CC-licensed material in their training data.

The attribution term in BY licenses requires that the user of the work retain the creator's identification, indicate whether the work is modified, and retain the Creative Commons license notice. This requirement can be satisfied in "any reasonable manner based on the medium, means, and context."[577] A training dataset could provide this information through suitable metadata, but many datasets do not.[578] If liability were a serious concern, and the availability of CC-licensed material sufficiently broad to justify it, it is possible that more datasets would bear these attributions, so that they would be fully allowed under CC-BY licenses.

This, however, is where attribution stops with current opaque generative-AI models. These models do not attempt to store information about the attribution of the works they were trained on.[579] To the extent that they copy from their CC-BY-licensed training data, these models are derivative works that do not bear proper attribution, so they fall outside the scope of the license. A model that does not retain attribution information cannot provide that information in its generations, so the generations also fall outside the license.

The non-commercial term in NC licenses prohibits uses "primarily intended for or directed towards commercial advantage or monetary compensation."[580] This definition roughly tracks the way in which commerciality is defined in fair use, as discussed above. It seems likely that the sale and licensing of datasets and models, and the provision of generations for money would be considered commercial. So this term would allow entirely open-source supply chains, but prohibit any commercial links in those chains.

The no-derivatives term in ND licenses allows the user to copy and share the work itself, but to "produce and reproduce, but not Share, Adapted Ma-terial."[581]   Adapted Material is defined as "material . . . that is derived from or based upon the Licensed Material and in which the Licensed Material is translated, altered, arranged, transformed, or otherwise modified in a manner requiring permission"[582] from the copyright owner.  In other words, it is any derivative work under copyright law.  An ND license therefore allows dataset curation (as datasets are compilations, not derivatives). But it probably prohibits model training, because a model is most likely a derivative work. So one could train models for research, but not share them.  The only way for models to escape from the ND term is for them not to be substantially similar to the copyrighted work, and thus escape from copyright law entirely.  Generations, too, are derivative works unless they are so substantially identical to a training example that they are memorized duplicates rather than generations, or unless they are so substantially dissimilar from the training example that they do not infringe at all.  The upshot is that an ND-license is effectively no license at all for models and generations.[583]

---

577. *Id.* § 3(a)(1)(A)(i).

578. Katherine Lee, Daphne Ippolito & A. Feder Cooper, The Devil is in the Training Data (2023) (unpublished manuscript), *in* Lee, Cooper, Grimmelmann & Ippolito, *supra* note 65, at 5.

579. *see supra* note 139 and accompanying text (for the challenges of attribution). Indeed, attribution is one of the motivations for using RAG: the hope is that the specific, retrieved examples will have a greater influence on the generation, thereby making attribution easier. *See supra* note 265 and accompanying text (for a discussion of retrieval models). In practice, however, this depends from generation to generation. *See generally* Longpre, Perisetla & Chen et al., *supra* note 498 (for an evaluation of how often generations are based on the retrieved context when the retrieved context is provided).

580. *Creative Commons Attribution-NonCommercial 4.0 International License* § 1(i) (2023), https://creativecommons.org/licenses/by-nc/4.0/.

The share-alike term in SA licenses does allow for the sharing of derivative works, but they must be placed under the same Creative Commons license that the underlying works were licensed under.[584] So a model trained on BY-SA works would itself need to be shared BY-SA — if it is shared at all.  A trainer who keeps the model in-house and uses it only to power a generation service, does not trigger the distribution threshold that causes the share-alike condition to kick in.  If the model is under an SA license, then most generations from it are derivative works of the model and themselves need to be shared SA. If the model is not SA, then only those generations that are derivative works of the original SA work need to be shared SA. Unlike with BY, this relicensing is feasible without individual attribution — a blanket BY-SA license applied to a dataset, a model, or a generation would suffice.

But note that it would probably not be possible to train a single model on both BY-SA and BY-NC-SA works.  Each license requires that any derivative works be released under *that license.*  And each license states that the licensee "may not offer or impose any additional or different terms or conditions" on the work.[585]

---

581. *Creative Commons Attribution-NoDerivatives 4.0 International License* § 2(a)(1)(B) (2023), https://creativecommons.org/licenses/by-nd/4.0/.

582. *Id.* § 1(a).

583. *See supra* Part I.B (concerning derivative works in the generative-AI supply chain).

584. *Creative Commons Attribution-ShareAlike 4.0 International License* § 3(b)(1) (2023), https://creativecommons.org/licenses/by-sa/4.0/.

585. *Id.* (3)(b)(3); *Creative Commons Attribution-NonCommercial-ShareAlike 4.0 International License* (3)(b)(3) (2023), https://creativecommons.org/licenses/by-nc-sa/4.0/.

Lastly, it is worth noting that generation-time plugins could pull in additional data that is not expressly licensed or further complicates our compatibility analysis above.

To summarize:

- An attribution requirement is a difficult technical problem, and no current systems do it effectively.[586]

- A non-commerciality requirement is feasible for most fully open-source supply chains, but difficult for many proprietary ones.

- A no-derivatives requirement effectively prohibits generative AI.

- A share-alike requirement is feasible and tries to compel AI developers to contribute their models to a share-alike commons, but may not reach all generation services, and may raise license-compatibility issues.

- Generation-time plugins could complicate licensing compatibility considerations.

The punch line is that BY is a common term in all of the six standard Creative Commons licenses. No current generative-AI model is licensed under any CC license.[587] Neither are any of their generations. All of the other license terms are irrelevant. For now, at least, CC licensing is a dead-end for generative AI.

### J. Implied Licenses

Implied copyright licenses arise when a copyright owner's conduct gives rise to an inference that they have consented to particular uses.[588] No particular formalities are required to create one.[589] Caselaw holds that the act of putting material online on the web typically creates an implied license for search engines to index it and for archives to maintain archival copies of it.[590] There is also some suggestion that this implied license only applies where the owner has not used a robots.txt file or exclusion headers to deny permission for bulk crawling.[591] The implied license probably does not apply to material behind a paywall or login form that a search engine accesses through surreptitious means.[592]  But it probably does apply to material that a website has specifically made available to a particular search engine.[593]

---

586. *See supra* note 139 and accompanying text. ; *see supra* note 265 and accompanying text.

587. *About the Licenses*, Creative Commons (2023), https://creativecommons.org/licenses/.

588. Effects Assocs., Inc. v. Cohen, 908 F.2d 555 (9th Cir. 1990).

589. Oddo v. Ries, 743 F.2d 630 (9th Cir. 1984).

590. Field v. Google Inc., 412 F. Supp. 2d 1106, 1115–17 (D. Nev. 2006).

591. *Id.* at 1117. The most prominent training dataset, the Common Crawl, respects the robots.txt protocol.*See Frequently Asked Questions*, Common Crawl (2023), https://commoncrawl.org/faq.

The relevant question, then, is what the scope of this implied license is.[594] If I put a photograph online with no further information, it is well-established that this act by itself does not grant permission to third parties to use the photograph in news articles or other publications.[595]  The implied license allows them to copy the photograph as part of viewing it on my page, but not to use it in other contexts.[596]

A training dataset seems broadly akin to the kind of archives that courts have held to be covered by the implied license in other cases.[597] User-supplied prompts, which could become future training data, could be covered by implied licenses, but also could involve express licenses when a user consents to use a particular service.

It is a little harder to say that model training fits within the implied license. This is a new use, one that did not exist when much of the data examples, which have recently been re-purposed for generative-AI training datasets, were first put online.[598]  With respect to re-purposing materials, there is a useful analogy here to the Google Books case.  Book scanning did not exist when most of the books in the corpus were published, so it is hard to say that authors and publishers consented to scanning when they published.[599]

---

**592.** Sites that use such barriers may also have express licensing in place for datasets based on their data.

**593.** *Cf. Structured Data for Subscription and Paywalled Content (CreativeWork)*, Google Search Cent. (May 23, 2023), https://developers.google.com/search/docs/appearance/structured-data/paywalled-content (describing how to make paywalled content accessible to Google's indexing bot).

**594.** *See generally* Christopher M. Newman, *"What Exactly Are You Implying?": The Elusive Nature of the Implied Copyright License*, 32 Cardozo Arts & Ent. L.J. 501 (2014).

**595.** This point is most clearly seen in the cases holding that news publishers cannot embed photographs posted to Instagram or other social networks *E.g.,* Sinclair v. Ziff Davis, LLC, 454 F.Supp.3d 342 (S.D.N.Y. 2020).

**596.** Agence Fr. Presse v. Morel, 769 F.Supp.2d 295, 302–03 (S.D.N.Y. 2011) (holding that the license a user granted to Twitter when he uploaded photographs did not run in favor of third-party publishers who downloaded the photographs from Twitter).

**597.** *E.g.,* Field v. Google Inc., 412 F. Supp. 2d 1106 (D. Nev. 2006) (Google Cache); Parker v. Yahoo!, Inc., 88 U.S.P.Q.2d 1779 (E.D. Pa. 2008) (Yahoo and Microsoft search). *But see* MidlevelU, Inc. v. ACI Info. Grp., 989 F.3d 120 (11th Cir. 2021) (accepting *Field* but holding, "Implied permission to enter through a front door (web crawler) does not also imply permission to enter through a back window (RSS feed).").

**598.** *See supra* Part I.B.4 (regarding web-scraped datasets); *supra* Part I.C.2 (regarding data creation); *supra* Part I.C.3 (regarding the creation and curation of training datasets from previously created data).

**599.** *See generally* Authors Guild v. Google, Inc., 804 F.3d 202 (2d Cir. 2015).

It is harder still to say that putting material online constitutes an implied license to use that material in AI generations.[600] It is certainly the case that many copyright owners strenuously object to this practice. And if a court is to say that generation is allowed, fair use (which applies whether or not the copyright owner consents) is a better fit for the facts than implied license (which applies only when the copyright owner consents).

This said, the fact that materials were voluntarily placed online can be relevant to the fair-use inquiry. As in *Sony*, which held that taping over-the-air television programs for time-shifting was a fair use, the choice to publish involves giving users access to a work.[601] Copyright owners did not need to license their works for broadcast; they had other alternatives that did not invite the public to view for free. One would not draw a similar inference from the choice to show a movie in theaters. So even if there is not an implied license as such for AI training, the fact that there is a broadly shared practice of putting material online, where any web user can view, helps to support a fair-use defense for AI systems and users.

In addition, other laws, such as trespass to chattels and the Computer Fraud and Abuse Act, may sometimes restrict the ability of dataset compilers to scrape data.[602] These other laws, however, typically only apply against the party that actually scrapes the data. They do not apply against others who come into possession of the data that was scraped, such as model trainers or application deployers. Only copyright runs with the data itself; because of these laws, only copyright is a right to own information as such. And even where these other laws apply, their scope can be quite limited. They typically allow the scraping of publicly accessible material unless there is some additional element of harm to the site being scraped, such as an impairment of its ability to serve others.[603]

---

**600**. *Cf.* Associated Press v. Meltwater U.S. Holdings, Inc., 931 F. Supp. 2d 537 (S.D.N.Y. 2013) (holding that excerpting of between 4.5% and 61% of news articles in a subscription news-monitoring service was not covered by implied license).

**601**. Sony Corp. of Am. v. Universal City Studios, Inc., 464 U.S. 417, 456 (1984) (" *Sony* demonstrated a significant likelihood that substantial numbers of copyright holders *who license their works for broadcast on free television* would not object to having their broadcasts time-shifted by private viewers.") (emphasis added).

**602**. *See generally* Benjamin L.W. Sobel, *A New Common Law of Web Scraping*, 25 Lewis & Clark L. Rev. 147 (2021).

**603**. *See, e.g.*, hiQ Labs, Inc. v. LinkedIn Corp., 31 F.4th 1180 (9th Cir. 2022); *see also* Internet Archive v. Shell, 505 F. Supp. 2d 755 (D. Colo. 2007) (rejecting racketeering claims against Internet Archive for scraping and archiving webpages).

### *K. Remedies*

The Copyright Act allows for a broad array of remedies against infringers.[604] Some of them could be highly significant in shaping the deployment of gener-ative-AI systems.[605]

### *Damages and Profits*

A successful copyright plaintiff is entitled to recover "the actual damages suffered by him or her as a result of the infringement."[606] This is a damage remedy measured by the victim's harm. It consists of the money the plaintiff *lost* as a result of the infringement, such as decreases in sales or cancelled licensing contracts with third parties. In *Harper & Row, Publishers, Inc. v. Nation Enterprises*, for example, *Time* cancelled a contract to publish excerpts of Gerald Ford's memoirs when *The Nation* published infringing excerpts ahead of the book's publication date.[607] These actual out-of-pocket losses, however, are rare and hard to prove, so the Copyright Act allows a variety of alternative theories to ground an award of damages.

The simplest such theory is that the plaintiff's damages can be measured by the lost licensing fee that the defendant saved by infringing.[608] This is a fair-market-value remedy; the plaintiff is awarded the licensing fee that a willing seller and willing buyer would have negotiated.[609] As with fair use, much depends on the existence of a licensing market for the kind of use at issue. If there is no such market, it can be hard for a court to estimate an appropriate royalty. So, for example, while there is a well-functioning market for licensing new editions of books, there is not a market for licensing AI training on books — because the use has not existed until now, neither has

---

**604.** *See generally* Douglas Laycock & Richard L. Hasen, Modern American Remedies: Cases and Materials (5th ed. 2018) (discussing types of remedies available under United States law).

**605.** *See generally* Pamela Samuelson, *How to Think About Remedies in the Generative AI Copyright Cases*, Lawfare (Feb. 15, 2024), https://www.lawfaremedia.org/article/how-to-think-about-remedies-in-the-generative-ai-copyright-cases (discussing potential remedies in the current generative-AI copyright lawsuit landscape).

**606.** 17 U.S.C. § 504(b).

**607.** Harper & Row, Publishers, Inc. v. Nation Enters., 471 U.S. 539 (1985).

**608.** *E.g.,* Dash v. Mayweather, 731 F.3d 303, 313 (4th Cir. 2013) ("Under the lost licensing fee theory, actual damages are generally calculated based on "what a willing buyer would have been reasonably required to pay to a willing seller for [the] plaintiffs' work.") (internal quotation omitted).

**609.** *Id.*

the market.[610]  In addition, it can be difficult for individual plaintiffs to show that their work in particular has a high licensing value.[611]  In *On Davis v. The Gap, Inc.*, for example, the plaintiff requested a $2,500,000 licensing fee for the unauthorized use of his eyewear in a Gap ad.[612]  The court held that his evidence supported a licensing fee of $50.[613]

Recognizing that this too may be an inadequate measure of damages, the Copyright Act also allows a successful plaintiff to recover "any profits of the infringer that are attributable to the infringement and are not taken into account in computing the actual damages."[614]  Instead of measuring the plaintiff's losses, this remedy measures the defendant's unfair gains.[615]  The Copyright Act has a burden-shifting provision for defendant's profits that on paper is quite generous to the copyright owner:

> In establishing the infringer's profits, the copyright owner is required to present proof only of the infringer's gross revenue, and the infringer is required to prove his or her deductible expenses and the elements of profit attributable to factors other than the copyrighted work.[616]

The hard part is determining how much of the defendant's profits are "attributable to factors other than the copyrighted work."  In a generative-AI context, we would ask, how much of a generation's value is due to a particular training work, as opposed to other training works and the training algorithm?  This is a hard question by itself; answering the same question for a model requires answering it for all generations the model is used to produce, and adding up the results.  In practice, the answer may depend on who bears the burden of persuasion on the relative value of different elements.

---

**610**. One reason for copyright owners to enter into licensing arrangements with some AI companies may thus be be to establish a baseline for calculating damages against others who do not agree to licensing arrangements.

**611**. *E.g., Dash*, 731 F.3d at 312–26 (rejecting licensing fee calculation in plaintiff's expert report).

**612**. On Davis v. The Gap, Inc., 246 F.3d 152, 156 (2d Cir. 2001).

**613**. *Id.* at 161.

**614**. 17 U.S.C. § 504(b).

**615**. That makes infringer's profits a *restitutionary* remedy rather than a compensatory remedy.  *See generally* Ward Farnsworth, Restitution: Civil Liability for Unjust Enrichment (2014) (discussing the theory of restitution).  The provision is phrased the way it is to avoid double-counting.  If the plaintiff loses one sale to the defendant, that sale would be "profits of the infringer" that *are* "taken into account in computing the [plaintiff's] actual damages."

**616**. 17 U.S.C. § 504(b).  *See generally* Frank Music Corp. v. Metro-Goldwyn-Mayer, Inc., 772 F.2d 505 (9th Cir. 1985) (performing apportionment calculation).

There is an illuminating passage in *On Davis*, where the court held that none of the Gap's overall profits were attributable to the use of the defendant's eyewear in one photograph.[617] Explaining its reasoning, the court wrote:

> Thus, if a publisher published an anthology of poetry which contained a poem covered by the plaintiff's copyright, we do not think the plaintiff's statutory burden would be discharged by submitting the publisher's gross revenue resulting from its publication of hundreds of titles, including trade books, textbooks, cookbooks, etc. In our view, the owner's burden would require evidence of the revenues realized from the sale of the anthology containing the infringing poem. The publisher would then bear the burden of proving its costs attributable to the anthology and the extent to which its profits from the sale of the anthology were attributable to factors other than the infringing poem, including particularly the other poems contained in the volume.[618]

On this analogy, a generation might be like an anthology. Once the plaintiff shows that a infringing generation has commercial value, the defendant bears the burden to show what portion of the value came from other sources — a burden that may be quite difficult to meet. So, to a first approximation, those who profit from infringing generations should expect to pay out their entire profits.

Also on this analogy, a generative-AI system (or model or training dataset) might be more like a full catalog. Any individual training work is utterly insignificant on the scale of the whole system.[619] A plaintiff who shows only that their work was included in the training dataset has not carried their burden to show that any of the resulting profits were attributable to infringement of their work.[620]

This point demonstrates the crucial importance of *mass* copyright litigation against the service hosts of and other participants in generative-AI systems. The answer may well be different if the plaintiff or plaintiffs own a large fraction of the works used as training data. Although individual apportionment may remain a difficult problem, it is much easier to show that the model's value collectively derives from the works that have been infringed.

---

**617**. *On Davis*, 246 F.3d at 160.

**618**. *Id.*

**619**. However, as we note above, some training data examples may have outsized influence on generations. *See generally* Koh & Liang, *supra* note 139; Akyurek, Bolukbasi & Liu et al., *supra* note 139; Grosse, Bae & Anil et al., *supra* note 139. (discussing influence functions).

**620**. *See supra* note 139 and accompanying text; *supra* note 265 and accompanying text.

This is one reason why so many of the current lawsuits against generative-AI companies have been brought as putative class actions.[621]  Getty's lawsuit against Stability AI is not a class action, but Getty controls the copyright to a large number of works in Stable Diffusion's training dataset.[622]  The *New York Times* has alleged that OpenAI trained more exensively on its "higher-quality" articles compared to other sources of training data.[623]

### Statutory Damages

Instead of recovering actual damages and/or profits, a successful copyright plaintiff may elect to recover statutory damages instead.[624]  This will typically be an appealing option.  First, the plaintiff can submit both theories to the court, see which one results in a larger award, and then choose that one.[625]  Second, the amount of statutory damages is fixed in the statute.  The base range is $750 to $30,000, "as the court considers just."[626]  This amount can be decreased to $200 for an innocent infringer who "was not aware and had no reason to believe that his or her acts constituted an infringement of copyright,"[627] but this defense is not available for works that were published with proper notice of copyright.[628]  The amount can also be increased up to $150,000 when the "infringement was committed willfully."[629]  Willful infringement consists either of actual knowledge or reckless disregard of infringement;[630] a defendant who has a reasonable and good-faith belief that their conduct is non-infringing is not a willful infringer.[631]  Under these ranges, an individual statutory-damage award could be a serious threat to an individual user, a moderate nuisance to a small company, or an insignificant bit of background noise to an OpenAI or a Google.

---

**621**. *E.g.,* Complaint, Kadrey v. Meta Platforms, Inc., No. 3:23-cv-03417 (N.D. Cal. July 7, 2023); Complaint, Doe 1 v. GitHub, Inc., No. 4:22-cv-06823 (N.D. Cal. Nov. 3, 2022); Complaint, Anderson v. Stability AI, Ltd., No. 3:23-cv-00201 (N.D. Cal. Jan. 13, 2023) (Doc. No. 1); Complaint, Tremblay v. OpenAI, Inc., No. 3:23-cv-03223 (N.D. Cal. June 28, 2023).

**622**. Complaint, Getty Images (US), Inc. v. Stability AI, Inc., No. 1:23-cv-00135 (D. Del. Feb. 3, 2023).

**623**. Complaint at ¶ 90, N.Y. Times Co. v. Microsoft, No. 2:24-cv-00711 (C.D. Cal. Dec. 27, 2023).

**624**. 17 U.S.C. § 504(c)(1).

**625**. Curet-Velazquez v. ACEMLA de P.R., Inc., 656 F.3d 47, 57–58 (1st Cir. 2011).

**626**. 17 U.S.C. § 504(c)(1).

**627**. *Id.* § 504(c)(2).

**628**. 17 U.S.C. § 401(d).

**629**. 17 U.S.C. § 504(c)(2).

**630**. Erickson Prods., Inc. v. Kast, 921 F.3d 822, 833 (9th Cir. 2019).

**631**. VHT, Inc. v. Zillow Grp., Inc., 918 F.3d 723, 748–49 (9th Cir. 2019).

Importantly, statutory damages are awarded *per work* infringed, regardless of how extensively each work was used. Again, the impact is clearest in mass copyright litigation. Statutory damages are a potentially existential threat to models trained on billions of works (and to the datasets that feed them and the services that incorporate them). Even without a finding of willfulness, the statutory damages for a billion infringed works could be as high as in the trillions of dollars — an impact that is no more survivable than the Chicxulub asteroid. Even at the minimum award for innocent infringement, the statutory damages for ten million infringed works would come to two hundred million dollars.[632]

One factor limiting statutory damage awards is that statutory damages are only available when the copyright owner registered the work with the Copyright Office before the infringement commenced.[633] This provision is designed to encourage authors to register their works promptly. It has the effect of making some generative-AI systems more vulnerable to copyright lawsuits than others. Books are typically registered as part of the publication process, so an LLM trained on hundreds of thousands of books could face hundreds of thousands of statutory-damage awards But many works of visual art and many websites are not registered unless and until the copyright owner needs to file a copyright lawsuit.[634] A model trained on a web scrape, then, may face a patchwork of statutory damage awards only for a small fraction of the works it was trained on. Differences in available damages based on the timing of registration may make it harder to assemble a plaintiff class with sufficiently common interests.[635]

### Attorney's Fees

Another remedy for copyright infringement is that a court may award "full costs" and "a reasonable attorney's fee to the prevailing party."[636] Costs are small potatoes; they include various court fees, printing fees, and other other required payments to the court.[637] But attorney's fees are a bigger deal, precisely because the expense of litigating a copyright case can be so

---

**632.** This sum is still high enough that it might deter a court from finding infringement against a smaller defendant that merely used a model someone else had trained.

**633.** 17 U.S.C. § 412(2).

**634.** Registration is a prerequisite to suit. § 411(a); Fourth Est. Pub. Corp v. Wall-St. com, LLC, 139 S. Ct. 881 (2019).

**635.** *See* Fed. R. Civ. P. 23(a)(2) (requiring " questions of law or fact common to the class"). The registration requirement cannot be circumvented through the use of a class action. *See* Reed Elsevier, Inc. v. Muchnick, 559 U.S. 154 (2010).

**636.** 17 U.S.C. § 505.

**637.** *See* Rimini St., Inc. v. Oracle USA, Inc., 139 S.Ct. 873 (2019) (interpreting "full costs").

high. Under the usual "American Rule" (so called because it is followed in the United States but not in many other countries), each party pays its own lawyers and decides how much the case is worth to them.[638] The Copyright Act's fee-shifting provision is one of a few exceptions to the American Rule. It provides an incentive to parties to bring meritorious cases — or to defend against unmeritorious ones — that would otherwise be financially unreasonable to pursue.[639] Like statutory damages, attorney's fees are only available for works that were registered before the infringement.[640]

While statutory damages are most important in mass litigation, the reverse is true of attorney's fees. A million dollars of expenses to litigate a class action with a hundred-million-dollar damage award is not the biggest deal. A fee award is a nice bonus, but it is not necessary to bring the suit in the first place. But a million dollars of expenses to litigate an individual claim leading to a $1,000 statutory damage award is completely unreasonable. Without an attorney's fee award, the lawyers involved could make more on a per-hour basis by busking on the subway.

Attorney's fees can also have a significant deterrent effect.[641] Because they are uncapped, a plaintiff can run up the total award a defendant faces. Indeed, the harder a defendant fights, the higher the plaintiff's attorney's fees will be. Along with statutory damages, attorney's fees can be used to coerce settlements from defendants who may have a strong defense on the merits.[642] Even though the defendant might be able to receive a fee award if they win — the fee-shifting rule is symmetrical[643] — they cannot run the risk of paying a massive fee award if they lose. This settlement pressure will be strongest against smaller and more risk-averse defendants: end users rather than well-capitalized AI companies, which can better absorb the cost of a fee shift. This difference helps to explain why several generative-AI companies have offered to indemnify their users against the copyright risks of using their systems.[644]

---

638. Fogerty v. Fantasy, Inc., 510 US 517, 533–34 (1994).

639. *Id.* at 524.

640. 17 U.S.C. § 412.

641. *See generally* Pamela Samuelson & Tara Wheatland, *Statutory Damages in Copyright Law: A Remedy in Need of Reform*, 51 Wm. & Mary L. Rev. 439 (2009); Talha Syed & Oren Bracha, *The Wrongs of Copyright's Statutory Damages*, 98 Tex. L. Rev. 1219 (2020).

642. *See, e.g.,* Mitch Stoltz, *Collateral Damages: Why Congress Needs To Fix Copyright Law's Civil Penalties*, Elec. Frontier Found. (July 24, 2014), https://www.eff.org/wp/collateral-damages-why-congress-needs-fix-copyright-laws-civil-penalties.

643. *Fogerty*, 510 US 517.

644. Brad Smith, *Microsoft Announces New Copilot Copyright Commitment for Customers*, Microsoft (Sept. 7, 2023), https://blogs.microsoft.com/on-the-issues/2023/09/07/copilot-copyright-commitment-ai-legal-concerns/; Bridget Johnston, *Introducing Indemnification for AI-Generated Images: An Industry First*, Shutter-

### Injunctions

A court may "grant temporary and final injunctions on such terms as it may deem reasonable to prevent or restrain infringement of a copyright."[645] An injunction is a court order commanding a person to take (or to avoid taking) some action. A party who fails to comply with an injunction can be punished for contempt of court with sanctions that include escalating fines and even imprisonment.

An injunction is an equitable remedy; a plaintiff is not automatically entitled to one.[646] Instead, a plaintiff seeking an junction must show:

> (1) that it has suffered an irreparable injury; (2) that remedies available at law, such as monetary damages, are inadequate to compensate for that injury; (3) that, considering the balance of hardships between the plaintiff and defendant, a remedy in equity is warranted; and (4) that the public interest would not be disserved by a permanent injunction.[647]

The first two factors are redundant; they mean exactly the same thing.[648] A damages award in a copyright case is inadequate when damages are hard to calculate. For all of the reasons discussed above, this will frequently be the case in generative-AI cases. Thus, most of the weight will fall on the third and fourth factors. The degree to which hardships fall on a defendant that provides generative-AI models or systems, and on third-party users, will depend substantially on the balance of infringing and noninfringing uses. An injunction is more appropriate against a system that (a court sees as) "good for nothing else but infringement,"[649] and less appropriate against one that is also "capable of substantial noninfringing uses."[650] (As these quotes suggest, there is substantial overlap between the substantive tests for infringement and the test for a permanent injunction.)

---

STOCK (July 11, 2023), https://www.shutterstock.com/blog/ai-generated-images-indemnification; Adobe, *Firefly Legal FAQs – Enterprise Customers* §§ 10–14 (June 12, 2023), https://www.adobe.com/content/dam/dx/us/en/products/sensei/sensei-genai/firefly-enterprise/Firefly_Legal_FAQs_Enterprise_Customers.pdf.

**645**. 17 U.S.C. § 502(a). We will discuss only permanent injunctions issued after a finding of infringement. Preliminary injunctions issued during the course of a lawsuit may be important for parties and litigators, but our focus is on the longer term.

**646**. eBay Inc. v. MercExchange, L.L.C., 547 U.S. 388, 392–93 (2006).

**647**. *Id.* at 391.

**648**. Douglas Laycock, *The Death of the Irreparable Injury Rule*, 103 HARV. L. REV. 687, 694 (1990).

**649**. Metro-Goldwyn-Mayer Studios Inc. v. Grokster, Ltd., 545 U.S. 913, 932 (2005).

**650**. *Id.* at 927.

Another factor weighing against generative-AI injunctions is the First Amendment interests of users and developers.[651] There is often a speech interest in using the speech of others verbatim;[652] these First Amendment interests are even stronger for novel generations. In individual cases against specific generations, users' speech rights are protected by the "traditional First Amendment safeguards" of fair use, particularly transformative fair use.[653] But an injunction against the use of a model or service can prevent these generations from being created; this is a speech harm too. So when a model is used to create expressive and noninfringing generations, there is a powerful argument that a court should not enjoin it in a way that would prevent these noninfringing uses.

And so we come to one of the most important features of an injunction: a court's ability to craft its specific terms. A court could enjoin the use of a model *entirely*, preventing the defendant from using it for any purpose. But a court could also enjoin the use of a model *to create infringing generations*, leaving it up to the defendant to implement appropriate content filters.[654] This type of injunction puts sharper teeth into the defendant's obligations, because the consequences for failing to comply with an injunction are swifter and more severe than for committing copyright infringement. Unfortunately for defendants (and for courts considering enjoining them), it is harder to "separat[e] the fair use sheep from the infringing goats" in a generative-AI system than it is on a content-hosting service like YouTube.[655] Even for a

---

**651**. Mark A Lemley & Eugene Volokh, *Freedom of Speech and Injunctions in Intellectual Property Cases*, 48 Duke L.J. 147 (1998).

**652**. *See* Rebecca Tushnet, *Copy This Essay: How Fair Use Doctrine Harms Free Speech and How Copying Serves It*, 114 Yale L.J. 535 (2004).

**653**. Eldred v. Ashcroft, 537 U.S. 186, 219–20 (2003).

**654**. For example, Copilot offers an option to check "code suggestions with their surrounding code of about 150 characters against public code on GitHub" and propose a different suggestion if the filter is triggered (*Configuring GitHub Copilot in your environment*, *supra* note 239). Unfortunately, while helpful, content filters like Copilot's are not enough by themselves to prevent the generation of potentially infringing content. For example, Copilot's filter would not be triggered if the generated code suggestion matched 149 characters of public code — which is long enough to at least raise copyright concerns. *See* Justin Hughes, *Size Matters (Or Should) in Copyright Law*, 74 Fordham L. Rev. 575 (2005) (discussing copyright protection of "microworks"). *See generally* Daphne Ippolito, Florian Tramèr & Milad Nasr et al., Preventing Verbatim Memorization in Language Models Gives a False Sense of Privacy (2023) (unpublished manuscript), https://arxiv.org/abs/2210.17546 (discussing how verbatim output filters are necessarily incomplete).

**655**. Campbell v. Acuff-Rose Music, 510 U.S. 569, 586 (1994).

defendant with a list of works to avoid, this type of filtering is a difficult and unsolved technical problem.[656]

### Destruction

Another equitable remedy is that the court may order "the destruction or other reasonable disposition of all [infringing] copies."[657] This is like a more severe version of an injunction, one that takes it out of the defendant's power to commit further infringements by taking away their copies. To the extent that a model is treated as an infringing copy, the destruction remedy does not add very much to a permanent injunction except for irreversibility. Actually deleting a model — as opposed to putting in in storage for future use if and when the law changes or copyright owners negotiate a license to allow it to be used — is an exceptionally harsh remedy that effectively means throwing away all of the compute used to train the model.

But there is a twist. As Elizabeth Joh observes,[658] the destruction remedy covers not just infringing copies but also "all plates, molds, matrices, masters, tapes, film negatives, or other articles by means of which such copies or phonorecords may be reproduced."[659] Even if a model is not itself treated as an infringing copy, if it is capable of producing infringing generations, it might be an "article[] by means of which" infringing copies "*may* be reproduced."[660] The courts have not restricted this remedy to items that themselves infringe or have been used to infringe.[661] Instead, they have allowed it to be used against dual-use technologies like computers and manufacturing equipment that can be used both to infringe and for noninfringing purposes.[662] Thus, the destruction remedy could reach not just models with multiple uses, but also the non-model portions of a generative-AI service. For example, a court could order the destruction of a style-transfer system that allows users to regenerate one image using the artistic style of another, on the theory that a user could prompt it with a copyrighted image and gen-erate an infringing derivative work.  Such an order would raise even more severe free-expression concerns.

---

656. *See supra* note 431 and accompanying text.

657. 17 U.S.C. § 503(b). *See generally* Elizabeth E. Joh, Equitable Legal Remedies and the Existential Threat to Generative AI (Aug. 27, 2023) (unpublished manuscript), https://papers.ssrn.com/sol3/papers.cfm?abstract_id=4553431. As with injunctions, there is also a preliminary version of destruction: a court may order the impoundment of infringing copies during the course of the litigation. 17 U.S.C. § 503(a)(1).

658. Joh, *supra* note 657.

659. 17 U.S.C. § 503(b).

660. *Id.* (emphasis added).

661. Mahan v. Roc Nation, LLC 720 Fed. Appx. 55 (2d Cir. 2018).

662. Anne-Marie Carstens, *Copyright's Deprivations*, 96 Wash. L. Rev. 1275 (2021).

### L. Copyright Management Information

Section 1202 of the Copyright Act, enacted like section 512 as part of the DMCA, deals with "**copyright management information** . . . conveyed in connection with copies . . . of a work" (CMI).[663] Types of CMI include a work's title, author, copyright owner, performers, and licensing information.[664] One prong of section 1202 prohibits providing "false" CMI;[665] another prohibits "remov[ing] or alter[ing]" CMI.[666]

The legislative history of section 1202 (and its passage as part of the *Digital* Millennium Copyright Act) suggests that it was designed to work in tandem with section 1201, which prohibits disabling digital rights management systems that protect copyrighted works.[667]  Where section 1201 guards the parts of the system that directly control access, section 1202 ensures that the metadata and watermarks attached to works are accurate and intact.[668]

But the language of section 1202 is not limited to digital metadata.  Unlike the World Intellectual Property Organization Copyright Treaty, which applies to "*electronic* rights management information,"[669] section 1202's text contains no such limitation.  As a result, courts have held that section 1202 can be violated when a magazine photo is reproduced online without the photographer's name from a "gutter credit" that appeared alongside it in print.[670]

Under these precedents, the assembly of works into datasets and the training of a model could result in the "remov[al]" of CMI through a similar decontextualization. Consider a diffusion model trained on one of the LAION image datasets. The dataset itself consists of URL links to images where they appear in context on webpages, *See supra* Part I.B.3 often with author, title, and copyright-owner credits of the type that qualify as protected CMI. This by itself is neither falsification, removal, or alteration.  But when the images *by themselves* are downloaded, the attached CMI is stripped in the same way

---

**663**. **1202 at (c)**.

**664**. *Id.*

**665**. *Id.* (a).

**666**. *Id.* (b).

**667**. **1201**; Severine Dusollier, *Some Reflections on Copyright Management Information and Moral Rights*, 25 Colum. J.L. & Arts 377 (2003).

**668**. IQ Grp., Ltd. v. Wiesner Pub., 409 F.Supp.2d 587, 593–97 (D.N.J. 2006); Textile Secrets Int'l, Inc. v. Ya-Ya Brand Inc., 524 F. Supp. 2d 1184, 1196–99 (C.D. Cal. 2007).

**669**. WIPO Copyright Treaty, art. 12(1)(i), 1996.

**670**. Murphy v. Millennium Radio Grp. LLC, 650 F.3d 295 (3d Cir. 2011); Mango v. BuzzFeed, 970 F.3d 167 (2d Cir. 2020).

as in the magazine cases. Training and generation do not repair the linkage once it has been severed; if a model outputs a similar image, it will not bear the original CMI.

Getty Images's complaint against Stability AI presents additional theories of section 1202 violation based on the Stable Diffusion models' treatment of the Getty watermarks on the images in its library.[671] First, to the extent that the training process learns features of training images without the watermark, Getty alleges removal and alteration of CMI.[672] Second, Getty shows that Stable Diffusion sometimes produces generations that include distorted versions of the watermark.[673] This, Getty argues, constitutes "false" CMI within the meaning of section 1202.[674]

The more serious doctrinal obstacle to section 1202 claims is that they require a nexus to copyright infringement. Falsification of CMI must be done "knowingly and with the intent to induce, enable, facilitate, or conceal infringement" to create liability,[675] and removal or alteration must be done "intentionally . . . knowing, or . . . having reasonable grounds to know, that it will induce, enable, facilitate, or conceal an infringement."[676] Most defendants in generative-AI cases will have the required intent to remove or alter the CMI. A developer training on the LAION dataset can hardly fail to know that the training process discards any information about the images on the webpages they came from.

Instead, it is not clear that a defendant's treatment of CMI at any stage of the generative-AI supply chain is intended to facilitate or conceal copyright infringement in any cases where copyright infringement would not already attach to the defendant. Getty objects that attaching a "modified version of the Getty Images watermark to bizarre or grotesque synthetic imagery,"[677] will harm its reputation. But that is a concern that sounds in trademark, not copyright.[678] Indeed, the "grotesque" nature of the images Getty includes in its complaint, if anything, cuts against infringement, by suggesting that the images are not suitable for any valuable purpose, let alone competing with Getty. The decontextualization of the training process might be said to help "conceal" infringement, but again the infringement itself is likely to be separately actionable.

---

**671**. Complaint, Getty Images (US), Inc. v. Stability AI, Inc., No. 1:23-cv-00135 (D. Del. Feb. 3, 2023).

**672**. *Id.* ¶¶ 81–86.

**673**. *Id.* ¶¶ 59–60.

**674**. *Id.* at 74–80.

**675**. **1202 at (a)**.

**676**. *Id.* (b).

**677**. Complaint at ¶ 59, Getty Images (US), Inc. v. Stability AI, Inc., No. 1:23-cv-00135.

**678**. *See id.* ¶¶ 87–99 (bringing claim for trademark infringement).

The real bite of the CMI claims may be remedial. A court is entitled to award statutory damages of $2,500 to $25,000 "for each violation of section 1202."[679] The liability is *per violation* rather than *per work* (as with ordinary copyright infringement). In theory, then, a defendant could face separate section 1202 liability for each variation of a dataset or model it creates, or each output bearing a watermark. On the other hand, while there is a $200 floor for ordinary copyright statutory damages in cases of innocent infringement,[680] a is entitled " in its discretion" to reduce section 1202 statutory damages or remit them entirely in cases of innocent violations.[681]

## *M. Right of Publicity*

A related but non-copyright form of IP is the **right of publicity**.[682] The right generally protects an individual's persona against commercial appropriation by others. Unlike copyright, which is almost entirely created by federal law, the right of publicity is almost entirely created by state law. As a result, its details vary substantially from state to state, including whether a state affords a right of publicity at all and, if it does, whether the right is regarded as a privacy or property right or both, what kinds of conduct it protects against, the scope of newsworthiness or expressive-use defenses, and whether and how long the right lasts after the subject's death.[683] As a result, the following summary is a broad overview, rather than a specific analysis of any state's (or each state's) law.

### Overview of the Right of Publicity

A typical statement of the right of publicity's subject matter is that it covers an individual's "name, voice, signature, photograph, or likeness"[684] — the aspects of a person's persona that are broadly recognizable by others. (We will collectively refer to these as "identity.") Courts have interpreted recognizability broadly, holding that race-car driver Lothar Motschenbacher's right of publicity was infringed by a commercial showing his car,[685] singer Bette Midler's by a commercial featuring a sound-alike vocalist,[686] and game-show host Vanna White's by a print ad showing a robot next to the game board from *The Price is Right*.[687]  The fact that an expressive work is recognizably *by* an author or artist does not mean that it implicates their right of publicity: a use must depict or summon up the person in the minds of viewers. Some states' rights of publicity terminate on death,[688] others last for a legislatively specified or judge-created term,[689] and Tennessee allows the right to continue indefinitely as long as it is being commercially exploited.[690]

---

679. **1203 at (c)(3)(B)**.

680. 17 U.S.C. § 504(c)(2).

681. **1203 at (c)(5)(A)**.

682. *See generally* Jennifer E. Rothman, The Right of Publicity: Privacy Reimagined for a Public World (2018) (providing a thorough history, analysis, and critique of the right of publicity as it exists in the United States today).

683. *See generally* Jennifer E. Rothman, *Rothman's Roadmap to the Right of Publicity* (2024), https://rightofpublicityroadmap.com (providing a detailed analysis of every state's right-of-publicity laws).

684. Cal. Civ. Code § 3344(a).

The right of publicity also applies only to *commercial* uses.  One core use is endorsement: using a person's identity to sell things.[691]  Another is merchandising: selling basketball jerseys with Steph Curry's name and number on them, or Dolly Parton Funko Pops.  And a third is what Eric Johnson calls "virtual impressment":[692] digitally recreating a person to perform in movies, video games, songs, and other audio or audiovisual media.[693]  Broadly speaking, endorsement issues can often be avoid with sufficient disclaimers to establish that the person has not endorsed the product in question or consented to appear in the advertising; but, to the extent that a plaintiff has a valid claim based on merchandising or virtual impressment, disclaimers will not save the defendant.

Although the right of publicity does not have a copyright-style, general-purpose, fair-use defense, some courts have recognized a narrower copyright-style, transformative-use defense when a person's likeness "is so transformed that it has become primarily the defendant's own expression rather than the celebrity's likeness."[694]  Some state statutes explicitly carve out uses affected with a strong public interest, such as a California's exception for "news, public affairs, or sports broadcast or account, or any political campaign."[695]  And sometimes sufficiently expressive uses are excluded entirely, as in California's

---

exception for "fictional or nonfictional entertainment, or a dramatic, literary, or musical work" after the person's death.[696]

The right of publicity has a close and complicated relationship with copyright. First, like all state-created IP rights, it is subject to federal **preemption**. The Copyright Act provides that "all legal or equitable rights that are equivalent to any of the exclusive rights within the general scope of copyright . . . are governed exclusively" by federal copyright.[697] To avoid preemption, a right of publicity claim must either protect different *subject matter* than copyright (e.g., a person's appearance is not a fixed work of authorship) or include an *additional element* not required for copyright infringement (e.g., using the plaintiff's likeness as advertising or promotion to sell another product).[698]

When the basis of a right-of-publicity claim is the distribution of a work either depicting a person or created by the person or both, courts frequently treat the right of publicity as having merged into the copyright in the work: they cannot further restrict the copyright owner's ordinary exploitation of the work. For example, in *Laws v. Sony Music Entertainment, Inc.*, the plaintiff Debra Laws's vocals from "Very Special" were used as a sample on Jennifer Lopez and L.L. Cool J's "All I Have."[699] The defendants had a copyright license from Laws's record label, but not a right of publicity license from Laws. The Ninth Circuit held that Laws's claim was preempted.[700] The case would have been different if the sample had been used for an advertisement rather than a new track; that would have been an extra element.[701] Difficult issues sometimes arise when footage or other works created for one project are reused in a related but different context, as in *Facenda v. NFL Films, Inc.*, where the NFL reused voice-over lines recorded by John Facenda as documentary narration for a 22-minute promotion for a video game, and the Third Circuit held that his estate's right of publicity claim was note preempted.[702]

---

696. Cal. Civ. Code § 3344.1(a)(2). Put another way, California's statutory right of publicity protects against virtual impressment of the living, but not of the dead.

697. 17 U.S.C. § 301(a).

698. *See generally* Jennifer E. Rothman, *Copyright Preemption and the Right of Publicity*, 36 U.C. Davis L. Rev. 199 (2002).

699. Laws v. Sony Music Ent., Inc., 448 F. 3d 1134 (9th Cir. 2006).

700. *Id.* at 1145.

701. *Id.* at 1141–42; *cf.* Downing v. Abercrombie & Fitch, 265 F. 3d 994 (9th Cir. 2001) (photographs used as advertisements).

702. Facenda v. NFL Films, Inc., 542 F. 3d 1007 (3d Cir. 2008).

### Incorporation and Advertising

The most famous generative-AI right-of-publicity lawsuit is both a legal non-starter and not actually about generative AI. In January 2024, the *Dudesy* podcast posted an hour-long episode titled "George Carlin: I'm Glad I'm Dead," which the podcast hosts claimed to feature both a script and audio that had been trained to imitate the late comedian George Carlin.[703] Carlin's estate sued, making claims under California's statutory and common-law rights of publicity.[704] But the statutory claim is a loser because California's postmortem right of publicity, as noted above, expressly excludes audiovisual entertainment, and the common-law claim is a loser because the courts have held that California's common-law right terminates at death.[705] Even more fundamentally, *Dudesy*'s hosts promptly admitted that the episode was entirely human-written.[706] The generative-AI veneer was just a publicity stunt.

The Carlin lawsuit, near-miss though it is, helpfully illustrates two ways in which the right of publicity can apply to generative AI. First, a technical artifact (a dataset, model, system, or generation) could *incorporate* a person's identity. "I'm Glad I'm Dead" imitated Carlin's distinctive voice. Second, a technical artifact could be *advertised* using a person's identity. "I'm Glad I'm Dead" was promoted using Carlin's name. Incorporation raises merchandising and virtual-impressment issues; advertising raises endorsement issues.

In most cases, generative AI will raise distinctive right of publicity issues only to the extent that it incorporates a person's identity. This is for two reasons. First, when it is legal to *sell* a product incorporating a person's identity or creative output, it is also generally legal to *promote* the product by truthfully describing the person's relationship to it.[707] Second, using a person's identity to sell generative-AI material that does not otherwise relate to them is a garden-variety case under the endorsement prong of the right of publicity. Whether it infringes on Salvador Dalí's right of publicity to name a family of image systems "DALL·E" has little to do with the fact that it is a generative-AI system. Almost the same issues would arise with calling a line of paintbrushes "DALL·E".

---

**703.** Christopher Kuo, *George Carlin's Estate Sues Podcasters Over A.I. Episode*, N.Y. Times, Jan. 29, 2024, C6 .

**704.** Main Sequence, Ltd. v. Dudesy, LLC, No. 2:24-cv-00711 (C.D. Cal.).

**705.** Lugosi v. Universal Pictures, 603 P.2d 425 (Cal. 1979).

**706.** Kuo, *supra* note 703.

**707.** Armstrong v. Eagle Rock Ent., Inc., 655 F. Supp. 2d 779 (E.D. Mich 2009) (defendant could use photograph of plaintiff on the cover and liner notes of a DVD concert video).

### Incorporation in the Generative-AI Supply Chain

Some AI generations are already being used for blatant right of publicity violations. Ads featuring a cloned version of Taylor Swift's voice have been used in fake giveaways for Le Creuset cookware;[708] a deepfake video of Tom Hanks has been used to advertise a dental plan.[709] Of course, fake celebrity endorsements are nothing new. The difference between an ad using an (actual) photograph of a celebrity and an ad using a (generated) video of them is a difference in degree, not in kind. Generative AI may make the deception more convincing by forging an explicit endorsement, but what makes these uses actionable is fundamentally the lack of permission. So the right of publicity violation is more about how the media is used, not how it is generated.

From the perspective of the system that is used to generate the media, or any other actors further upstream in the generative-AI supply chain, this is ultimately a secondary-liability question that is quite similar to the secondary-liability question for copyright.[710] The law of secondary liability in right of publicity is both less developed (because the cases are fewer) and more fragmented (because the sources of law are more numerous) than in copyright.[711] It is not obvious that there are any material differences between the two.

That said, the statutory safe harbor potentially applicable to the right of publicity is both less and more complicated than in copyright. On the one hand, the right of publicity is not subject to the safe harbor notice-and-takedown regime of section 512, which applies only to copyright. On the other, a different immunity, "section 230," protects Internet immediacies from liability from third-party information provided by another."[712] Section 230 has an exception for "intellectual property,"[713] and courts are split on whether this includes the state-created right of publicity or not.[714] And if section 230 does apply to the right of publicity, there is deep disagreement

---

**708**. Tiffany Hsu & Yiwen Lu, *No, That's Not Taylor Swift Peddling Le Creuset Cookware*, N.Y. Times, Jan. 9, 2024, B1 .

**709**. Derrick Bryson Taylor, *Tom Hanks Warns of Dental Ad Using A.I. Version of Him*, N.Y. Times, Oct. 2, 2023, https://www.nytimes.com/2023/10/02/technology/tom-hanks-ai-dental-video.html.

**710**. *See supra* Part II.F.

**711**. *See* Perfect 10, Inc. v. Cybernet Ventures, Inc., 213 F. Supp. 2d 1146, 1183–87 (C.D. Cal. 2002); J. Thomas McCarthy, The Rights of Publicity and Privacy § 3:17 to 3:20 (2d ed. 2023) (surveying the limited caselaw).

**712**. 47 U.S.C. § 230(c)(1).

**713**. 47 U.S.C. § 230(e)(2).

**714**. Perfect 10, Inc. v. CCBill LLC, 488 F. 3d 1102, 1118–19 (9th Cir. 2007) (section 230 immunity applies to the right of publicity) *with* Hepp v. Facebook, 14 F.4th 204 (3d Cir. 2021) (no it doesn't).

(and an utter absence of caselaw) on how it applies to generative AI because it is unsettled whether and when AI generations should be regarded as third-party content.[715]

A different theory of a right-of-publicity variation is the use of a generative-AI system to produce outputs in the style of particular artists or authors. Styling these claims as right-of-publicity violations rather than under copyright[716] introduces a few twists. Most fundamentally, there is copyright preemption. To the extent that these claims mirror copyright claims based on the imitation of one's style as embodied in fixed creative works — e.g., a photograph in the style of Cindy Sherman — they are preempted unless there is some extra element. One candidate for such an element is the prompt. At least as to commercial services, there is an argument that if a service produces a generation in response to the prompt `"a photograph in the style of cindy sherman"`, then this constitutes a use by the service of Sherman's name. The doctrinal hurdle here, however, is that it is not clear that the user's prompt should be attributed to the service, which is not using Sherman's name to advertise.[717]

A stronger version of this theory is that services based on models which have been developed to specifically imitate an artist's style or person's appearance or voice are more clearly selling that person's identity. One way of framing this situation is that a commercial model provider is directly violating Drake's identity by using Drake's name to sell its models (of Drake): a form of advertising. Another way is to say that it is the models that are the problem: a form of merchandising. And a third is to say that the model is a kind of toolkit for anyone to engage in virtual impressment of Drake, so the seller is engaged in contributory virtual impressment.

All of these theories are subject to the usual right-of-publicity defenses. The transformative-use argument is particularly strong, for the same reasons it is strong in copyright. Indeed, because the earlier stages of the generative-AI supply chain generally cannot be the basis of right-of-publicity claims — they do not by themselves involve recognizable uses of a person's identity — the transformative-use defense is needed at all only in the later stages, where the transformation is the most pronounced. And the general public-interest

---

**715.** *See generally* Peter J. Benson & Valerie C. Brannon, Congressional Research Service (Congressional Research Service Legal Sidebar LSB11097 Dec. 28, 2023) (surveying caselaw and commentary).

**716.** *See supra* Part II.C.

**717.** To the extent that a service does (or does not) analyze prompts to detect problematic or prohibited requests, there is a question of whether this analysis constitutes use sufficient to trigger the right of publicity or to avoid preemption. A similar issue will arise under other use-based bodies of law, such as trademark.

defense will apply so clearly to some generations (it is not hard to imagine news programs making generative-AI illustrations), and others will be clearly non-infringing private non-commercial uses, so most systems will have substantial noninfringing uses.

## N.  Hot News Misappropriation

One final relevant copyright-like form of IP liability is **hot news misappropriation.** The common-law cause of action for misappropriation has a long history; it is a species of unfair competition law, which prohibits businesses from "reaping the fruits" of their competitors' investments.[718]  Its most famous statement is in the 1918 Supreme Court case *International News Service v. Associated Press.*[719]  The Associated Press (AP) and the International News Service (INS) were competing wire services that reported and transmitted news stories to their member newspapers.  AP alleged that INS was copying news stories from early editions of AP papers so that INS papers could report on them in their later editions.

It is important to note why this practice was not copyright infringement, and is not to this day.  The Associated Press could potentially have a copyright in the articles its employees wrote,[720] but the facts it reported were uncopyrightable. As Justice Brandeis wrote in dissent, "[T]he noblest of human productions—knowledge, truths ascertained, conceptions, and ideas— become, after voluntary communication to others, free as the air to common use."[721]

Justice Pitney's majority opinion, then, focused on the "novelty and freshness" of the news reported by the AP.[722] It held that the AP had a kind of "*quasi* property" as against competitors like the INS.[723] While it could not prevent readers and other members of the general public from freely discussing and writing about the news, it could prevent the INS from engaging in a systematic process of copying the news "precisely at the point where the profit is to be reaped" — that is, while the news was still fresh and there was value in being first to report it in a given newspaper market.[724]

---

**718**. Int'l News Serv. v. Associated Press, 248 U.S. 215, 241 (1918).

**719**. *Id.*

**720**. In addition, under the 1909 Copyright Act, copyright was too encumbered with formalities to provide the AP with effective relief. The 1976 Copyright Act, in which copyright attaches on fixation, overcomes this procedural barrier.

**721**. *Int'l News Serv.*, 248 U.S. at 250 (Brandeis, J., dissenting).

**722**. *Id.* at 238.

**723**. *Id.* at 236.

**724**. *Id.* at 240.

The courts have held that the essential core of a hot-news claim is free-riding on a competitor's costly production of information in a way that undermines the incentives to produce that information at all.[725]  For example, in *National Basketball Ass'n v. Motorola*, the court explained that while it would be misappropriation for one real-time sports-score business to retransmit scores distributed by another, it was legal for the defendant to have its own reporters watch games to keep the scores updated.[726]

For unrelated reasons, there is no longer a federal cause of action for misappropriation.[727]  Instead, it is now governed entirely by state law, and as such it is subject to copyright preemption.  The plaintiff must allege either that the information being copied does not fall within the general scope of copyright or that the cause of action contains an extra element.

The *New York Times* has brought hot-news misappropriation claims against Microsoft and OpenAI, in addition to its copyright claims[728]  The *Times* is a closer fit for misappropriation than many other generative-AI copyright plaintiffs, because it "gathers information, which often takes the form of time-sensitive breaking news, for its content at a substantial cost."[729]  It is a news organization, much like the AP.

Still, these claims are unlikely to succeed.  The training and deployment of most generative-AI systems take place on such a drawn-out time scale that any breaking-news value in the training data will have been long since exhausted by the time anyone uses them.  Hot news is ice cold six months later.  (By contrast, INS papers reported news the same day as AP papers.)  Mere competition with the *Times* for readership is likely not enough to generate a misappropriation claim that can survive preemption.

The *Times* also brings a claim that the defendants are misappropriating shopping recommendations from its Wirecutter subsite.[730]  Here, the complaint emphasizes that the removal deprives the *Times* of affiliate revenue.  But here too copyright preemption gives the *Times*'s theory of liability a difficult hill to climb.  Wirecutter reviews are clearly fixed works of authorship, and there is no element here — creation at cost, duplication by competitors, value-capture by defendants — that is not also present in a typical copyright-infringement claim.

---

A continuously-updated, real-time generative-AI system might raise more difficult hot-news issues. But it is not clear that these issues would be different in kind from those already presented by news aggregators and search engines, both of which already provide rapid access to news reported by others, often without payment and with less attribution than news organizations would prefer. If hot-news misappropriation is not the answer there, it is unlikely to be the answer here, either.[731]

### III. Which Way from Here?

The generative-AI supply chain is extremely complex. So is copyright law. Putting the two of them together multiplies the intricacy. Two unsettling conclusions follow from this radiating complexity.

First, because of the complexity of the *supply chain*, it is not possible to make accurate sweeping statements about the copyright legality of generative AI. Too much depends on the details of the specific system in question. All the pieces matter, from the curatorial choices in the training dataset, to the training algorithm, to the deployment environment, to the prompt supplied by the user. Courts will inevitably have to work through these details in numerous lawsuits, as they develop doctrines to distinguish among different systems and uses.

Second, because of the complexity of *copyright law*, there is enormous play in the joints. In particular, substantial similarity, indirect infringement, fair use, and remedies all have open-ended tests that can reach different results depending on the facts a court emphasizes and the conclusions it draws. This complexity gives courts the flexibility to deal with the many variations in the supply chain. Paradoxically, it also gives courts the freedom to reach any of several different plausible conclusions about a generative-AI system.

In this Part, we explore some of the ways that courts might try to use their discretion to apply copyright law to generative AI,[732] and then discuss some of the considerations that courts should keep in mind as they do.[733]

### A. Possible Outcomes

Although the details of which generative-AI systems fall into which boxes may vary, there are a few boxes that courts may find it appealing to sort them

---

**731**. *See generally* Joseph A. Tomain, *First Amendment, Fourth Estate and Hot News: Misappropriation is Not a Solution to the Journalism Crisis*, 2012 Mich. St. L. Rev. 769.

**732**. *See infra* Part III.A.

**733**. *See infra* Part III.B.

into. In this section, we sketch a few of the possible copyright regimes that might result.

### No Liability

First, courts might settle on a regime of no liability for services and their users. Anything produced by a generative-AI system would be categorically legal, under a combination of no substantial similarity and fair use. The result would be that models and services would also be categorically legal — there would be no primary liability for them to be indirectly liable for, and intermediate nonexpressive fair use would shield them in any event. Training datasets would also usually be legal as well (except perhaps in cases of blatant infringement like Books3).[734] They would be fair -use inputs to non-infringing downstream stages of the supply chain.

This regime is clear and simple. It would also be unstable. While such an outcome might make sense for some generative-AI systems, it seems both unworkable and undesirable for others, including systems trained specifically to emulate the styles of particular creators, and systems that use retrieval-augmented generation, which find matching works and reproduce them nearly exactly.[735] If all generative AI were categorically legal, then developers would plausibly start adding generative components to other systems in order to launder copyrighted works through them. The endpoint could be the effective collapse of copyright. On the assumption that this is not an outcome that courts would willingly preside over, then, a blanket no-liability regime seems unlikely. Instead, courts would be more likely to find at least some infringement — so the question becomes where to draw the line.

### Liability for Generations Only

Second, courts could draw a line between generative-AI services and the users of those services. In this regime, only generations would be treated as infringing, and then only when a user made some external use of them.[736] In this world, generative-AI systems would be creative tools like Photoshop.[737]

---

734. Knibbs, *supra* note 565; Reisner, *supra* note 565; Complaint, Kadrey v. Meta Platforms, Inc., No. 3:23-cv-03417 (N.D. Cal. July 7, 2023).

735. *See supra* note 265 and accompanying text.

736. Here, we use the term "user" broadly. A user could be a customer using a web application to produce a generation, a developer using an API to produce a generation in their own code, a developer using an API to produce a generation for a company, etc.

737. Sometimes literally so. *See* Adobe, *Experience the Future of Photoshop With Generative Fill* (July 27, 2023), https://helpx.adobe.com/photoshop/using/generative-fill.html.

The user would be responsible for making sure that anything they create with the tools is noninfringing, but the tools would be shielded under something like a strong *Sony* rule, assembled out of a combination of no substantial similarity, no indirect infringement, and/or fair use. This result might be unfair to users whose infringements resulted from systems producing generations that reproduce material in the underlying model's training dataset, through no choice or fault of their own. But this is arguably the same kind of situation that some courts currently countenance when they hold that users can be liable for embedding images from Instagram even though Instagram is not liable for hosting those images.[738] And this is also precisely the type of situation that indemnification of users could help address.

The main difficulty with this regime would be policing against systems designed specifically for infringement. Something like the *Grokster* rule, carefully followed, might suffice. The providers of a service that was geared to produce infringing outputs could be held liable. So could the publishers or deployers of a model that had been trained or fine-tuned to optimize its effectiveness specifically for infringing uses. So could the curator of a dataset that included only or primarily infringing works, or was intentionally organized to meet the needs of a model known to be intentionally trained for infringement. At every stage, a party would be held responsible only for its own actions specifically directed towards increasing the use of a system for infringement, with no substantial noninfringing purpose.

### *Notice and Removal*

Third, courts could treat generative-AI services as generally legal in themselves, but require them to respond to knowledge of specific infringements under a *Napster*-like rule. One plausible doctrinal route to this regime would be to treat infringing generations as creating direct liability for users and only indirect liability for service providers. Another would use fair use to shield service providers as long as they took reasonable overall precautions, including responding when they had sufficient knowledge of infringement. And a third would be to find liability but craft an injunction that only required services to act against infringement they were aware of.

Regardless of which of these doctrinal routes a court took, there would be an inevitable gravitational force pulling the provider's duties towards the duties of a service provider under section 512(c) or (d). This is not because Section 512 applies to generative-AI services. In most cases, it almost certainly does not.[739] Instead, the Section 512 doctrines may be a convergence point because courts have now had two decades of experience — which means two decades of precedents — with the Section 512 safe harbors. These precedents have come to set expectations — among copyright owners, in the technology industry, in the copyright bar, and in the judiciary — for what legally "responsible" behavior by an online intermediary looks like. A generative-AI service operator that does not appear to be making a good-faith effort to achieve something like this system may strike a court as intending to induce infringement, not making a good-faith effort to comply with an injunction, etc.

---

738. *E.g.,* Sinclair v. Ziff Davis, LLC, 454 F.Supp.3d 342 (S.D.N.Y. 2020).
739. *See supra* Part II.G.

If courts do end up recreating a notice-and-takedown regime, they would likely settle on familiar elements: a way for copyright owners to give notice of infringement, block infringing generations on notice, block infringing generations on actual knowledge, block infringing generations on red-flag knowledge, avoid having a business model that directly ties income to infringement, and terminate the abilities of repeat infringers to continue making generations. These would probably not be notices directed to specific generations by named users, which would be difficult to detect and track. Instead, they would involve copyright owners identifying copyrighted works and demanding that the generative-AI service operator prevent generations that are substantially similar to those works. Some of those works might be identified based on known outputs that are recognizably similar to suspected inputs. But others might simply involve copyright owners handing over to service operators large catalogs of works to block, much as they currently do with ContentID on YouTube.

This is a very difficult technical problem. It would be much harder for a generative-AI system to implement than it is for a hosting platform to implement Section 512 compliance. The reason is that a notice directed to a hosting provider under Section 512(c) must include "Identification of the material that is claimed to be infringing . . . and information reasonably sufficient to permit the service provider to locate the material."[740] A valid notice is a roadmap; it tells the hosting provider exactly what to take down to comply. That material already exists, and the hosting provider can compare it to the copyrighted work to verify that they are substantially similar. But a notice to a generative-AI system is a notice against future generations, which may be different from each other and resemble the copyrighted work in different ways. Filtering for this kind of much more inexact match is much harder technically.

That said, matching material against a catalog of copyrighted works is a problem that has been very approximately solved by major social networks,

---

**740**. 17 U.S.C. § 512(c)(3)(A)(i)(i)(i).

which use perceptual hashing to prevent the upload of various kinds of identified content. Generative-AI companies could at least add similar perceptual-hash-driven filtering to the outputs of their models, but clearly this would only solve part of the problem.[741] The challenges of implementing removal for models are even harder. A service can add filters on the input and output sides — monitoring prompts and scanning outputs. It can also fine-tune or align the model, or provide it with an overall prompt that instructs the model to respond in ways that reduce its propensity to infringe.

But a model by itself does not implement these controls. The model cannot control how it is prompted or what the user does with the output. The model cannot stop anyone from fine-tuning it to remove its guardrails. Further, there is no simple analogue for takedown in generative-AI models. It remains an active and unsolved area of research to figure out how to remove a particular training example's influence from a model's parameters.[742] Absent the ability to do so, the safest bet is to retrain the model from scratch. Due to the time and expense required to retrain a model, it will often be infeasible to retrain it simply to remove infringing works, and completely unworkable to retrain on each new notice.

Courts could respond to this difficulty in one of two ways. If they have sympathy for model trainers, they could apply the *Sony* rule, and hold that it is not infringement to distribute a trained model as a set of parameters (as Stability AI's releases have been). The fact that the model is used by others for infringing purposes would be counterbalanced by the substantial non-infringing uses, leading to immunity under *Sony*. This might not always be an attractive business model, because it might be hard for buyers to monetize these models and because of the ease of copying and further redistributing the models, but it could at least exist legally. And truly open-source models would generally be allowed.

But if courts had less sympathy for model trainers, they might hold that the difficulty of complying with removal notices is not an excuse. On this view, the model trainer chose to create a model that could be used for substantial infringement, and to hopelessly commingle infringing and noninfringing material. If so, then it would generally not be legal to distribute a model that was trained on unlicensed works and had infringing outputs, at least once those works they were based on were pointed out. It would be legal to train a model, but the trainer would need to take care that the

---

model was only deployed in a safe environment with sufficient guardrails to prevent infringement. (This is the approach generally taken by OpenAI, for example.[743])

In this world, open-source models would be extremely risky. As a result, there would likely be a split between two classes of models. Some proprietary models might train on unlicensed works and be deployed only in closed services with carefully designed guardrails. Open-source models would be trained only on public-domain and openly-licensed works, or be trained using very conservative methods to attempt ensure that extremely little copyrighted material was memorized.

A notice-and-removal regime also has implications for training datasets. A dataset provider cannot pull back these works for which it receives a notice from others who have already used those works for training. But it can delete the works from the dataset it makes available to others going forward. (For an open-source dataset, or one that has been leaked, this second option may be futile, as others will still have copies of the dataset that they can share.) Compared with a model, it is much easier to remove a work from a training dataset; one searches for the work and removes it. Indeed, one could use exact hashing rather than perceptual hashing and still get substantial efficacy in removing a large number of identified works from the dataset — or, for datasets compiled from web crawls or other sources, remove works by tracing their provenance through into the part of the dataset they have ended up in. This makes datasets comparatively more attractive as removal targets, both because they are upstream from many models and because it is easier to define and enforce enforceable removal obligations.

### Infringing Models

A fourth possibility is that courts would hold that some or all generative-AI services are illegal because the models themselves infringe. This outcome is an existential threat to many model trainers and service providers; it essentially makes their operations *per se* copyright infringement. It is also the outcome being sought by the class-action plaintiffs in high-profile lawsuits against OpenAI, Stability AI, and some of their partners. In this regime, the

---

**743**. The sufficiency of OpenAI's guardrails is currently hotly contested, due to the frequency of successful adversarial behaviors and security attacks that are able to circumvent these guardrails. *See supra* Part II.E (for an example of a user circumventing mechanisms to prevent the generation of potentially copyright-infringing illustrations of Calvin and Hobbes). Nasr, Carlini & Hayase et al., *supra* note 1 (for an attack on ChatGPT that breaks alignment and gets the system to regurgitate training data at relatively enormous rates).

most important component of copyright law would quickly become licensing. Models could only be trained on data that had been licensed from the copyright owners, and the terms under which those models and their generations could be used would have to be negotiated as part of the licensing agreement. Each model would have a fully licensed training dataset, and the question of infringement would not arise except in cases where there were infringing works in the dataset itself or some other failure of quality control somewhere along the supply chain.

## B. Lessons

Having discussed what courts and policymakers could do, we now consider what they should do. In keeping with our bottom line — *the generative-AI supply chain is too complicated to make sweeping rules prematurely* — we offer a few general observations about the overall shape of copyright and generative AI that courts and policymakers should keep in mind as they proceed.

First, *copyright touches every part of the generative-AI supply chain.* Every stage from training data to alignment can make use of copyrighted works. Generative AI raises many other legal issues: Can a generative-AI system commit defamation?[744] Can a generative-AI system do legal work,[745] and should they be allowed to?[746] But these issues mainly have to do with the outputs of a generative-AI system. In contrast, copyright pervades every step of the process; copyright is present every time anyone anywhere in the supply chain makes a decision. Copyright cannot be ignored.[747]

---

[744]. Eugene Volokh, *Large Libel Models? Liability for AI Output*, 3 J. Free Speech L. 489 (2023); Jon Garon, *An AI's Picture Paints a Thousand Lies: Designating Responsibility for Visual Libel*, 3 J. Free Speech L. 425 (2023); Nina Brown, *Bots Behaving Badly: A Products Liability Approach to Chatbot-Generated Defamation*, 3 J. Free Speech L. 389 (2023); Derek Bambauer & Mihai Surdeanu, *Authorbots*, 3 J. Free Speech L. 375 (2023); Peter Henderson, *Tatsunori Hashimoto, and Mark Lemley, Where's the Liability in Harmful AI Speech?*, 3 J. Free Speech L. 589 (2023).

[745]. Jonathan H. Choi, Kristen E. Hickman, Amy Monahan & Daniel Schwarcz, *ChatGPT Goes to Law School*, 2023 J. Legal Educ. (forthcoming 2023).

[746]. Mata v. Avianca, No. 22-cv-1461 (S.D.N.Y. June 22, 2023).

[747]. Copyright is not the only socially relevant concept that pervades the supply chain. The supply-chain framing illuminates other legal and ethical challenges as well, such as developer responsibility for harmful uses, *see, e.g.,* David Gray Widder & Dawn Nafus, *Dislocated Accountabilities in the "AI Supply Chain": Modularity and Developers' Notions of Responsibility*, June 15, 2023 volume Big Data & Soc'y 1, or for the environmental and labor considerations involved in AI training, *see, e.g.,* David Gray Widder & Richmond Wong, Thinking Upstream: Ethics and Policy Opportunities in AI Supply Chains (2023) (unpublished manuscript), https://arxiv.org/abs/2303.07529.

Second, and relatedly, *copyright concerns cannot be localized* to a single link in the supply chain. We have argued, time and time again, that decisions made by one actor can affect the copyright liability of another, potentially far away actor in the supply chain. Whether an output looks like Snoopy or like a generic beagle depends on what images were collected in a dataset, which model architecture and training algorithms are used, how trained models are fine-tuned and aligned, how models are embedded in deployed services, what the user prompts with, etc. Every single one of these steps could be under the control of a different person, company, or organization.

Third, *design choices matter*. Every actor in the generative-AI supply chain is in a position to make choices that affect their copyright exposure, and others'. These are obvious choices about copyright, like whether to train on unlicensed data (which can affect downstream risks), and how to respond to notices that a system is producing infringing outputs (which can affect upstream risks). But subtler architectural choices matter, too. Different settings on a training algorithm can affect how much the resulting model will memorize specific works. Different deployment environments can affect whether users have enough control over a prompt to steer a system towards infringing outputs. Copyright law will necessarily have to engage with these choices — as will AI policy more generally.

Fourth, *fair use is not a silver bullet*. For a time, it seemed that training and using AI models would often constitute fair use. In such a world, AI development is generally a low-risk activity, at least from a copyright perspective. Yes, training datasets and models and systems may all include large quantities of copyrighted works — but they will never be shown to users. Generative AI scrambles this assumption. The serious possibility that some generations will infringe means that the fair-use analysis at every previous stage of the supply chain is up for grabs again.

Fifth, *generative AI does not make the ordinary business of copyright law irrelevant*. Courts will still need to make plenty of old-fashioned, retail judgments about individual works — e.g., how much does this image resemble Elsa and Anna in particular, rather than generic tropes of fantasy princesses? To decide these cases, courts will need to avoid getting distracted by the shininess of new technologies and chasing after inappropriately categorical new rules. Similarity is similarity, proof of copying is proof of copying, transformation in content is transformation in content. Courts *must* leave themselves room to continue making these retail judgments on a case-by-case basis, responding to the specific facts before them, just as they always have. Perhaps, in the fullness of time, as society comes to understand what uses generative AI can be put to and with what consequences, it will reconsider the very fundamentals of copyright law. But until that day, we must live with the copyright system we have. And that system cannot function unless courts are able to say that some generative-AI systems and generations infringe, and others do not.

Sixth, *analogies can be misleading.* There are plenty of analogies for generative AI ready to hand. A generative-AI model or system is like a search engine, or like a website, or like a library, or like an author, or like any number of other people and things that copyright has a well-developed framework for dealing with.[748] These analogies are useful, but we wish to warn against treating any of them as definitive. As we have seen, generative AI is and can consist of many things. It is also literally a generative technology: it can be put to an amazingly wide variety of uses.[749] And one of the things about generative technologies is that they cause convergence;[750] precisely because they can emulate many other technologies, they blur the boundaries between things that were formerly distinct. Generative AI can be like a search engine, and also like a website, a library, an author, and so on. Prematurely accepting one of these analogies to the exclusion of the others would mean ignoring numerous relevant similarities — precisely the opposite of what good analogical reasoning is supposed to do.

## IV. Conclusion

Our conclusion is simple. "Does generative AI infringe copyright?" is not a question that has a yes-or-no answer. There is currently no blanket rule that determines which participants in the generative-AI supply chain are copyright infringers. The underlying technologies and systems are too diverse to be treated identically, and copyright law has too many open decision points to provide clear answers. Our hope is that the supply-chain framing provides a clear and precise mechanism for understanding this diversity and, in turn, for reasoning about the various legal consequences.

Copyright is not the only, or the best, or the most important way of confronting the policy challenges that generative AI poses. But copyright is here, and it is asking good questions about how generative-AI systems are created, how they work, how they are used, and how they are updated. These ques-

---

**748**. *See supra* Part I.A (for why generations are not like collages).

**749**. Jonathan Zittrain, The Future of the Internet – And How to Stop It (2008) (developing theory of generative technologies); Cooper, Lee, Grimmelmann & Ippolito et al., *supra* note 22 (connecting Zittrain's theory of generative technologies with generative AI).

**750**. *See generally* Tejas N. Narechania, *Convergence and a Case for Broadband Rate Regulation,* 37 Berkeley Tech. L.J. 339 (2022) (discussing convergence caused by the Internet).

tions deserve good answers, or failing that, the best answers our copyright system is equipped to give.

Amit Alfassy, Anna Rogers, Ariel Kreisberg Nitzav, Canwen Xu, Chenghao Mou, Chris Emezue, Christopher Klamm, Colin Leong, Daniel van Strien, David Ifeoluwa Adelani, Dragomir Radev, Eduardo González Ponferrada, Efrat Levkovizh, Ethan Kim, Eyal Bar Natan, Francesco De Toni, Gérard Dupont, Germán Kruszewski, Giada Pistilli, Hady Elsahar, Hamza Benyamina, Hieu Tran, Ian Yu, Idris Abdulmumin, Isaac Johnson, Itziar Gonzalez-Dios, Javier de la Rosa, Jenny Chim, Jesse Dodge, Jian Zhu, Jonathan Chang, Jörg Frohberg, Joseph Tobing, Joydeep Bhattacharjee, Khalid Almubarak, Kimbo Chen, Kyle Lo, Leandro Von Werra, Leon Weber, Long Phan, Loubna Ben allal, Ludovic Tanguy, Manan Dey, Manuel Romero Muñoz, Maraim Masoud, María Grandury, Mario Šaško, Max Huang, Maximin Coavoux, Mayank Singh, Mike Tian-Jian Jiang, Minh Chien Vu, Mohammad A. Jauhar, Mustafa Ghaleb, Nishant Subramani, Nora Kassner, Nurulaqilla Khamis, Olivier Nguyen, Omar Espejel, Ona de Gibert, Paulo Villegas, Peter Henderson, Pierre Colombo, Priscilla Amuok, Quentin Lhoest, Rheza Harliman, Rishi Bommasani, Roberto Luis López, Rui Ribeiro, Salomey Osei, Sampo Pyysalo, Sebastian Nagel, Shamik Bose, Shamsuddeen Hassan Muhammad, Shanya Sharma, Shayne Longpre, Somaieh Nikpoor, Stanislav Silberberg, Suhas Pai, Sydney Zink, Tiago Timponi Torrent, Timo Schick, Tristan Thrush, Valentin Danchev, Vassilina Nikoulina, Veronika Laippala, Violette Lepercq, Vrinda Prabhu, Zaid Alyafeai, Zeerak Talat, Arun Raja, Benjamin Heinzerling, Chenglei Si, Davut Emre Taşar, Elizabeth Salesky, Sabrina J. Mielke, Wilson Y. Lee, Abheesht Sharma, Andrea Santilli, Antoine Chaffin, Arnaud Stiegler, Debajyoti Datta, Eliza Szczechla, Gunjan Chhablani, Han Wang, Harshit Pandey, Hendrik Strobelt, Jason Alan Fries, Jos Rozen, Leo Gao, Lintang Sutawika, M Saiful Bari, Maged S. Al-shaibani, Matteo Manica, Nihal Nayak, Ryan Teehan, Samuel Albanie, Sheng Shen, Srulik Ben-David, Stephen H. Bach, Taewoon Kim, Tali Bers, Thibault Fevry, Trishala Neeraj, Urmish Thakker, Vikas Raunak, Xiangru Tang, Zheng-Xin Yong, Zhiqing Sun, Shaked Brody, Yallow Uri, Hadar Tojarieh, Adam Roberts, Hyung Won Chung, Jaesung Tae, Jason Phang, Ofir Press, Conglong Li, Deepak Narayanan, Hatim Bourfoune, Jared Casper, Jeff Rasley, Max Ryabinin, Mayank Mishra, Minjia Zhang, Mohammad Shoeybi, Myriam Peyrounette, Nicolas Patry, Nouamane Tazi, Omar Sanseviero, Patrick von Platen, Pierre Cornette, Pierre François Lavallée, Rémi Lacroix, Samyam Rajbhandari, Sanchit Gandhi, Shaden Smith, Stéphane Requena, Suraj Patil, Tim Dettmers, Ahmed Baruwa, Amanpreet Singh, Anastasia Cheveleva, Anne-Laure Ligozat, Ar-jun Subramanian, Aurélie Névéol, Charles Lovering, Dan Garrette, Deepak Tunuguntla, Ehud Reiter, Ekaterina Taktasheva, Ekaterina Voloshina, Eli Bogdanov, Genta Indra Winata, Hailey Schoelkopf, Jan-Christoph Kalo, Jekaterina Novikova, Jessica Zosa Forde, Jordan Clive, Jungo Kasai, Ken Kawamura, Liam Hazan, Marine Carpuat, Miruna Clinciu, Najoung Kim, Newton Cheng, Oleg Serikov, Omer Antverg, Oskar van der Wal, Rui Zhang, Ruochen Zhang, Sebastian Gehrmann, Shachar Mirkin, Shani Pais, Tatiana Shavrina, Thomas Scialom, Tian Yun, Tomasz Limisiewicz, Verena Rieser, Vitaly Protasov, Vladislav Mikhailov, Yada Pruksachatkun, Yonatan Belinkov, Zachary Bamberger, Zdeněk Kasner, Alice Rueda, Amanda Pestana, Amir Feizpour, Ammar Khan, Amy Faranak, Ana Santos, Anthony Hevia, Antigona Unldreaj, Arash Aghagol, Arezoo Abdollahi, Aycha Tammour, Azadeh HajiHosseini, Bahareh Behroozi, Benjamin Ajibade, Bharat Saxena, Carlos Muñoz Ferrandis, Daniel McDuff, Danish Contractor, David Lansky, Davis David, Douwe Kiela, Duong A. Nguyen, Edward Tan, Emi Baylor, Ezinwanne Ozoani, Fatima Mirza, Frankline Ononiwu, Habib Rezanejad, Hessie Jones, Indrani Bhattacharya, Irene Solaiman, Irina Sedenko, Isar Nejadgholi, Jesse Passmore, Josh Seltzer, Julio Bonis Sanz, Livia Dutra, Mairon Samagaio, Maraim Elbadri, Margot Mieskes, Marissa Gerchick, Martha Akinlolu, Michael McKenna, Mike Qiu, Muhammed Ghauri, Mykola Burynok, Nafis Abrar, Nazneen Rajani, Nour Elkott, Nour Fahmy, Olanrewaju Samuel, Ran An, Rasmus Kromann, Ryan Hao, Samira Alizadeh, Sarmad Shubber, Silas Wang, Sourav Roy, Sylvain Viguier, Thanh Le, Tobi Oyebade, Trieu Le, Yoyo Yang, Zach Nguyen, Abhinav Ramesh Kashyap, Alfredo Palasciano, Alison Callahan, Anima Shukla, Antonio Miranda-Escalada, Ayush Singh, Benjamin Beilharz, Bo Wang, Caio Brito, Chenxi Zhou, Chirag Jain, Chuxin Xu, Clémentine Fourrier, Daniel León Periñán, Daniel Molano, Dian Yu, Enrique Manjavacas, Fabio Barth, Florian Fuhrimann, Gabriel Altay, Giyaseddin Bayrak, Gully Burns, Helena U. Vrabec, Imane Bello, Ishani Dash, Jihyun Kang, John Giorgi, Jonas Golde, Jose David Posada, Karthik Rangasai Sivaraman, Lokesh Bulchandani, Lu Liu, Luisa Shinzato, Madeleine Hahn de Bykhovetz, Maiko Takeuchi, Marc Pàmies, Maria A Castillo, Marianna Nezhurina, Mario Sänger, Matthias Samwald, Michael Cullan, Michael Weinberg, Michiel De Wolf, Mina Mihaljcic, Minna Liu, Moritz Freidank, Myungsun Kang, Natasha Seelam, Nathan Dahlberg, Nicholas Michio Broad, Nikolaus Muellner, Pascale Fung, Patrick Haller, Ramya Chandrasekhar, Renata Eisenberg, Robert Martin, Rodrigo Canalli, Rosaline Su, Ruisi